\documentclass[10pt,twocolumn,letterpaper]{article}

\usepackage[pagenumbers]{cvpr}
\usepackage{iftex}
\ifXeTeX\usepackage{fontspec}
\fi

\usepackage{amsmath,amssymb,amsfonts}
\usepackage{graphicx}
\usepackage{booktabs}
\usepackage{longtable}
\usepackage{multirow}
\usepackage{makecell}
\usepackage{array}
\usepackage{tabularx}
\usepackage{xcolor}
\usepackage{textcomp}      %
\usepackage{caption}
\usepackage{xspace}
\usepackage{tikz}
\usetikzlibrary{arrows.meta,positioning,fit,backgrounds,calc,shapes.callouts,shapes.geometric,shapes.misc}
\usepackage{fontawesome5}
\usepackage{pifont}
\usepackage{enumitem}
\usepackage{placeins}\usepackage{needspace}
\usepackage[pagebackref,breaklinks,colorlinks,citecolor=blue,linkcolor=blue]{hyperref}

\usepackage{stfloats}%
\usepackage{colortbl}
\usepackage[most]{tcolorbox}
\definecolor{bestgreen}{HTML}{DDF0E2}\definecolor{dGood}{HTML}{C6E5E1}\definecolor{dBad}{HTML}{FBD9B5}%
\DeclareRobustCommand{\swatch}[2]{{\setlength{\fboxsep}{1pt}\colorbox{#1}{#2}}}%
\definecolor{rqTeal}{HTML}{123F4A}\definecolor{rqPale}{HTML}{E6F1F3}
\definecolor{vsViseek}{HTML}{77AADD}\definecolor{vsRag}{HTML}{CC6677}\definecolor{vsClosed}{HTML}{CBB8E0}
\definecolor{vsRagSFT}{HTML}{BBCC33}\definecolor{vsZS}{HTML}{EEDD88}\definecolor{vsSFT}{HTML}{EE8866}
\definecolor{vsOwn}{HTML}{44BB99}\definecolor{vsOneBox}{HTML}{FFAABB}\definecolor{vsMask}{HTML}{AAAA00}
\definecolor{vsInk}{HTML}{2B2B2B}\definecolor{vsLight}{HTML}{F4F0FA}
\definecolor{vsCOne}{HTML}{E8956F}\definecolor{vsCTwo}{HTML}{5E8FC7}\definecolor{vsCThree}{HTML}{2E9E80}
\definecolor{vsGrayCtrl}{HTML}{D9B8A0}\definecolor{vsRandCtrl}{HTML}{9F86C0}\definecolor{vsExt}{HTML}{D4A373}
\definecolor{vsViseekP}{HTML}{E7F0FA}%
\providecommand{\sw}[1]{{\setlength{\fboxsep}{0pt}\setlength{\fboxrule}{0.3pt}\fcolorbox{black!45}{#1}{\rule{0pt}{1.25ex}\rule{1.25ex}{0pt}}}\,}
\definecolor{vsMCS}{HTML}{AA4499}\definecolor{vsWQ}{HTML}{44AA99}\definecolor{vsSAGE}{HTML}{DDAA33}%

\newtcolorbox{insight}{enhanced, colback=rqPale, colframe=rqTeal, boxrule=0pt, leftrule=2.2pt, sharp corners, left=4pt,
  right=3pt, top=1.5pt, bottom=1.5pt, before skip=4pt, after skip=5pt, fontupper=\small}

\def\confName{CVPR}
\def\confYear{2027}

\title{ViBES: Learning Budgeted Multimodal Evidence Selection\\for Agentic Reasoning}

\author{Shreyan Ganguly$^{*}$ \quad Tirtho Roy$^{*}$ \quad Zhanhong Jiang \quad Adarsh Krishnamurthy \quad Soumik Sarkar$^{\dagger}$\\
Iowa State University, Ames, IA, USA \quad {\small $^{*}$Equal contribution \quad $^{\dagger}$Corresponding author: \texttt{soumiks@iastate.edu}}\\
{\small Project page: \href{https://vibes-lens.github.io/}{\texttt{https://vibes-lens.github.io}}}}

\usepackage{fvextra}
\colorlet{tMCS}{vsMCS!9}\colorlet{tWQ}{vsWQ!9}\colorlet{tSAGE}{vsSAGE!12}%

\definecolor{vsMMQA}{HTML}{4477AA}%
\colorlet{dqMCS}{vsMCS!85!black}\colorlet{dqWQ}{vsWQ!80!black}\colorlet{dqSAGE}{vsSAGE!75!black}\colorlet{dqMMQA}{vsMMQA!90!black}
\newlength{\dqtw}
\newlength{\dqth}
\newlength{\dqcardh}
\newcommand{\dqid}[1]{{\tiny\color{black!50}\texttt{#1}}}
\newcommand{\dqtile}[1]{\includegraphics[width=\dqtw,height=\dqth]{figs/dataqa/#1}}
\newcommand{\dqtext}[2]{{\setlength{\fboxsep}{0pt}\colorbox{#1!10}{\parbox[b][\dqth][c]{\dqtw}{\centering
  \parbox{\dimexpr\dqtw-6pt\relax}{\centering\tiny\itshape #2}}}}}
\newcommand{\dqtiles}[4]{\noindent#1\hfill#2\\[-2pt]%
  \makebox[\dqtw][c]{\tiny\color{black!55}#3}\hfill\makebox[\dqtw][c]{\tiny\color{black!55}#4}\par\vspace{2pt}}
\newcommand{\dqline}[2]{\noindent\textbf{#1}\ \,#2\par\vspace{2pt}}
\newcommand{\dqcard}[5]{%
\begin{tcolorbox}[enhanced, nobeforeafter, width=0.246\textwidth, height=\dqcardh, valign=top, sharp corners,
  colback=black!2, colframe=#1, boxrule=0pt, toprule=1.3pt, boxsep=0pt, left=3.5pt, right=3.5pt, top=2.5pt, bottom=2pt,
  fontupper=\footnotesize]
{\scriptsize\textbf{\textcolor{#1}{#2}}\hfill{\tiny\textcolor{black!55}{#3}}}\par\vspace{2.5pt}
#4\raggedright #5
\end{tcolorbox}}

\def\confName{CVPR}
\def\confYear{2027}

\begin{document}
\raggedbottom %
\makeatletter\renewcommand\paragraph{\@startsection{paragraph}{4}{\z@}{1.0ex \@plus 0.3ex \@minus 0.2ex}{-0.8em}{\normalfont\normalsize\bfseries}}\makeatother
\twocolumn[{%
\renewcommand\twocolumn[1][]{#1}%
\maketitle
\vspace{-2.8em}
\begin{center}
\resizebox{\textwidth}{!}{\definecolor{fInk}{HTML}{2B2B2B}\definecolor{fHead}{HTML}{3F4A5A}\definecolor{fZs}{HTML}{EEDD88}\definecolor{fZsB}{HTML}{6B7280}\definecolor{fSft}{HTML}{EE8866}
\definecolor{fRet}{HTML}{2E9E80}\definecolor{fIns}{HTML}{8F7A16}\definecolor{fInsI}{HTML}{C9B44A}\definecolor{fCrop}{HTML}{AA4499}\definecolor{fAgt}{HTML}{7A7A7A}
\definecolor{fOwn}{HTML}{77AADD}\definecolor{fLrn}{HTML}{3F78B4}\definecolor{fOk}{HTML}{1F7A4F}\definecolor{fErr}{HTML}{B03A2E}
\definecolor{fTool}{HTML}{F4F6F8}\definecolor{fEval}{HTML}{ECECEC}\definecolor{fShare}{HTML}{EEF1F4}
\begin{tikzpicture}[font=\sffamily\fontsize{6.3}{7.3}\selectfont, x=1cm, y=1cm, line cap=round,
  panel/.style={draw=black!25, rounded corners=4pt, line width=0.6pt, inner sep=0pt, anchor=north west},
  hd/.style={font=\sffamily\bfseries\small, anchor=west, inner sep=0pt, text=white},
  colhd/.style={font=\sffamily\bfseries\footnotesize, anchor=base west, inner sep=0pt, text=white},
  agent/.style={draw=black!50, line width=0.6pt, rounded corners=4pt, fill=white},
  tool/.style={draw=black!18, line width=0.5pt, rounded corners=4pt, fill=fTool},
  icon/.style={anchor=center, inner sep=0pt, font=\normalsize},
  sic/.style={anchor=center, inner sep=0pt, font=\scriptsize},
  cnt/.style={anchor=north, inner sep=0pt, font=\sffamily\bfseries\fontsize{5.6}{6}\selectfont, text=black!60},
  txt/.style={anchor=north west, inner sep=0pt, align=left, text=fInk,
              execute at begin node={\hyphenpenalty=10000\exhyphenpenalty=10000}},
  mid/.style={anchor=west, inner sep=0pt, align=left, text=fInk,
              execute at begin node={\hyphenpenalty=10000\exhyphenpenalty=10000}},
  evaltag/.style={fill=fEval, text=black!70, rounded corners=1.5pt, inner sep=1.4pt, anchor=west},
  img/.style={inner sep=0pt, anchor=west},
  pic/.style={inner sep=0pt, anchor=north west},
  arr/.style={-{Triangle[length=1.1mm,width=1.4mm]}, black!45, line width=0.6pt},
  nn/.style={circle, inner sep=0pt, minimum size=1.1mm, draw=fLrn, line width=0.5pt, fill=white},
  pt/.style={anchor=west, inner sep=0pt, text=fInk, font=\sffamily\fontsize{6.8}{8}\selectfont},
  pl/.style={anchor=west, inner sep=0pt, text=fInk, font=\sffamily\fontsize{5.8}{7}\selectfont},
  pa/.style={anchor=east, inner sep=0pt, text=black!55, font=\sffamily\fontsize{5.4}{6}\selectfont},
  pc/.style={anchor=north, inner sep=0pt, text=fInk, font=\sffamily\fontsize{5.9}{7}\selectfont},
  pv/.style={anchor=south, inner xsep=0pt, inner ysep=0.3pt, fill=white, text=black!70, font=\sffamily\fontsize{4.8}{6}\selectfont},
  pvb/.style={anchor=south, inner xsep=0pt, inner ysep=0.3pt, fill=white, text=fInk, font=\sffamily\bfseries\fontsize{4.8}{6}\selectfont},
  pd/.style={anchor=south, inner sep=0pt, text=fLrn, font=\sffamily\bfseries\fontsize{5.9}{7}\selectfont},
]
\providecommand{\nt}{}\renewcommand{\nt}[1]{\textcolor{black!45}{note:}~\textit{#1}}
\providecommand{\err}{}\renewcommand{\err}[1]{\textcolor{fErr}{\faIcon{times-circle}~#1}}
\node[panel, fill=white, minimum width=14.0cm, minimum height=7.84cm] at (0,0) {};
\begin{scope}[yshift=0.44cm]
\node[icon, text=black!60] at (0.38,-1.14) {\faIcon{user}};
\draw[agent, draw=black!35] (0.65,-0.52) rectangle (10.05,-1.76);
\node[mid, font=\sffamily\bfseries\fontsize{7.6}{9}\selectfont] at (0.8,-1.14) {Which disease does the soybean plant in the input image have?};
\node[pic] at (9.15,-0.57) {\includegraphics[height=1.14cm]{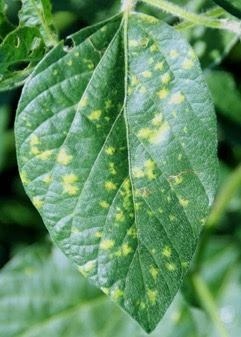}};
\begin{scope}[yshift=-0.06cm]
\fill[fZs, rounded corners=3pt] (0.1,-1.76) rectangle (4.75,-2.04);
\node[colhd, text=fInk] at (0.2,-1.975) {A.\ Zero-shot agent};
\fill[fOwn, rounded corners=3pt] (4.9,-1.76) rectangle (9.4,-2.04);
\node[colhd, text=fInk] at (5.0,-1.975) {B.\ ViBES};
\fill[fLrn, rounded corners=3pt] (9.55,-1.76) rectangle (13.95,-2.04);
\node[colhd] at (9.65,-1.975) {C.\ ViBES-Lens};
\def\xa{0.62}\def\xb{4.7}\def\xi{0.36}
\draw[tool] (\xa,-2.1) rectangle (\xb,-2.92);
\node[icon, text=fRet] at (\xi,-2.3) {\faIcon{cog}}; \node[cnt] at (\xi,-2.49) {1/10};
\node[txt] at (0.72,-2.16) {Tool: \textcolor{fRet}{\textbf{Retrieve}}\\ question + input image\\ \textcolor{fRet}{\faIcon{file-alt}}~Septoria brown spot entry, photos};
\node[img] at (3.3,-2.4) {\includegraphics[height=0.42cm]{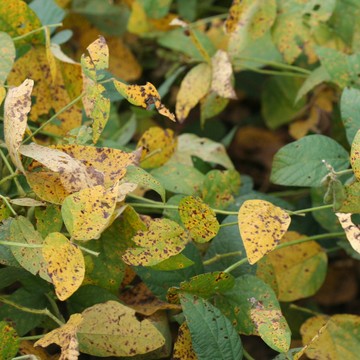}};
\node[img] at (3.77,-2.4) {\includegraphics[height=0.42cm]{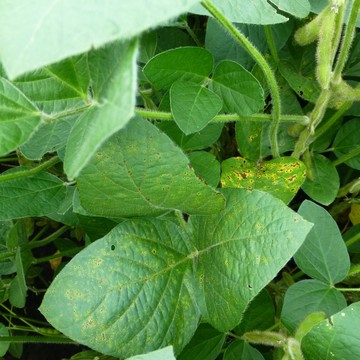}};
\draw[agent] (\xa,-2.98) rectangle (\xb,-3.56);
\node[icon, text=fAgt] at (\xi,-3.27) {\faIcon{robot}};
\node[txt, text width=3.9cm] at (0.72,-3.05) {\nt{\ldots{} `Septoria brown spot' with symptoms matching the input image \ldots}};
\draw[tool] (\xa,-3.62) rectangle (\xb,-5.1);
\node[icon, text=fInsI] at (\xi,-4.26) {\faIcon{cog}}; \node[cnt] at (\xi,-4.45) {2--7};
\node[txt, text width=3.9cm] at (0.72,-3.68) {Tools 2--7: \textcolor{fIns}{\textbf{Inspect}} 3 Septoria photos, \textcolor{fCrop}{\textbf{Crop}} the input 3 times};
\node[pic] at (0.72,-4.18) {\includegraphics[height=0.84cm]{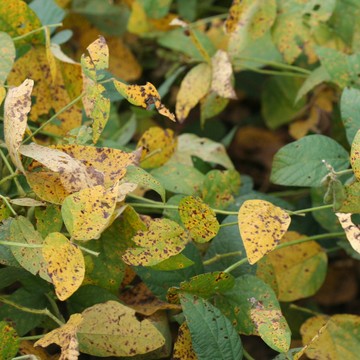}};
\node[pic, draw=fZsB, line width=0.8pt] at (1.66,-4.18) {\includegraphics[height=0.84cm]{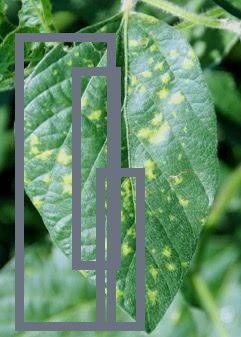}};
\node[txt, text width=2.2cm, text=black!60] at (2.4,-4.24) {Inspect and Crop alternate; no Retrieve by image};
\draw[agent] (\xa,-5.16) rectangle (\xb,-5.72);
\node[icon, text=fAgt] at (\xi,-5.44) {\faIcon{robot}};
\node[txt, text width=3.9cm] at (0.72,-5.23) {\nt{\ldots{} spots on the leaf, consistent with Septoria brown spot.}};
\draw[tool] (\xa,-5.78) rectangle (\xb,-6.68);
\node[icon, text=fRet] at (\xi,-6.07) {\faIcon{cog}}; \node[cnt] at (\xi,-6.26) {8/10};
\node[txt, text width=3.9cm] at (0.72,-5.84) {Tool: \textcolor{fRet}{\textbf{Retrieve}} ``Soybean disease with angular brown spots on leaves''\\ \textcolor{fRet}{\faIcon{file-alt}}~Septoria entry, Brown Spot photos};
\draw[tool] (\xa,-6.74) rectangle (\xb,-7.38);
\node[icon, text=fInsI] at (\xi,-6.94) {\faIcon{cog}}; \node[cnt] at (\xi,-7.13) {9/10};
\node[txt] at (0.72,-6.82) {Tool: \textcolor{fIns}{\textbf{Inspect}} a Brown Spot photo\\ \err{rejected: budget exhausted}};
\draw[agent] (\xa,-7.53) rectangle (\xb,-8.12);
\node[icon, text=fAgt] at (\xi,-7.825) {\faIcon{robot}};
\node[mid] at (0.72,-7.69) {\textcolor{black!45}{answer:}~\textbf{Septoria brown spot}};
\node[evaltag, anchor=east] at (4.64,-7.69) {\textcolor{fErr}{\ding{55}~wrong}};
\node[mid, text=black!55] at (0.72,-7.96) {9/10 calls};
\fill[fShare, rounded corners=4pt] (4.9,-2.06) rectangle (13.95,-5.42);
\def\xa{5.42}\def\xb{13.9}\def\xi{5.16}
\draw[tool] (\xa,-2.1) rectangle (\xb,-2.54);
\node[sic, text=fRet] at (\xi,-2.23) {\faIcon{cog}}; \node[cnt] at (\xi,-2.37) {1/10};
\node[mid] at (5.52,-2.32) {Tool: \textcolor{fRet}{\textbf{Retrieve}}, question + input image};
\node[img] at (9.55,-2.32) {\includegraphics[height=0.38cm]{figs/teaser/fig1s_r1a.jpg}};
\node[img] at (9.98,-2.32) {\includegraphics[height=0.38cm]{figs/teaser/fig1s_r1b.jpg}};
\node[mid] at (10.5,-2.32) {\textcolor{fRet}{\faIcon{file-alt}}~Septoria brown spot entry};
\draw[tool] (\xa,-2.6) rectangle (\xb,-3.04);
\node[sic, text=fRet] at (\xi,-2.73) {\faIcon{cog}}; \node[cnt] at (\xi,-2.87) {2/10};
\node[mid] at (5.52,-2.82) {Tool: \textcolor{fRet}{\textbf{Retrieve}} by input image};
\node[img] at (8.6,-2.82) {\includegraphics[height=0.38cm]{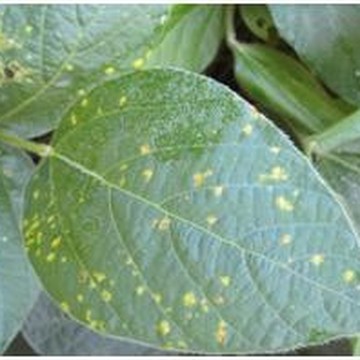}};
\node[img] at (9.03,-2.82) {\includegraphics[height=0.38cm]{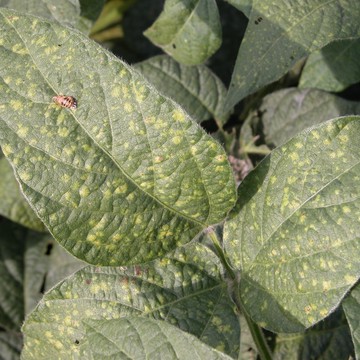}};
\node[img] at (9.46,-2.82) {\includegraphics[height=0.38cm]{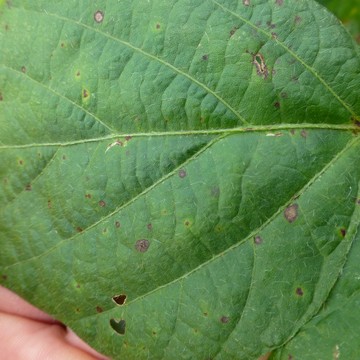}};
\draw[tool] (\xa,-3.10) rectangle (\xb,-3.52);
\node[sic, text=fInsI] at (\xi,-3.22) {\faIcon{cog}}; \node[cnt] at (\xi,-3.36) {3/10};
\node[mid] at (5.52,-3.31) {Tool: \textcolor{fIns}{\textbf{Inspect}} a Downy Mildew photo};
\node[img] at (9.3,-3.31) {\includegraphics[height=0.34cm]{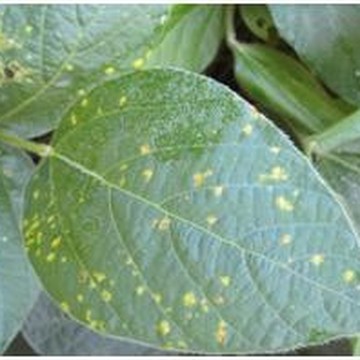}};
\draw[agent] (\xa,-3.58) rectangle (\xb,-3.88);
\node[icon, text=fAgt, font=\footnotesize] at (\xi,-3.73) {\faIcon{robot}};
\node[mid] at (5.52,-3.73) {\nt{The inspected `Soybean - Downy Mildew' image shows yellowish spots \ldots}};
\draw[tool] (\xa,-3.94) rectangle (\xb,-4.56);
\node[sic, text=fRet] at (\xi,-4.13) {\faIcon{cog}}; \node[cnt] at (\xi,-4.27) {4/10};
\node[txt] at (5.52,-3.975) {Tool: \textcolor{fRet}{\textbf{Retrieve}} ``Soybean - Downy Mildew''};
\node[mid, draw=fOk, line width=0.7pt, rounded corners=2pt, inner sep=1.0pt, fill=white] at (5.54,-4.40) {\textcolor{fOk}{\faIcon{file-alt}}~Downy Mildew entry};
\node[evaltag, inner sep=1.0pt] at (7.9,-4.40) {eval.: gold entry found};
\draw[tool] (\xa,-4.62) rectangle (\xb,-5.02);
\node[sic, text=fInsI] at (\xi,-4.74) {\faIcon{cog}}; \node[cnt] at (\xi,-4.88) {5/10};
\node[mid] at (5.52,-4.82) {Tool: \textcolor{fIns}{\textbf{Inspect}} the Downy Mildew entry};
\node[mid, text=black!55, anchor=east] at (13.74,-4.82) {\textit{B and C share every turn up to the first crop}};
\draw[agent] (\xa,-5.08) rectangle (\xb,-5.38);
\node[icon, text=fAgt, font=\footnotesize] at (\xi,-5.23) {\faIcon{robot}};
\node[mid] at (5.52,-5.23) {\nt{\ldots{} I will now crop the input image to examine the leaf spots in detail \ldots}};
\draw[arr] (7.15,-5.42) -- (7.15,-5.50);
\draw[arr] (11.75,-5.42) -- (11.75,-5.50);
\def\xa{5.42}\def\xb{9.35}\def\xi{5.16}
\draw[tool, draw=fOwn, line width=0.9pt] (\xa,-5.48) rectangle (\xb,-6.56);
\node[icon, text=fCrop] at (\xi,-5.92) {\faIcon{cog}}; \node[cnt] at (\xi,-6.11) {6--8};
\node[txt] at (5.52,-5.54) {Tools 6--8: \textcolor{fCrop}{\textbf{Crop}} $\times$3, own boxes};
\node[pic] at (5.52,-5.78) {\includegraphics[height=0.7cm]{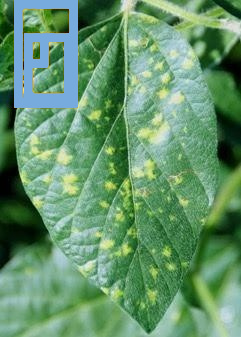}};
\draw[arr] (6.06,-6.13) -- (6.19,-6.13);
\node[pic, draw=fOwn, line width=0.8pt] at (6.25,-5.78) {\includegraphics[height=0.7cm]{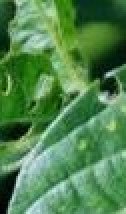}};
\node[pic, draw=fOwn, line width=0.8pt] at (6.75,-5.78) {\includegraphics[height=0.7cm]{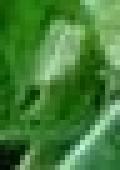}};
\node[pic, draw=fOwn, line width=0.8pt] at (7.33,-5.78) {\includegraphics[height=0.7cm]{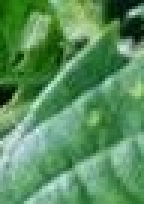}};
\node[txt, text width=1.35cm, text=black!55] at (7.95,-5.92) {the crops the agent sees};
\draw[agent] (\xa,-6.62) rectangle (\xb,-7.16);
\node[icon, text=fAgt] at (\xi,-6.89) {\faIcon{robot}};
\node[txt, text width=3.8cm] at (5.52,-6.68) {\nt{The cropped image is too blurry to provide useful details.}};
\draw[tool] (\xa,-7.21) rectangle (\xb,-7.48);
\node[icon, text=black!45, font=\scriptsize] at (\xi,-7.345) {\faIcon{cog}};
\node[mid, text=black!65] at (5.52,-7.345) {Tools 9--10: \textbf{inspect}, \textbf{crop} again};
\draw[agent] (\xa,-7.53) rectangle (\xb,-8.12);
\node[icon, text=fAgt] at (\xi,-7.825) {\faIcon{robot}};
\node[mid] at (5.52,-7.69) {\textcolor{black!45}{answer:}~\textbf{Downy Mildew}};
\node[evaltag, anchor=east] at (9.29,-7.69) {\textcolor{fOk}{\ding{51}~correct}};
\node[mid, text=black!55] at (5.52,-7.96) {10/10 calls};
\def\xa{10.07}\def\xb{13.9}\def\xi{9.81}
\draw[tool, draw=fLrn, line width=0.9pt] (\xa,-5.48) rectangle (\xb,-6.56);
\node[icon, text=fCrop] at (\xi,-5.92) {\faIcon{cog}}; \node[cnt] at (\xi,-6.11) {6--8};
\node[txt] at (10.17,-5.54) {Tools 6--8: \textcolor{fCrop}{\textbf{Crop}} $\times$3 from attention};
\node[pic] at (10.17,-5.76) {\includegraphics[height=0.62cm]{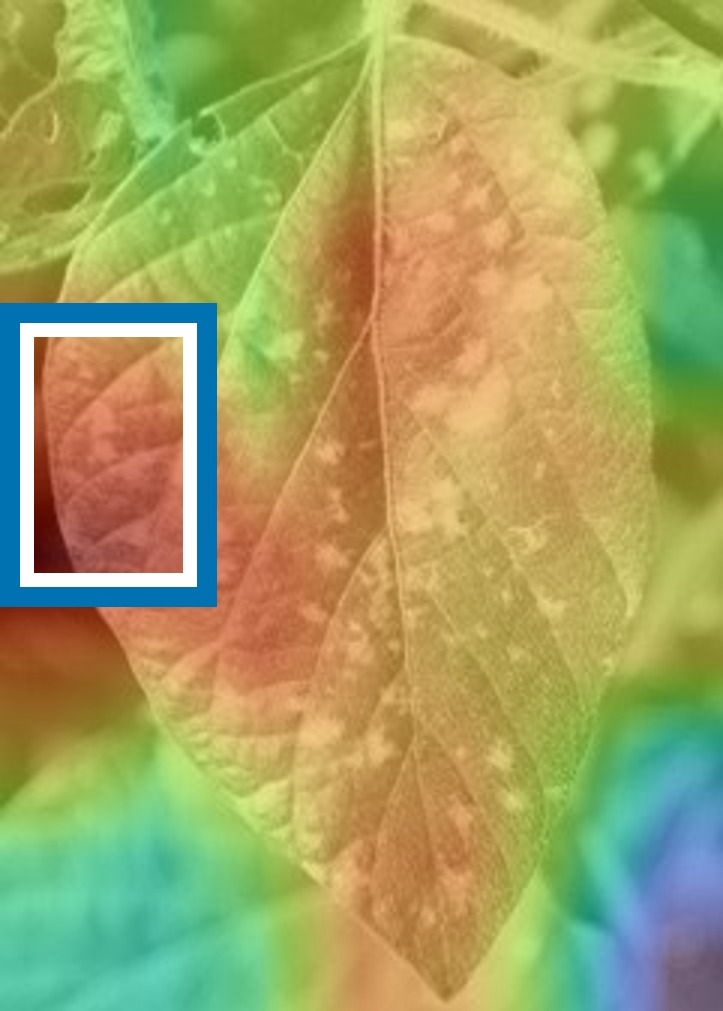}};
\draw[arr] (10.65,-6.07) -- (10.77,-6.07);
\node[pic, draw=fLrn, line width=0.8pt] at (10.81,-5.76) {\includegraphics[height=0.62cm]{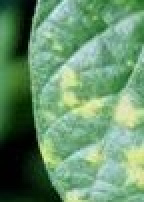}};
\node[pic, draw=black!35, line width=0.4pt] at (11.42,-5.76) {\includegraphics[height=0.62cm]{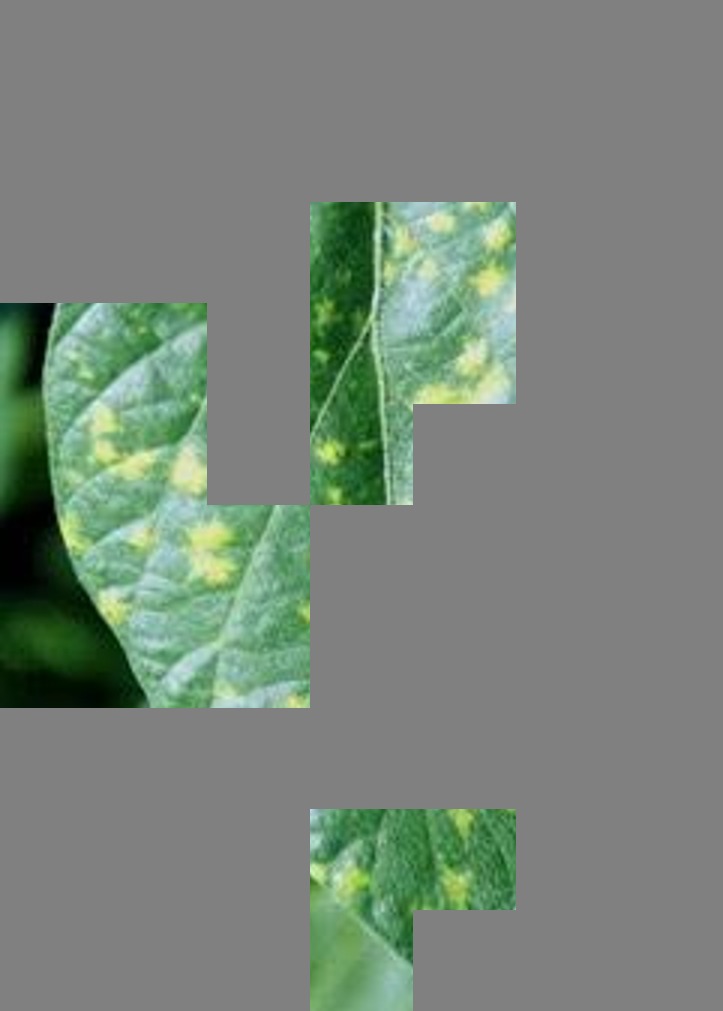}};
\draw[black!45, line width=0.6pt, rounded corners=1.5pt, -{Triangle[length=1.0mm,width=1.3mm]}] (10.39,-6.39) -- (10.39,-6.48) -- (11.64,-6.48) -- (11.64,-6.4);
\draw[fLrn, line width=0.7pt, -{Triangle[length=1.1mm,width=1.4mm]}] (11.91,-6.07) -- (12.04,-6.07);
\node[mid, text=fLrn, font=\sffamily\bfseries\fontsize{5.8}{6.6}\selectfont] at (12.08,-6.07) {step\\ confidence};
\draw[agent] (\xa,-6.62) rectangle (\xb,-7.16);
\node[icon, text=fAgt] at (\xi,-6.89) {\faIcon{robot}};
\node[txt, text width=3.6cm] at (10.17,-6.68) {\nt{The cropped image shows yellowish spots on the leaf \ldots}};
\draw[agent] (\xa,-7.53) rectangle (\xb,-8.12);
\node[icon, text=fAgt] at (\xi,-7.825) {\faIcon{robot}};
\node[mid] at (10.17,-7.69) {\textcolor{black!45}{answer:}~\textbf{Downy Mildew}};
\node[evaltag, anchor=east] at (13.82,-7.69) {\textcolor{fOk}{\ding{51}~correct}};
\node[mid, text=black!55] at (10.17,-7.96) {8/10 calls};
\node[mid, draw=fLrn, line width=0.7pt, rounded corners=2pt, inner sep=1.0pt, fill=white] at (11.3,-7.955) {confidence \textbf{0.93}};
\end{scope}
\end{scope}
\node[panel, fill=white, minimum width=5.25cm, minimum height=7.84cm] at (14.15,0) {};
\begin{scope}[shift={(14.15,0.44)}]
\node[pt] at (0.12,-0.72) {\textbf{Test accuracy (\%), 8B}};
\fill[fZs] (0.12,-1.0) rectangle (0.32,-1.16); \node[pl] at (0.38,-1.08) {Zero-shot agent};
\fill[fSft] (1.98,-1.0) rectangle (2.18,-1.16); \node[pl] at (2.24,-1.08) {Agent + SFT};
\fill[fOwn] (3.55,-1.0) rectangle (3.75,-1.16); \node[pl] at (3.8099999999999996,-1.08) {ViBES};
\draw[black!12, line width=0.4pt] (0.62,-3.460) -- (5.13,-3.460); \node[pa] at (0.56,-3.460) {0};
\draw[black!12, line width=0.4pt] (0.62,-2.849) -- (5.13,-2.849); \node[pa] at (0.56,-2.849) {20};
\draw[black!12, line width=0.4pt] (0.62,-2.237) -- (5.13,-2.237); \node[pa] at (0.56,-2.237) {40};
\draw[black!12, line width=0.4pt] (0.62,-1.626) -- (5.13,-1.626); \node[pa] at (0.56,-1.626) {60};
\fill[fZs] (0.787,-3.46) rectangle (1.147,-2.326);
\node[pv] at (0.967,-2.306) {37.1};
\fill[fSft] (1.177,-3.46) rectangle (1.537,-2.301);
\node[pv] at (1.357,-2.281) {37.9};
\fill[fOwn] (1.567,-3.46) rectangle (1.927,-2.155);
\node[pvb] at (1.747,-2.135) {42.7};
\node[pc] at (1.372,-3.52) {MC-Search};
\fill[fZs] (2.290,-3.46) rectangle (2.650,-1.656);
\node[pv] at (2.470,-1.636) {59.0};
\fill[fSft] (2.680,-3.46) rectangle (3.040,-1.589);
\node[pv] at (2.860,-1.569) {61.2};
\fill[fOwn] (3.070,-3.46) rectangle (3.430,-1.430);
\node[pvb] at (3.250,-1.410) {66.4};
\node[pc] at (2.875,-3.52) {WebQA};
\fill[fZs] (3.793,-3.46) rectangle (4.153,-2.338);
\node[pv] at (3.973,-2.318) {36.7};
\fill[fSft] (4.183,-3.46) rectangle (4.543,-2.448);
\node[pv] at (4.363,-2.428) {33.1};
\fill[fOwn] (4.573,-3.46) rectangle (4.933,-2.161);
\node[pvb] at (4.753,-2.141) {42.5};
\node[pc] at (4.378,-3.52) {SAGE};
\draw[black!45, line width=0.5pt] (0.62,-3.46) -- (5.13,-3.46);
\node[pt] at (0.12,-4.3) {\textbf{AUROC of the answer confidence, 8B}};
\fill[fOwn] (0.12,-4.58) rectangle (0.32,-4.74); \node[pl] at (0.38,-4.66) {ViBES};
\fill[fLrn] (1.3,-4.58) rectangle (1.5,-4.74); \node[pl] at (1.56,-4.66) {ViBES-Lens};
\draw[black!12, line width=0.4pt] (0.62,-7.500) -- (5.13,-7.500); \node[pa] at (0.56,-7.500) {0.5};
\draw[black!12, line width=0.4pt] (0.62,-6.492) -- (5.13,-6.492); \node[pa] at (0.56,-6.492) {0.6};
\draw[black!12, line width=0.4pt] (0.62,-5.484) -- (5.13,-5.484); \node[pa] at (0.56,-5.484) {0.7};
\fill[fOwn] (0.882,-7.5) rectangle (1.342,-6.497);
\node[pv] at (1.112,-6.477) {0.60};
\fill[fLrn] (1.372,-7.5) rectangle (1.832,-6.201);
\node[pv] at (1.602,-6.181) {0.63};
\node[pd] at (1.372,-5.941) {$+$0.029};
\node[pc] at (1.372,-7.56) {MC-Search};
\fill[fOwn] (2.385,-7.5) rectangle (2.845,-6.851);
\node[pv] at (2.615,-6.831) {0.56};
\fill[fLrn] (2.875,-7.5) rectangle (3.335,-6.485);
\node[pv] at (3.105,-6.465) {0.60};
\node[pd] at (2.875,-6.225) {$+$0.036};
\node[pc] at (2.875,-7.56) {WebQA};
\fill[fOwn] (3.888,-7.5) rectangle (4.348,-6.135);
\node[pv] at (4.118,-6.115) {0.64};
\fill[fLrn] (4.378,-7.5) rectangle (4.838,-5.442);
\node[pv] at (4.608,-5.422) {0.70};
\node[pd] at (4.378,-5.182) {$+$0.069};
\node[pc] at (4.378,-7.56) {SAGE};
\draw[black!45, line width=0.5pt] (0.62,-7.5) -- (5.13,-7.5);
\node[pl, text=black!55] at (0.12,-7.92) {0.5 = chance; splits as in Table~\ref{tab:confidence}};

\end{scope}
\end{tikzpicture}
}
\vspace{-0.4em}

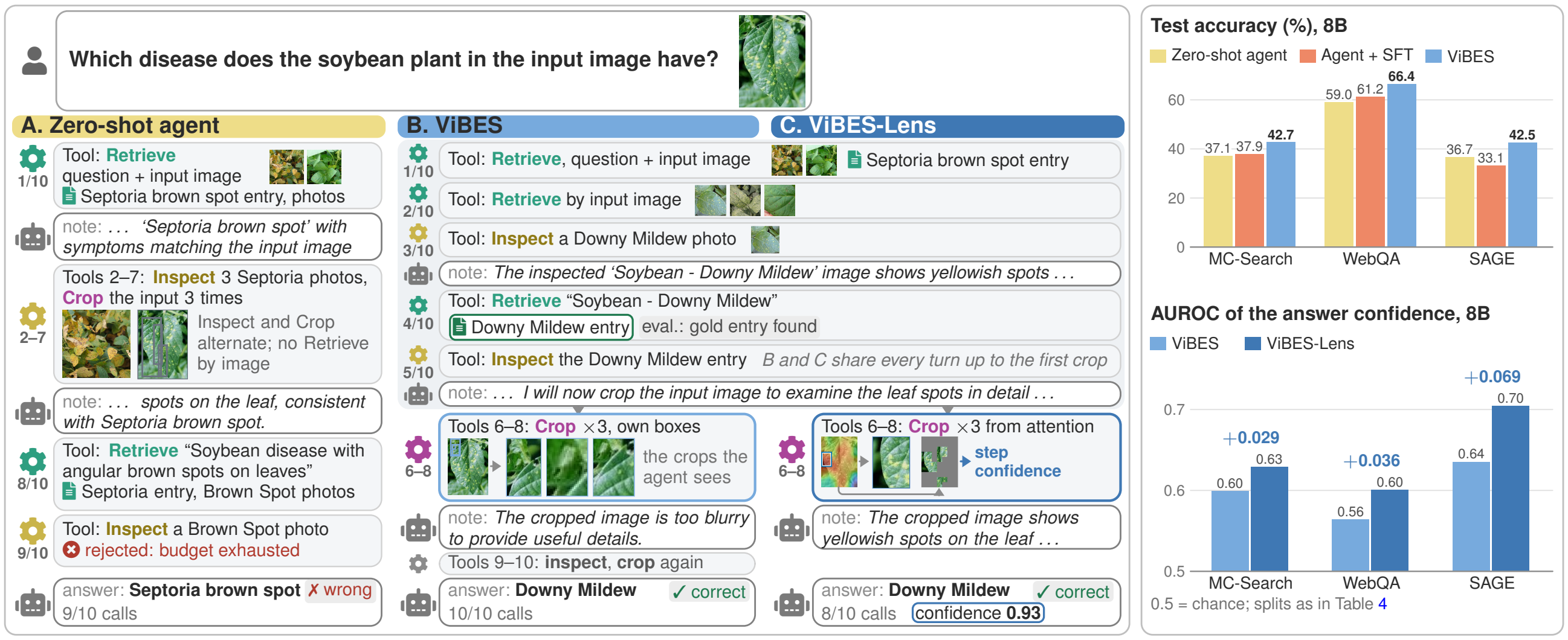
\captionof{figure}{\textbf{ViBES at a glance.} Left: three runs on one SAGE test question. (A) The zero-shot agent answers with the first disease retrieved; (B) ViBES and (C) ViBES-Lens are correct; in (C), a probe of the attention view gives a confidence. Right: accuracy and confidence AUROC (Tables~\ref{tab:indomain},~\ref{tab:confidence}).}
\label{fig:teaser}
\end{center}
\vspace{0.2em}
}]

\begin{abstract}
Recent advances in agentic retrieval-augmented generation (RAG) enable models to iteratively search and reason over external evidence, while multimodal agents further extend these capabilities through visual inspection. However, selecting relevant image–text records and fine-grained visual evidence under limited tool budgets remains challenging, and how these decisions affect the predictability of answer correctness is not well understood. We introduce ViBES, an open-weight multimodal agent for evidence-grounded reasoning over indexed image–text collections that integrates iterative retrieval, record inspection, and targeted visual cropping. We train the agent through supervised fine-tuning and reinforcement learning using a reward that combines answer correctness with recall of annotated supporting records. We further develop ViBES-Lens, a reasoning-conditioned visual region selector that combines the agent's reasoning state with image-patch features to identify informative regions without bounding-box annotations. We evaluate ViBES on two general-purpose benchmarks, MC-Search and WebQA, and SAGE, a specialized plant-disease benchmark that allows us to examine the value of fine-tuning smaller models for domain-specific reasoning. At the 8B scale, ViBES improves average answer accuracy by 13.6 percentage points and gold-evidence recall by 20.2 points over matched one-shot RAG, with gains largely retained across two external agent harnesses without retraining. Notably, on SAGE, our fine-tuned 8B agent outperforms all evaluated external agents, including a 32B model. ViBES-Lens's single-box variant improves trajectory-level correctness-discrimination AUROC by 0.061 on average across six dataset–model configurations while maintaining comparable accuracy.
\end{abstract}

\section{Introduction}
\label{sec:intro}

Retrieval-augmented generation (RAG) has emerged as an effective approach for enhancing language models with external knowledge~\cite{guu2020realm,karpukhin2020dpr,lewis2020rag}. Building upon this paradigm, multimodal retrieval-augmented generation (MM-RAG) extends knowledge retrieval to heterogeneous collections containing both textual and visual information~\cite{chen2022murag,faysse2025colpali,yu2025visrag,racm3_2023}. Recent advances in agentic reasoning have further extended retrieval-based methods from single-step retrieval to iterative information-seeking processes, where models formulate queries, examine retrieved results, and refine subsequent retrieval actions based on previously acquired evidence~\cite{geng2026webwatcher,searchr1_2025,ning2026mcsearch,wu2026mmsearch}.

Search-based methods, such as MMSearch-R1~\cite{wu2026mmsearch} and WebWatcher~\cite{geng2026webwatcher}, train VLMs to interact with external search tools and perform multi-step information retrieval. For indexed multimodal knowledge bases, MC-Search~\cite{ning2026mcsearch} investigates structured multi-hop retrieval over image--text collections and introduces Search-Align to improve evidence acquisition. In parallel, active visual perception methods have demonstrated the effectiveness of selectively inspecting image regions during reasoning. DeepEyes~\cite{deepeyes2025} employs reinforcement learning (RL) to enable adaptive image cropping, while SenseSearch~\cite{chng2026sensesearch} and DeepMMSearch-R1~\cite{narayan2025deepmmsearch} integrate visual inspection with multimodal web search. More recently, MCite-RL~\cite{zhao2026mciterl} combines iterative retrieval and region-level inspection over visually rich documents to improve answer grounding and citation quality. These methods establish the utility of iterative retrieval and targeted visual exploration within multimodal agents. However, interaction between fine-grained visual inspection and iterative evidence acquisition over indexed image--text collections remains important setting for further investigation.

Reasoning over such collections presents several challenges. First, the evidence required to answer a question may not appear in the initial retrieval results, requiring multiple retrieval steps guided by previously acquired information. Second, relevant visual details may occupy only a small portion of an image, making targeted inspection necessary to expose information that may be overlooked in the original representation. Third, the information needed for reasoning may be distributed across multiple visual regions, requiring the agent to identify and examine complementary details. These decisions must also be made under a limited tool-call budget, making it important to acquire relevant evidence within the available interactions. Moreover, because evidence is progressively acquired through a sequence of retrieval and inspection decisions, errors can arise at different stages of the reasoning process. An agent may fail to retrieve relevant records, inspect an uninformative region, or generate an incorrect answer despite accessing sufficient evidence. This raises an additional question beyond improving answer accuracy: whether the resulting execution trajectory provides sufficient information to distinguish correct answers from incorrect ones.

\begin{figure*}[t]
\centering
\resizebox{\textwidth}{!}{\input{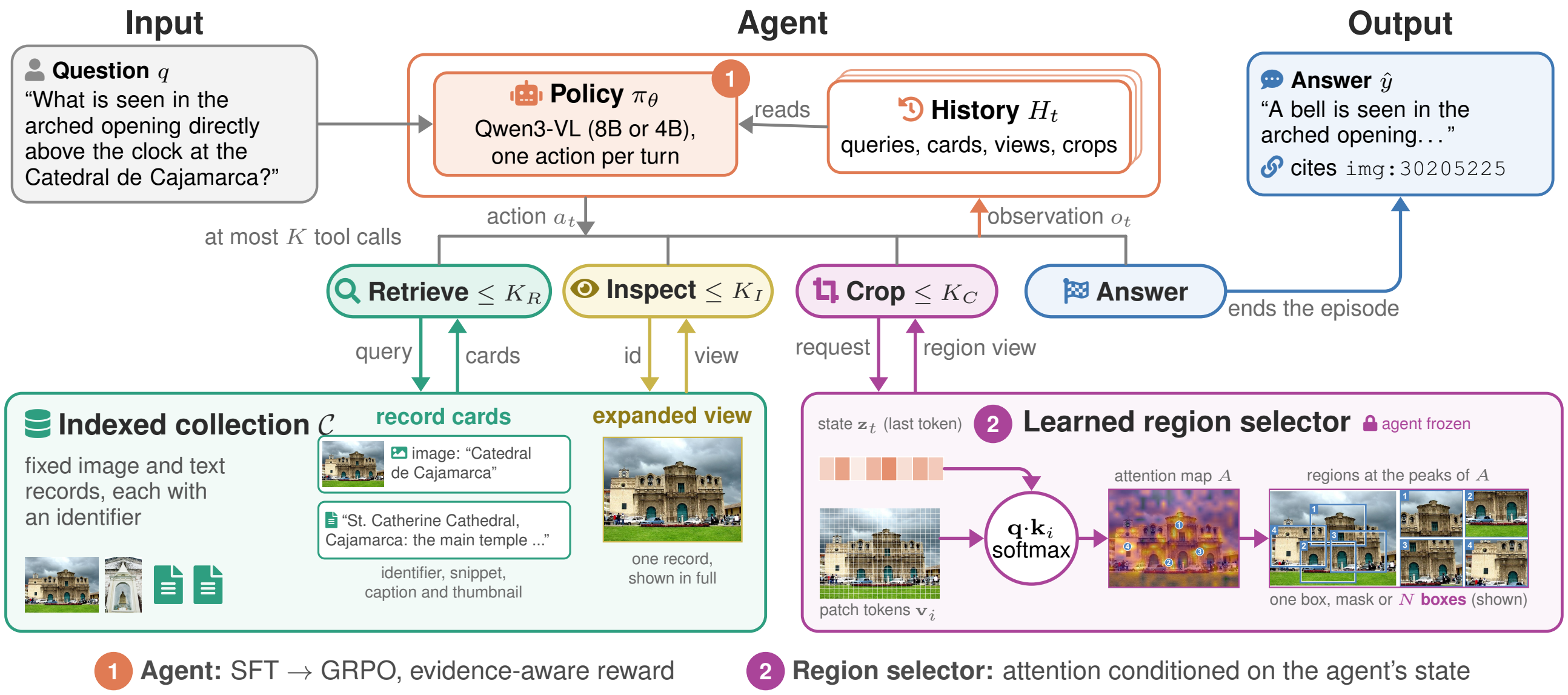}}
\caption{\textbf{Overview of ViBES and ViBES-Lens.} The agent iteratively retrieves and inspects multimodal evidence. At a crop request, the reasoning-conditioned region selector ViBES-Lens identifies informative image regions using attention over visual patches. The agent and selector are trained sequentially.}
\label{fig:overview}
\end{figure*}

In this paper, we introduce \textbf{Vi}sual \textbf{B}udgeted \textbf{E}vidence \textbf{S}election (\textbf{ViBES}) for multimodal agentic reasoning, an MM-RAG framework that integrates iterative multimodal retrieval, evidence inspection, and targeted visual cropping over indexed image--text collections (Figure~\ref{fig:overview}). A VLM-based policy selects retrieval, inspection, cropping, and answering actions based on the question and its observation history, allowing subsequent decisions to build upon previously acquired evidence. We train the policy through supervised fine-tuning (SFT) followed by Group Relative Policy Optimization (GRPO)~\cite{grpo2024}, using a reward that combines final-answer correctness with the recall of annotated gold evidence, encouraging the acquisition of relevant evidence within fixed tool-call budget. To support fine-grained visual inspection, we further introduce \textbf{ViBES-Lens}, a reasoning-conditioned, attention-guided region selector that combines agent's reasoning representation with patch-level visual features to identify informative image regions.

To study the aforementioned question, we compare trajectories generated using ViBES-Lens with those obtained using ViBES, building on trajectory-based confidence estimation~\cite{zhang2026agenticconfidence}. This comparison separates two aspects of visual tool use: whether a cropping mechanism helps an agent produce a correct answer, and whether the resulting trajectory makes that correctness easier to predict, which we measure with the AUROC, ECE and Brier score.

We evaluate ViBES on two generic benchmarks, MC-Search~\cite{ning2026mcsearch} and WebQA~\cite{webqa2022}, and on SAGE~\cite{sage2026}, a plant-disease benchmark that represents a deep domain-specific problem. With two model sizes, the trained agent improves final-answer accuracy by 13.6 points and gold-evidence recall by 20.2 points over matched one-shot MM-RAG baselines, and this advantage persists in two external agent harnesses. On SAGE, ViBES and a fine-tuned one-shot RAG baseline, both at 8B, outperform every external agent we evaluate, including a 32B model, showing the utility of fine-tuning small models for domain-specific problems. Under the shared 10-call limit, ViBES also retrieves more gold evidence within its first three tool calls than the comparable baselines retrieves after ten calls on WebQA and SAGE. Building on ViBES, ViBES-Lens raises the AUROC of final-answer correctness prediction with mean $+0.061$.

Our main contributions are summarized as follows:
\begin{enumerate}\itemsep.1em
\item We introduce ViBES, an agentic MM-RAG framework for evidence acquisition and visual inspection over indexed image--text collections, trained with SFT and RL on rewards for answer correctness and evidence acquisition under a limited tool-call budget.
\item We propose ViBES-Lens, a reasoning-conditioned region selector that returns a single box, a mask or several boxes and is trained without bounding-box annotations. The single-box mode of ViBES-Lens achieves accuracy comparable to that of ViBES and produces trajectories whose final-answer correctness is easier to predict.
\item We study trajectory-level correctness prediction in agentic multimodal RAG on three benchmarks and two model sizes. ViBES-Lens improves on ViBES in the AUROC of final-answer correctness prediction and achieves comparable answer accuracy.
\end{enumerate}

\section{Related Work}
\label{sec:related}

\noindent\textbf{Multimodal Retrieval-Augmented Generation.} Retrieval-augmented generation (RAG) enhances language models by incorporating information retrieved from external knowledge sources~\cite{guu2020realm,karpukhin2020dpr,lewis2020rag}. Multimodal RAG extends this paradigm to collections containing both textual and visual evidence. Early approaches, including MuRAG~\cite{chen2022murag} and RA-CM3~\cite{racm3_2023}, integrate retrieved images and text into multimodal generation, while ColPali~\cite{faysse2025colpali}, VisRAG~\cite{yu2025visrag}, and ViDoRAG~\cite{wang2025vidorag} investigate retrieval over visually rich documents. Methods such as Search-R1~\cite{searchr1_2025}, MMSearch-R1~\cite{wu2026mmsearch}, WebWatcher~\cite{geng2026webwatcher}, and VSearcher~\cite{zhang2026vsearcher} investigate multi-step retrieval over textual and web-based resources, while MC-Search~\cite{ning2026mcsearch} studies structured multi-hop retrieval over indexed image--text collections and introduces Search-Align to improve evidence acquisition.

Beyond evidence retrieval, recent studies have explored active visual inspection to support fine-grained reasoning. DeepEyes~\cite{deepeyes2025} enables VLMs to adaptively crop image regions through reinforcement learning, while Visual-ARFT~\cite{liu2025visualarft} investigates code-based visual manipulation, including cropping and rotation, during reasoning. SenseSearch~\cite{chng2026sensesearch} and DeepMMSearch-R1~\cite{narayan2025deepmmsearch} integrate visual inspection with multimodal web search, whereas MCite-RL~\cite{zhao2026mciterl} combines iterative document retrieval, recursive cropping, and citation-oriented rewards to improve answer grounding. Our work builds upon these advances by integrating iterative retrieval and reasoning-conditioned region selection over indexed image--text collections, while further investigating the implications of visual inspection for trajectory-level correctness prediction.

\noindent\textbf{Confidence Estimation in Agentic Systems.} Estimating the correctness of language-model outputs has received increasing attention, particularly as models are deployed in multi-step, tool-using environments. Liu \etal~\cite{liu2024uncertaintycalibration} investigate uncertainty calibration in tool-using language agents. Duan \etal~\cite{duan2025uprop} study uncertainty propagation in multi-step agentic decision-making. More recently, Zhang \etal~\cite{zhang2026agenticconfidence} introduce Agentic Confidence Calibration and propose Holistic Trajectory Calibration (HTC), which extracts process-level features from complete execution trajectories to predict agent success and failure. However, they do not directly characterize how different visual evidence inspection mechanisms affect the informativeness of trajectories for predicting final-answer correctness.

\begin{table}[t]
\centering\scriptsize
\def\Y{\textcolor{black!80}{\ding{51}}}\def\N{\textcolor{black!45}{\ding{55}}}
\caption{\textbf{Comparison of multimodal search and visual-tool agents.} Methods differ in evidence sources, iterative retrieval, visual operations, and training strategies. Non-web evidence sources are fixed collections.}
\label{tab:prior}
\setlength{\tabcolsep}{2pt}\renewcommand{\arraystretch}{1.0}
\begin{tabularx}{\columnwidth}{@{}l l c >{\raggedright\arraybackslash}X l@{}}
\toprule
Method & Evidence & Iter. & Visual operation & Training \\
\midrule
RA-CM3~\cite{racm3_2023} & Image--text KB & \N & None & Pretraining \\
Search-R1~\cite{searchr1_2025} & Text index & \Y & None & RL \\
MMSearch-R1~\cite{wu2026mmsearch} & Live web & \Y & None & RL \\
WebWatcher~\cite{geng2026webwatcher} & Live web & \Y & OCR & SFT + RL \\
VSearcher~\cite{zhang2026vsearcher} & Live web & \Y & None & SFT + RL \\
Visual-ARFT~\cite{liu2025visualarft} & Live web & \Y & Code (crop, rotate) & RL \\
\makecell[tl]{DeepMMSearch-R1\\\cite{narayan2025deepmmsearch}} & Live web & \Y & Crop for image search, 1~grounded box & SFT + RL \\
SenseSearch~\cite{chng2026sensesearch} & Live web & \Y & Crop, 1 model box & SFT + RL \\
DeepEyes~\cite{deepeyes2025} & None & \N & Crop, 1 model box & RL \\
Search-Align~\cite{ning2026mcsearch} & Image--text KB & \Y & None & SFT \\
MCite-RL~\cite{zhao2026mciterl} & Document pages & \Y & Crop, 1 box (box-supervised) & SFT + RL \\
\midrule
\rowcolor{vsViseekP}
\sw{vsViseek}\textbf{ViBES (ours)} & Indexed collection & \Y & Crop, own box & SFT + RL \\
\rowcolor{vsViseekP}
\sw{vsViseek}\textbf{ViBES-Lens (ours)} & Indexed collection & \Y & Crop, learned regions; multimodal UQ & SFT + RL \\
\bottomrule
\end{tabularx}
\end{table}

\noindent\textbf{Confidence in Multimodal RAG.} CLIMB~\cite{gao2026climb} introduces an evidence-grounded confidence estimation mechanism that guides iterative answer refinement using complementary retrieved evidence. However, existing approaches primarily employ confidence to guide evidence selection or answer refinement, rather than examining how the mechanisms used to acquire and inspect visual evidence influence subsequent confidence estimation. In contrast, our work investigates the relationship between visual inspection strategies and trajectory-level correctness discrimination.

\section{Method}
\label{sec:delta}

ViBES has three components (Figure~\ref{fig:overview}). An agent policy parameterized by $\theta\in\mathbb R^d$, $\pi_\theta$, gathers evidence from an indexed collection by
retrieving, inspecting and cropping records (Section~\ref{sec:formulation}). The policy is trained with supervised
fine-tuning and GRPO, under a reward that scores both the answer and the annotated evidence the episode retrieves
(Section~\ref{sec:policy}). The learned region selector, ViBES-Lens, replaces the agent's crop box with one box, a mask or $N$ boxes,
chosen from an attention map over image patches that is conditioned on the agent's state (Section~\ref{sec:module}).
Finally, Section~\ref{sec:confidence} defines how the correctness of a trajectory is predicted at its crop steps, and how
this prediction is compared between ViBES-Lens and the agent's own box.

\subsection{Problem Formulation}
\label{sec:formulation}

\noindent\textbf{Task.} Given a question $q$, optionally with an image $x$, the agent answers by gathering evidence
from an indexed collection $\mathcal C$ of records, each with an identifier and an image, a text, or both. The evidence
a question needs may be spread over several records, and a decisive detail may occupy a small part of an image. Over
several turns, the agent therefore decides what to retrieve, which records to inspect in full, and which image regions
to crop before it commits to an answer. What the agent sees, and so what it can answer, depends on this sequence of
actions.

\noindent\textbf{Observation and action spaces.} The agent accesses $\mathcal C$ only through four tools.
\emph{Retrieve} takes a text, image or text-plus-image query and returns a short list of record cards; a card shows the
record identifier and a text snippet, plus a caption and thumbnail for an image record. \emph{Inspect} returns an
expanded view of a record in the history. \emph{Crop} takes an image in the history (the input image $x$, an inspected
image or an earlier crop) and a box $b$, and returns that region as a new view. \emph{Answer} ends the episode with the
final answer $\hat y$ and the cited record identifiers. The history $H_t$ stores every query, card, view and crop box;
it starts as $H_0=\emptyset$, or holds skill instructions when the agent is given any. At turn $t\ge1$ the policy
samples one action $a_t\sim\pi_\theta(\cdot\mid q,x,H_{t-1})$, the executor returns an observation $o_t$, and
$H_t=H_{t-1}\oplus(a_t,o_t)$. Per-tool budgets $K_R,K_I,K_C$ and a total budget $K$ bound the episode: a tool that
reaches its budget is disabled, and an episode that exhausts $K$ ends with a forced answer (Algorithm~1).

\noindent\textbf{Trajectories and correctness.} An episode yields a trajectory $\tau=(a_1,o_1,\dots,a_T)$ whose last
action $a_T$ is \emph{Answer}; we write $\hat y(\tau)$ for the final answer it returns and $y^\star$ for the gold
answer. The correctness of $\tau$ is $\delta(\tau)=\mathrm{Acc}_{\mathcal D}\big(\hat y(\tau),y^\star\big)\in\{0,1\}$,
where $\mathrm{Acc}_{\mathcal D}$ is the accuracy protocol of dataset $\mathcal D$ (see Appendix). We
study two properties of a system: its accuracy, $\mathbb E[\delta(\tau)]$, and how well a confidence score $c(\tau)$
computed from the trajectory separates correct from incorrect answers, measured by the area under the ROC curve
(AUROC) of $c$ against $\delta$.

\subsection{Policy Learning with an Evidence-Aware Reward}
\label{sec:policy}

\noindent\textbf{Cold-start fine-tuning.} We first fine-tune low-rank adapters on teacher episodes. The teacher is the
base model running the same loop, with a private description of the gold records added to its system prompt. We keep
only teacher episodes that answer correctly with valid citations. The student never receives the teacher's prompt.

\noindent\textbf{Evidence-aware reward.} The benchmarks annotate the supporting records $\mathcal G(q)\subseteq\mathcal C$
of each question, and we write $\mathcal G^\star(q)$ for the subset the index can return. Because $\mathcal C$ is fixed,
whether an episode saw the supporting evidence can be measured separately from whether it answered correctly; the
annotations need not cover every record that supports an answer. Let $S(\tau_i)$ be the set of records shown to episode
$\tau_i$ in any card or view. The reward adds the recall of retrievable gold records to the answer's token F1,
\begin{equation}
R_i=\mathrm{F1}\big(\hat y(\tau_i),y^\star\big)+\lambda\,\frac{|S(\tau_i)\cap\mathcal G^\star(q)|}{|\mathcal G^\star(q)|}.
\label{eq:reward}
\end{equation}
The recall term scores the set of records an episode has seen. It does not depend on their rank or on the step at
which they appeared, unlike rank-based retrieval rewards such as nDCG~\cite{zhao2026mciterl}, and it needs no box
annotation. Gold identifiers enter only the reward and scorer, never the policy's input.

\noindent\textbf{Optimization.} We train with GRPO~\cite{grpo2024}. For each question we sample $G$ episodes and
normalize the rewards within the group, $A_i=(R_i-\bar R)/\max(\sigma_R,0.1)$. Log-probabilities are taken only over
tokens the policy generates, so tool observations are masked. We take one optimizer step per rollout batch such that the
importance ratio is one and the update follows $\frac1G\sum_i A_i\nabla_\theta\log\pi_\theta(\tau_i)$.

\subsection{Reasoning-Conditioned Visual Region Selection}
\label{sec:module}

With the agent frozen, the learned region selector ViBES-Lens replaces the box the agent writes (the \emph{own box}) at a crop
request on the input image or an inspected image.

\noindent\textbf{Reasoning-conditioned attention.} At crop request $t$, one forward pass of the frozen agent gives the
last-token final-layer state $\mathbf z_t$ and the patch tokens $\mathbf v_1,\dots,\mathbf v_P$ of the target image
($32{\times}32$-pixel patches). After PCA, small networks map them to a query $\mathbf q$ and keys
$\mathbf k_i$ of dimension $d$:
\begin{equation}
\ell_i=\mathbf q^{\top}\mathbf k_i/\sqrt d,\qquad A_i=\mathrm{softmax}_i(\ell_i).
\label{eq:attn}
\end{equation}
Because $\mathbf z_t$ encodes the question and everything read so far, the same image receives different maps at
different turns.

\noindent\textbf{Visual region selection.} Based on the reasoning-conditioned attention map, ViBES-Lens has three
output modes $m$, each a selector $g_{\phi,m}$ with parameters $\phi$ that returns the region $R_t$
shown to the agent. \emph{One box} selects a single region centered at the highest-scoring
attention patch, with a learned size predictor determining its spatial extent. A skip head keeps the own box when it
predicts no crop. \emph{Mask} retains the top $k\%$ of visual patches while masking the remaining image regions in
gray. \emph{$N$ boxes} extends region selection to multiple
regions by greedily selecting attention peaks outside previously covered areas and returning the corresponding crops
of fixed side $s_{\mathcal D}$ in a single tool call. All three modes share the
same attention-guided formulation but differ in their visual outputs and optimization objectives. The one box and the
mask act at every crop request, and the $N$ boxes at the first crop request of an episode.

\noindent\textbf{Training the region selectors.} The agent policy remains frozen while the region selectors are
optimized with GRPO on crop requests of training-split episodes, without bounding-box annotations. The single-box and
mask variants use a teacher-forced answer-likelihood improvement objective: the standardized gain over no crop in the
frozen agent's teacher-forced score of the gold answer, minus $0.1$ times the box's share of the image for the one box
(at 4B, the mask uses the decoded answer's score except on WebQA). The $N$ box variant is optimized using final-answer
correctness after the agent continues reasoning with the selected regions. These variants are trained
separately, with checkpoints, $k$ and $N$ chosen on development data (see Appendix).

\subsection{Correctness Prediction at the Crop Step}
\label{sec:confidence}

A crop mechanism changes what the agent sees only at crop requests, so we read the agent's confidence there. For a
trajectory produced with a visual region-selection mechanism $m$ (Section~\ref{sec:module}), let $\mathcal T(\tau)$ be
the crop requests at which its selector $g_{\phi,m}$ is called (on the input image or an inspected image). At each $t\in\mathcal T(\tau)$, we append one user turn to the cached conversation. The turn shows a
region view $V_t$ and asks another question $u$:
``Is the information needed to answer the original question visible in this view? Reply with only Yes or No.'' The
frozen agent's log-odds, with context $\kappa_t=(q,x,H_{t-1},V_t,u)$ is as follows:
\begin{equation}
\psi(V_t)=\log p_\theta(\texttt{Yes}\mid\kappa_t)-\log p_\theta(\texttt{No}\mid\kappa_t),
\label{eq:probe}
\end{equation}
which is teacher-forced and do not change the episode. The trajectory's confidence is the mean,
$c(\tau)=\frac{1}{|\mathcal T(\tau)|}\sum_{t\in\mathcal T(\tau)}\psi(V_t)$. Like P(True)
self-evaluation~\cite{kadavath2022know}, the probe needs no fitted calibrator for AUROC; ECE~\cite{guo2017calibration}
and Brier score~\cite{brier1950verification} use a cross-fitted Platt scaling~\cite{platt1999probabilistic} of $c$.

\noindent\textbf{Matched comparison.} For a learned mechanism, $V_t=M(A_t)$, where $A_t=(A_1,\dots,A_P)$ is its attention map (Eq.~\eqref{eq:attn}) at
request $t$ and the mask $M$ keeps the 25\% of patches with the highest $3{\times}3$-smoothed logits $\ell_i$, crops to their
bounding box and grays the other patches. For the own box,
$\mathcal T(\tau)$ holds all of the agent's crop requests of the same kind and $V_t$ is the agent's own crop. The probe views differ by design: the learned selector provides targeted patch-level attention, whereas the agent's own cropping provides only a bounding box. Both systems run the same frozen agent on the same questions and use the probe of Eq.~\eqref{eq:probe} with the same weights. Each
trajectory is scored against its own correctness label. On the questions where both systems
issue such a crop request, we report
\begin{equation}
\Delta\mathrm{AUROC}=\mathrm{AUROC}(c^{L},\delta^{L})-\mathrm{AUROC}(c^{O},\delta^{O}),
\label{eq:dauroc}
\end{equation}
where $c^{S}=c(\tau^{S})$ and $\delta^{S}=\delta(\tau^{S})$, and $\tau^{L}$ and $\tau^{O}$ are the learned-mechanism
and own-box trajectories. This separates whether a mechanism changes accuracy,
$\mathbb E[\delta]$, from whether it makes correctness easier to predict.

\begin{figure}[t]
\setlength{\fboxsep}{2pt}%
\newcommand{\nw}[1]{#1}%
\definecolor{cnA}{HTML}{E07B54}\definecolor{cnB}{HTML}{4F86C2}\definecolor{cnC}{HTML}{2E9E80}%
\newcommand{\cn}[1]{\tikz[baseline=-0.6ex]\node[circle,fill=\ifcase#1\or cnA\or cnB\or cnC\fi,text=white,font=\sffamily\bfseries\tiny,inner sep=0pt,minimum size=3.2mm]{#1};}%
\noindent\fbox{\begin{minipage}{\dimexpr\columnwidth-2\fboxsep-2\fboxrule\relax}\footnotesize
\textbf{Algorithm 1: ViBES agent execution, visual region selection, and correctness probing.}
\begin{tabbing}
\hspace{1.4em}\=\hspace{0.8em}\=\hspace{0.8em}\=\hspace{0.8em}\=\kill
\textbf{Input:} $q$, $x$, collection $\mathcal{C}$, policy $\pi_\theta$, selector $g_{\phi,m}$ (optional),\\
\>mode $m\in\{\text{one box},\text{mask},N\text{ boxes}\}$,\\
\>budgets $K_R, K_I, K_C, K$ \hfill ($\oplus$: append)\\
1\>$H \leftarrow H_0$ \hfill empty, or skill instructions\\
2\>\textbf{while} fewer than $K$ tool calls \textbf{do}\\
3\>\>$a \sim \pi_\theta(\cdot \mid q, x, H)$ \hfill \cn{1}\\
4\>\>\textbf{if} retrieve: $H \leftarrow H \oplus \textsc{Retrieve}_{\mathcal C}(a.\text{query}, a.\text{mode})$\\
5\>\>\textbf{if} inspect, $a.\text{id}\in H$: $H \leftarrow H \oplus \textsc{View}(a.\text{id})$\\
6\>\>\textbf{if} crop, $a.\text{id}\in H$:\\
7\>\>\>$\psi_t \leftarrow \psi(V_t)$, Eq.~\eqref{eq:probe} (teacher-forced) \hfill \cn{3}\\
8\>\>\>$R_t \leftarrow g_{\phi,m}(\mathbf z_t,\{\mathbf v_i\})$ if called, else $a.\text{box}$ \hfill \cn{2}\\
9\>\>\>$H \leftarrow H \oplus \textsc{Crop}(a.\text{id}, R_t)$ \hfill crops, or masked view\\
10\>\>\textbf{if} answer: \textbf{return} answer, cited ids,\\
\>\>\>$c(\tau)=\mathrm{mean}_t\,\psi_t$\\
11\>\textbf{end while}; \textbf{return} forced answer from $H$\\[2pt]
\cn{1} \textbf{Agent:} SFT on teacher episodes,\\
\>then GRPO with Eq.~\eqref{eq:reward}\\
\cn{2} \textbf{Selector:} GRPO on crop requests, per mode $m$; $\pi_\theta$ frozen\\
\cn{3} \textbf{Confidence:} $\Delta$AUROC of $c(\tau)$, learned vs.\ own box
\end{tabbing}
\end{minipage}}
\end{figure}

\section{Experimental Setup}
\label{sec:setupmain}
\begin{table*}[t]
\centering
\caption{\textbf{End-to-end answer accuracy (\%) on three benchmarks.} In-house baselines and released external checkpoints. 8B and 4B: our execution loop; Pi and OC: 8B models in the pi and OpenCode harnesses. --: no model of that size exists for the setting. $^{\ast}$32B model shown in the 8B column. Each cell is one run; bold: best, underline: second best.}
\label{tab:indomain}\label{tab:external}
\resizebox{\textwidth}{!}{\footnotesize\setlength{\tabcolsep}{9pt}\renewcommand{\arraystretch}{1.0}\colorlet{tMCS}{vsMCS!9}\colorlet{tWQ}{vsWQ!9}\colorlet{tSAGE}{vsSAGE!12}\setlength{\aboverulesep}{0pt}\setlength{\belowrulesep}{0pt}\setlength{\extrarowheight}{1.2pt}
\begin{tabular}{@{}c l|rrrr|rrrr|rrrr@{}}
\toprule
& & \multicolumn{4}{>{\columncolor{tMCS}}c|}{\textbf{MC-Search}} & \multicolumn{4}{>{\columncolor{tWQ}}c|}{\textbf{WebQA}} & \multicolumn{4}{>{\columncolor{tSAGE}}c}{\textbf{SAGE}} \\
\cmidrule(lr){3-6}\cmidrule(lr){7-10}\cmidrule(lr){11-14}
& & \multicolumn{2}{>{\columncolor{tMCS}}c}{Size} & \multicolumn{2}{>{\columncolor{tMCS}}c|}{Harness (8B)} & \multicolumn{2}{>{\columncolor{tWQ}}c}{Size} & \multicolumn{2}{>{\columncolor{tWQ}}c|}{Harness (8B)} & \multicolumn{2}{>{\columncolor{tSAGE}}c}{Size} & \multicolumn{2}{>{\columncolor{tSAGE}}c}{Harness (8B)} \\
\cmidrule(lr){3-4}\cmidrule(lr){5-6}\cmidrule(lr){7-8}\cmidrule(lr){9-10}\cmidrule(lr){11-12}\cmidrule(lr){13-14}
\textbf{Group} & \textbf{System} & \multicolumn{1}{>{\columncolor{tMCS}}r}{8B} & \multicolumn{1}{>{\columncolor{tMCS}}r}{4B} & \multicolumn{1}{>{\columncolor{tMCS}}r}{Pi} & \multicolumn{1}{>{\columncolor{tMCS}}r|}{OC} & \multicolumn{1}{>{\columncolor{tWQ}}r}{8B} & \multicolumn{1}{>{\columncolor{tWQ}}r}{4B} & \multicolumn{1}{>{\columncolor{tWQ}}r}{Pi} & \multicolumn{1}{>{\columncolor{tWQ}}r|}{OC} & \multicolumn{1}{>{\columncolor{tSAGE}}r}{8B} & \multicolumn{1}{>{\columncolor{tSAGE}}r}{4B} & \multicolumn{1}{>{\columncolor{tSAGE}}r}{Pi} & \multicolumn{1}{>{\columncolor{tSAGE}}r}{OC} \\
\midrule
\multirow{5}{*}{\rotatebox[origin=c]{90}{\footnotesize\textbf{In-house}}} & Closed-book & 6.6 & 6.6 & 4.9 & 1.1 & 42.9 & 40.0 & 35.6 & 34.2 & 24.9 & 21.4 & 28.8 & 22.6 \\ %
 & One-shot RAG & 28.9 & 27.0 & 28.9 & 27.8 & 55.7 & 58.3 & 55.8 & 55.8 & 26.2 & 24.5 & 23.7 & 20.2 \\ %
 & RAG-SFT & 27.9 & 24.0 & 28.9 & 28.9 & 63.5 & 61.4 & 62.8 & 63.2 & 41.3 & \underline{37.9} & 24.2 & 20.6 \\ %
 & Agent, zero-shot & 37.1 & \underline{36.0} & 41.5 & 39.9 & 64.3 & 59.8 & 63.0 & 60.8 & 36.7 & 28.6 & 31.7 & 33.6 \\ %
 & Agent + SFT & 37.9 & 32.9 & \underline{42.3} & \textbf{40.7} & 65.9 & 58.2 & 64.6 & 62.4 & 33.1 & 31.2 & 28.1 & 30.0 \\ %
\midrule
\multirow{6}{*}{\rotatebox[origin=c]{90}{\footnotesize\textbf{External}}} & DeepEyes + cards~\cite{deepeyes2025} & 20.8 & -- & 23.2 & 21.3 & 28.3 & -- & 54.0 & 51.2 & 22.0 & -- & 22.1 & 19.7 \\ %
 & Qwen2.5-VL~\cite{qwen25vl} & 23.8 & -- & 22.4 & 23.0 & 44.3 & -- & 44.0 & 42.2 & 19.9 & -- & 19.0 & 18.2 \\ %
 & Qwen2.5-VL-32B-AWQ (4-bit)~\cite{qwen25vl} & 24.4$^{\ast}$ & -- & 17.4 & 31.3 & 38.7$^{\ast}$ & -- & 28.8 & 48.6 & 21.5$^{\ast}$ & -- & 21.8 & 21.1 \\ %
 & DeepEyes weights~\cite{deepeyes2025} & 31.3 & -- & 25.3 & 22.2 & 53.1 & -- & 51.4 & 49.2 & 25.6 & -- & 22.5 & 23.2 \\ %
 & MMSearch-R1 weights~\cite{wu2026mmsearch} & 27.8 & -- & 24.3 & 25.6 & 52.2 & -- & 44.4 & 44.0 & 22.3 & -- & 20.9 & 20.9 \\ %
 & SenseNova-MARS weights~\cite{chng2026sensesearch} & \underline{40.8} & -- & 37.9 & 36.5 & \textbf{69.4} & -- & \underline{67.7} & 65.5 & 35.0 & -- & 31.7 & 30.5 \\ %
\midrule
 & \textbf{ViBES} (ours) & \textbf{42.7} & \textbf{38.6} & \textbf{43.1} & \underline{40.1} & 66.4 & \underline{65.0} & 67.2 & \underline{67.2} & \textbf{42.5} & 36.9 & \underline{42.7} & \underline{42.4} \\ %
 &\textbf{ViBES-Lens} (ours) & \textbf{42.7} & \textbf{38.6} & \textbf{43.1} & \underline{40.1} & \underline{66.8} & \textbf{65.4} & \textbf{68.4} & \textbf{67.4} & \underline{42.2} & \textbf{39.1} & \textbf{43.6} & \textbf{42.9} \\ %
\bottomrule
\end{tabular}}
\end{table*}

\noindent\textbf{Datasets.} We evaluate on four image--text collections. MC-Search~\cite{ning2026mcsearch} asks
long-form, multi-fact questions (see Appendix). WebQA~\cite{webqa2022} asks short-answer
questions; we use its image questions for the ViBES cells, the matched comparisons and the crop analyses.
SAGE~\cite{sage2026} asks for exact disease label of a plant photograph, a domain-specific task that general-purpose agents are not trained for, and MultiModalQA~\cite{talmor2021mmqa}
contains image--text questions without tables.

\noindent\textbf{Implementation.} The policy is Qwen3-VL-Instruct~\cite{qwen3vl2025} with 8B or 4B parameters.
Qwen3-VL-Embedding-2B indexes every record by its text, its image and both. This index is the only source of evidence
for every system, including the external methods; no system uses web search. In all experiments, $H_0$ holds one
seeded retrieval with $(q,x)$, which is also the input of one-shot RAG. The budgets are $K_R{=}3$ (including the seeded
retrieval), $K_I{=}3$, $K_C{=}4$ and $K{=}10$. The system prompt states these limits and every observation reports the
remaining budget, for every system run in our loop. The tools, the scoring, statistics and confidence-evaluation protocols, and the deviations from preregistration are
described in the Appendix.

\noindent\textbf{Training recipe.} SFT trains LoRA adapters ($r{=}16$)~\cite{hu2022lora} on the filtered teacher
episodes of Section~\ref{sec:policy}. GRPO~\cite{grpo2024} then samples $G{=}8$ episodes per question at temperature
0.8, with 32 questions per step, a learning rate of $2{\times}10^{-5}$, $\lambda{=}0.5$ and no Kullback--Leibler term.

\noindent\textbf{Baselines.} Each in-house baseline removes one component of ViBES (Table~\ref{tab:indomain}).
\emph{Closed-book} answers from question and input image alone. \emph{One-shot RAG} answers in a single call from
the five records of the seeded retrieval, and \emph{RAG-SFT} fine-tunes this answer step on the same training
questions, which separates answer training from iterative search. \emph{Agent, zero-shot} runs the base model with our
tools and budgets, and \emph{Agent + SFT} stops before GRPO.

\noindent\textbf{External agent harnesses.} We also serve the 8B checkpoint to two general-purpose agent harnesses,
pi~\cite{zechner2026pi} and OpenCode~\cite{opencode2026}. Their tools call our executor, so retrieval and budgets are
unchanged. Each harness keeps its own system prompt and reads the answer from a final line, so harness answers carry no
citations.
\section{Results}
\label{sec:results}\label{sec:ablations}
We first examine agentic evidence acquisition (Section~\ref{sec:mainres}), then visual region selection
(Section~\ref{sec:q3}) and its effect on correctness prediction (Section~\ref{sec:confres}), and close with tool use,
cost and portability (Section~\ref{sec:portability}).

\begin{table*}[t]
\centering
\caption{\textbf{Effect of visual region selection on answer accuracy (\%).} The agent is fixed while varying the cropping mechanism. MC-Search 8B uses the zero-shot agent with cropping enabled. Each cell reports one run.}
\label{tab:crop_ablation}
\resizebox{\textwidth}{!}{\footnotesize\setlength{\tabcolsep}{5.2pt}\renewcommand{\arraystretch}{1.08}\colorlet{tMCS}{vsMCS!9}\colorlet{tWQ}{vsWQ!9}\colorlet{tSAGE}{vsSAGE!12}\setlength{\aboverulesep}{0pt}\setlength{\belowrulesep}{0pt}\setlength{\extrarowheight}{1.2pt}
\begin{tabular}{@{}ll|rrrrrr|rrrrrr|rrrrrr@{}}
\toprule
& & \multicolumn{6}{>{\columncolor{tMCS}}c|}{\textbf{MC-Search}} & \multicolumn{6}{>{\columncolor{tWQ}}c|}{\textbf{WebQA}} & \multicolumn{6}{>{\columncolor{tSAGE}}c}{\textbf{SAGE}} \\
\cmidrule(lr){3-8}\cmidrule(lr){9-14}\cmidrule(lr){15-20}
& & \multicolumn{3}{>{\columncolor{tMCS}}c}{8B, zero-shot} & \multicolumn{3}{>{\columncolor{tMCS}}c|}{4B} & \multicolumn{3}{>{\columncolor{tWQ}}c}{8B} & \multicolumn{3}{>{\columncolor{tWQ}}c|}{4B} & \multicolumn{3}{>{\columncolor{tSAGE}}c}{8B} & \multicolumn{3}{>{\columncolor{tSAGE}}c}{4B} \\
\cmidrule(lr){3-5}\cmidrule(lr){6-8}\cmidrule(lr){9-11}\cmidrule(lr){12-14}\cmidrule(lr){15-17}\cmidrule(lr){18-20}
\textbf{Method} & \textbf{Crop regions} & \multicolumn{1}{>{\columncolor{tMCS}}r}{Loop} & \multicolumn{1}{>{\columncolor{tMCS}}r}{Pi} & \multicolumn{1}{>{\columncolor{tMCS}}r}{OC} & \multicolumn{1}{>{\columncolor{tMCS}}r}{Loop} & \multicolumn{1}{>{\columncolor{tMCS}}r}{Pi} & \multicolumn{1}{>{\columncolor{tMCS}}r|}{OC} & \multicolumn{1}{>{\columncolor{tWQ}}r}{Loop} & \multicolumn{1}{>{\columncolor{tWQ}}r}{Pi} & \multicolumn{1}{>{\columncolor{tWQ}}r}{OC} & \multicolumn{1}{>{\columncolor{tWQ}}r}{Loop} & \multicolumn{1}{>{\columncolor{tWQ}}r}{Pi} & \multicolumn{1}{>{\columncolor{tWQ}}r|}{OC} & \multicolumn{1}{>{\columncolor{tSAGE}}r}{Loop} & \multicolumn{1}{>{\columncolor{tSAGE}}r}{Pi} & \multicolumn{1}{>{\columncolor{tSAGE}}r}{OC} & \multicolumn{1}{>{\columncolor{tSAGE}}r}{Loop} & \multicolumn{1}{>{\columncolor{tSAGE}}r}{Pi} & \multicolumn{1}{>{\columncolor{tSAGE}}r}{OC} \\
\midrule
ViBES & own box & \underline{38.9} & 40.2 & \underline{38.5} & \textbf{41.8} & \textbf{37.0} & 37.7 & \underline{66.4} & 67.2 & \underline{67.2} & \underline{65.0} & \textbf{63.8} & \underline{62.6} & \underline{42.5} & 42.7 & 42.4 & 36.9 & 32.6 & \underline{36.7} \\
ViBES-Lens & one box & 38.6 & \underline{42.1} & 38.2 & 41.1 & 35.8 & 37.9 & 66.0 & 66.4 & 65.8 & 64.4 & 62.6 & 62.2 & \textbf{43.1} & \underline{43.4} & \underline{42.5} & \underline{38.4} & \underline{34.6} & \textbf{36.9} \\
ViBES-Lens & mask & \textbf{39.5} & 41.4 & 38.0 & \underline{41.4} & 35.4 & \textbf{39.5} & 66.2 & \underline{67.8} & 65.4 & \underline{65.0} & 62.4 & 61.6 & 41.7 & 40.8 & 42.2 & 36.9 & 32.6 & 35.8 \\
ViBES-Lens & $N=4$ boxes & 38.5 & \textbf{42.7} & \textbf{39.5} & \textbf{41.8} & \underline{36.3} & \underline{38.2} & \textbf{66.8} & \textbf{68.4} & \textbf{67.4} & \textbf{65.4} & \underline{63.0} & \textbf{63.4} & 42.2 & \textbf{43.6} & \textbf{42.9} & \textbf{39.1} & \textbf{34.8} & \textbf{36.9} \\
\bottomrule
\end{tabular}}
\end{table*}
\subsection{Agentic Evidence Acquisition}
\label{sec:mainres}\label{sec:q2}\label{sec:rq1}
\textbf{ViBES improves accuracy over one-shot RAG.} On matched questions in our loop, ViBES is 13.6 accuracy points
above one-shot RAG (Table~\ref{tab:indomain}); the MC-Search agents of this
table never crop, so the ViBES-Lens row repeats ViBES there. The zero-shot agent already
recovers about half of this gain, so the interactive loop helps before any training, and GRPO beats the SFT agent
in every loop column, by 0.5 to 9.4 points at 8B.

\noindent\textbf{Fine-tuning small models matters most in the specialized domain.} On SAGE, ViBES and RAG-SFT at 8B exceed the best
external agent, the SenseNova-MARS weights, by 7.5 and 6.3 points and the 32B model by about 20, whereas on WebQA the
SenseNova-MARS weights remain ahead (Table~\ref{tab:indomain}).

\noindent\textbf{Accuracy follows the retrieved evidence.} ViBES raises gold-evidence recall by 20.2 points over
one-shot RAG (see Appendix for the recall of every system), and the gain is largest where
the one-shot retrieval misses the gold records. On MC-Search and SAGE, its accuracy
rises from at most one in five questions without any gold record to about three in four with all of them (see
Appendix). A fixed-observation replay points the same
way: with the answerer fixed, the observations collected by the GRPO agent raise SAGE accuracy more than changing the
answerer does (see Appendix). These findings identify iterative evidence acquisition as a major
contributor to answer accuracy, and motivate a separate look at how the agent selects and inspects visual regions.

\subsection{Visual Region Selection}
\label{sec:q3}
\textbf{Controlled comparison.} Table~\ref{tab:crop_ablation} keeps the agent fixed and changes only the
region-selection mechanism of Section~\ref{sec:module}; the 8B MC-Search cells use the zero-shot agent, because the
trained 8B agent has no crop tool. Averaged over the nine columns of each model size, ViBES-Lens with one box is level with the own box, with a
mask it is 0.3 and 0.4 points below it, and with $N$ boxes it is 0.7 points above it at 8B and 0.5 points at
4B, with the best accuracy in 13 of the 18 columns (see Appendix).

\noindent\textbf{Coverage and controls.} With $N$ boxes, ViBES-Lens shows more of the image: the union of its boxes covers a median of
about two thirds of it (see Appendix). On WebQA and SAGE, it also scores above gray and size-matched random
boxes in every cell. Larger coverage does not by itself indicate better
localization. The best source still changes across
datasets and harnesses, and recall stays within two points of the own box, so
no learned mechanism gives a clear and consistent accuracy gain over the agent's own box. We next ask whether the mechanisms change how well a trajectory confidence signal predicts answer correctness.

\begin{table}[!h]
\centering
\caption{\textbf{Correctness prediction across visual inspection mechanisms.} Learned minus own box in AUROC, ECE and Brier score (0--1 scale) of the visibility probe, on same set of questions, each scored against its own correctness. \swatch{dGood}{Teal}: learned better; \swatch{dBad}{orange}: worse. Mean: over datasets. Each cell uses one split; see Appendix for both systems' values.}
\label{tab:confidence}
\resizebox{\columnwidth}{!}{\footnotesize\setlength{\tabcolsep}{3.5pt}%
\begin{tabular}{@{}l|ccc|ccc@{}}
\toprule
& \multicolumn{3}{c|}{\textbf{8B}} & \multicolumn{3}{c}{\textbf{4B}} \\
\cmidrule(lr){2-4}\cmidrule(l){5-7}
\textbf{Dataset} & $\Delta$AUROC$\uparrow$ & $\Delta$ECE$\downarrow$ & $\Delta$Brier$\downarrow$ & $\Delta$AUROC$\uparrow$ & $\Delta$ECE$\downarrow$ & $\Delta$Brier$\downarrow$ \\
\midrule
\multicolumn{7}{@{}l}{\textit{ViBES-Lens, one box}} \\
\quad WebQA & \cellcolor{dGood}$+$0.036 & \cellcolor{dBad}$+$0.024 & \cellcolor{dBad}$+$0.004 & \cellcolor{dGood}$+$0.022 & \cellcolor{dBad}$+$0.022 & \cellcolor{dBad}$+$0.002 \\
\quad SAGE & \cellcolor{dGood}$+$0.069 & \cellcolor{dGood}$-$0.049 & \cellcolor{dGood}$-$0.020 & \cellcolor{dGood}$+$0.073 & \cellcolor{dGood}$-$0.010 & \cellcolor{dGood}$-$0.029 \\
\quad MC-Search & \cellcolor{dGood}$+$0.029 & \cellcolor{dGood}$-$0.018 & \cellcolor{dGood}$-$0.005 & \cellcolor{dGood}$+$0.136 & \cellcolor{dBad}$+$0.020 & \cellcolor{dBad}$+$0.004 \\
\quad \textit{Mean} & \cellcolor{dGood}$+$0.045 & \cellcolor{dGood}$-$0.014 & \cellcolor{dGood}$-$0.007 & \cellcolor{dGood}$+$0.077 & \cellcolor{dBad}$+$0.010 & \cellcolor{dGood}$-$0.008 \\
\midrule
\multicolumn{7}{@{}l}{\textit{ViBES-Lens, mask}} \\
\quad WebQA & \cellcolor{dGood}$+$0.052 & \cellcolor{dBad}$+$0.011 & \cellcolor{dGood}$-$0.001 & \cellcolor{dGood}$+$0.062 & \cellcolor{dBad}$+$0.014 & 0.000 \\
\quad SAGE & \cellcolor{dGood}$+$0.065 & \cellcolor{dBad}$+$0.058 & \cellcolor{dGood}$-$0.003 & \cellcolor{dGood}$+$0.016 & \cellcolor{dGood}$-$0.015 & \cellcolor{dGood}$-$0.013 \\
\quad MC-Search & \cellcolor{dGood}$+$0.108 & \cellcolor{dGood}$-$0.003 & \cellcolor{dGood}$-$0.015 & \cellcolor{dGood}$+$0.018 & \cellcolor{dGood}$-$0.041 & \cellcolor{dBad}$+$0.002 \\
\quad \textit{Mean} & \cellcolor{dGood}$+$0.075 & \cellcolor{dBad}$+$0.022 & \cellcolor{dGood}$-$0.006 & \cellcolor{dGood}$+$0.032 & \cellcolor{dGood}$-$0.014 & \cellcolor{dGood}$-$0.004 \\
\midrule
\multicolumn{7}{@{}l}{\textit{ViBES-Lens, $N=4$ boxes}} \\
\quad WebQA & \cellcolor{dGood}$+$0.011 & \cellcolor{dBad}$+$0.008 & 0.000 & \cellcolor{dGood}$+$0.014 & \cellcolor{dBad}$+$0.011 & \cellcolor{dGood}$-$0.003 \\
\quad SAGE & \cellcolor{dBad}$-$0.008 & \cellcolor{dGood}$-$0.016 & \cellcolor{dBad}$+$0.001 & \cellcolor{dGood}$+$0.041 & \cellcolor{dGood}$-$0.042 & \cellcolor{dGood}$-$0.023 \\
\quad MC-Search & \cellcolor{dGood}$+$0.003 & \cellcolor{dGood}$-$0.037 & \cellcolor{dGood}$-$0.007 & \cellcolor{dBad}$-$0.010 & \cellcolor{dGood}$-$0.002 & \cellcolor{dGood}$-$0.005 \\
\quad \textit{Mean} & \cellcolor{dGood}$+$0.002 & \cellcolor{dGood}$-$0.015 & \cellcolor{dGood}$-$0.002 & \cellcolor{dGood}$+$0.015 & \cellcolor{dGood}$-$0.011 & \cellcolor{dGood}$-$0.010 \\
\bottomrule
\end{tabular}}
\end{table}
\begin{figure}[t]
\centering
\includegraphics[width=0.75\columnwidth]{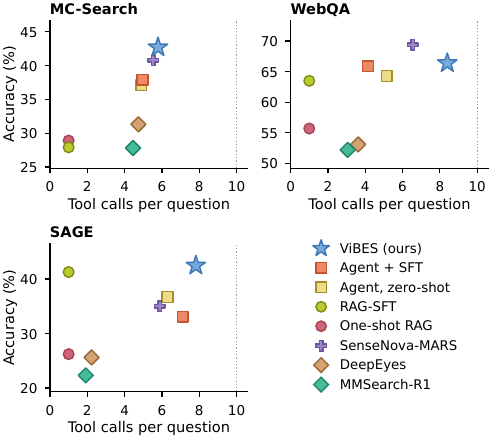}
\caption{\textbf{Answer accuracy versus tool usage.} Test accuracy and mean tool calls per question for 8B-class systems under a shared 10-call limit. WebQA evaluations use different question subsets across systems.}
\label{fig:budget}
\end{figure}
\begin{figure}[t]
\centering
\includegraphics[width=0.95\columnwidth]{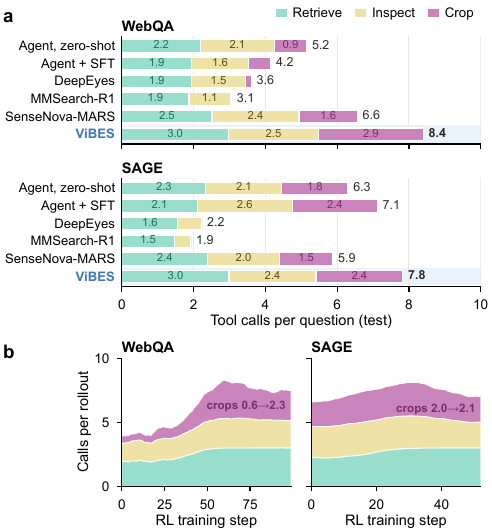}
\caption{\textbf{Tool-use behavior during inference and training.} (a) Mean retrieval, inspection, and cropping calls per question for 8B-class systems. (b) Evolution of tool usage during RL training of the 8B agent. Results are on WebQA and SAGE.}
\label{fig:tooluse}
\end{figure}
\begin{figure}[t]
\centering
\includegraphics[width=0.99\columnwidth]{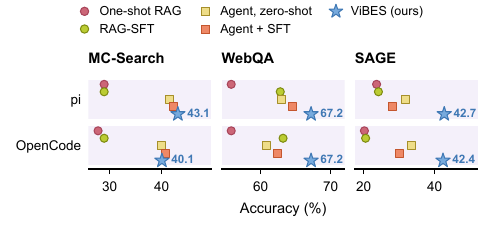}
\caption{\textbf{Transfer across agent execution harnesses.} Test accuracy of 8B systems in pi and OpenCode using the same model checkpoints. Axis ranges differ across benchmarks.}
\label{fig:harness}
\end{figure}
\subsection{Visual Inspection and Correctness Prediction}
\label{sec:confres}
\textbf{Protocol.} We compute the AUROC, ECE and Brier score of the probe of Section~\ref{sec:confidence} on
questions where both systems crop, each against its own correctness (Table~\ref{tab:confidence}).

\noindent\textbf{ViBES-Lens with one box gives the strongest and most consistent AUROC gain.} It raises AUROC over the own box in
all six dataset--size pairs, with a mean $\Delta$AUROC of $+0.061$ and a range of $+0.022$ to $+0.136$, while its
accuracy stays level with the own box (Section~\ref{sec:q3}).

\noindent\textbf{Mechanisms with similar accuracy differ in correctness discrimination.} The mask mode also raises
AUROC in all six pairs (mean $+0.054$, largest $+0.108$ on MC-Search at 8B), whereas the $N$-box mode does so in four
of the six pairs, with a mean of only $+0.009$. The AUROC advantage therefore comes from the one-box and
mask modes, not from the $N$-box mode, even though the $N$-box mode has the highest average accuracy in
Table~\ref{tab:crop_ablation}.

\noindent\textbf{Calibration changes are small and mixed.} Mean changes lie between $-0.015$ and $+0.022$ for ECE and
between $-0.010$ and $-0.002$ for Brier score; the one-box mode lowers both on SAGE but raises both on WebQA.

\noindent\textbf{Interpretation.} The one-box and mask modes discriminate correctness better without an accuracy
gain or a consistent calibration gain, so visual inspection strategies should be evaluated on both accuracy and
correctness discrimination.

\subsection{Tool Use, Cost and Portability}
\label{sec:portability}\label{sec:external}
\textbf{ViBES uses more of its tool budget and reaches the evidence earlier.} All systems in our loop share the same limits of 10 tool calls and 3 retrievals and are shown them in the prompt, but only ViBES and Agent + SFT were trained under them (see Appendix). ViBES makes the most tool calls per question on every dataset, 5.8 to 8.4 of the 10 allowed, and is the most accurate system on MC-Search and SAGE (Figure~\ref{fig:budget}). The DeepEyes and MMSearch-R1 weights stop after about 2 calls on SAGE and 3 to 5 calls on MC-Search and WebQA, and score within 4 points of one-shot RAG. The extra calls reach the evidence early: after three tool calls ViBES has retrieved 64.4\% of the WebQA gold records and 54.3\% on SAGE, more than the zero-shot agent retrieves after ten (61.7\% and 48.6\%; see Appendix). During reinforcement learning, retrievals per rollout rise to the limit of 3 on WebQA and SAGE, and crops per rollout rise from 0.6 to 2.3 on WebQA (Figure~\ref{fig:tooluse}). The cost is high: ViBES processes 10 to 19 times as many prompt tokens per question as one-shot RAG and more than every external model in our loop (see Appendix), and it keeps calling tools after the evidence is complete because the reward does not penalize calls. On WebQA, the SenseNova-MARS weights are more accurate than ViBES with fewer calls when evaluated on all questions. Each point in Figure~\ref{fig:budget} is a different system under the same limits, so the figure does not show how accuracy changes with the budget, and the call counts of the external weights describe their behavior in our loop, not under their own protocols.

\noindent\textbf{ViBES remains more accurate in two other agent harnesses.} We moved the 8B weights and tools into pi
and OpenCode without retraining (Figure~\ref{fig:harness}; Table~\ref{tab:indomain}, Pi and OC). ViBES is more accurate
than every in-house baseline in five of the six harness--dataset cells (Agent + SFT is 0.6 points higher in OpenCode on
MC-Search), 15.1 points above one-shot RAG and 12.3 above RAG-SFT in the same harness on average. Each harness also
changes system prompt, tool protocol and image placement, so these runs show portability to two integrations,
not harness invariance.

\section{Conclusions}

We presented ViBES, an agentic multimodal retrieval-augmented generation (RAG) framework that integrates iterative evidence retrieval, record inspection, and visual cropping under a fixed tool-call budget. ViBES improves answer accuracy by 13.6\% and gold-evidence recall by 20.2\% over matched one-shot RAG baselines across three benchmarks. Our analysis indicates that improved evidence acquisition accounts for a large part of these gains, which largely persist across alternative agent execution harnesses without retraining. We also introduced ViBES-Lens, a reasoning-conditioned visual region selector, and investigated trajectory-level confidence estimation in agentic multimodal RAG. We compared visual inspection mechanisms and showed that ViBES-Lens improves the correctness-discrimination AUROC relative to ViBES in all six evaluated dataset–model configurations. The tool interface of ViBES, consisting of evidence retrieval, record inspection, and visual cropping, can be applied to other collections that pair text records with images (for example, scientific articles, technical manuals, or clinical reports). ViBES-Lens is trained without bounding-box annotations, so same training procedure can be used in domains that lack such annotations. These findings show that agentic multimodal reasoning should be evaluated beyond answer accuracy, and they can be used to study effect of visual evidence inspection on the reliability of trajectory-level correctness prediction.

\paragraph{Limitations.} We construct the supervised training data only from successful teacher-generated trajectories, so these data may underrepresent challenging evidence-acquisition scenarios and limit the diversity of learned reasoning behaviors. ViBES-Lens improves trajectory-level correctness discrimination but does not consistently improve answer accuracy. We will investigate constructing more diverse training trajectories and improving visual region selection to increase answer accuracy.
\clearpage
\section*{LLM Usage Statement}
Large language models were used to support the development and preparation of this work, including answer scoring, debugging baseline training code, optimizing data-loading scripts, and improving manuscript clarity and grammar. Any AI-assisted code was reviewed and validated through unit tests, numerical checks, and empirical comparisons with established baselines. The experimental design was developed by the authors, who take full responsibility for the correctness of the manuscript, experimental results, and codebase.

{\small
\bibliographystyle{ieeenat_fullname}
\bibliography{references}

\begin{thebibliography}{36}
\providecommand{\natexlab}[1]{#1}
\providecommand{\url}[1]{\texttt{#1}}
\expandafter\ifx\csname urlstyle\endcsname\relax
  \providecommand{\doi}[1]{doi: #1}\else
  \providecommand{\doi}{doi: \begingroup \urlstyle{rm}\Url}\fi

\bibitem[Arshad et~al.(2026)Arshad, Roy, Shen, Elango, Chiranjeevi, Singh, Ganapathysubramanian, Hegde, Singh, and Sarkar]{sage2026}
Muhammad~Arbab Arshad, Tirtho Roy, Yanben Shen, Dinakaran Elango, Shivani Chiranjeevi, Asheesh~K. Singh, Baskar Ganapathysubramanian, Chinmay Hegde, Arti Singh, and Soumik Sarkar.
\newblock \href{https://arxiv.org/abs/2605.09768}{{SAGE}: Scalable Agentic Grounded Evaluation for Crop Disease Diagnosis}.
\newblock In \emph{CVPR 2026 Workshop on Emerging Directions in Data for Multimodal Foundation Models}, 2026.
\newblock Non-archival track; also available as arXiv:2605.09768.

\bibitem[Bai et~al.(2025)Bai, Chen, Liu, et~al.]{qwen25vl}
Shuai Bai, Keqin Chen, Xuejing Liu, et~al.
\newblock \href{https://arxiv.org/abs/2502.13923}{Qwen2.5-VL Technical Report}.
\newblock \emph{arXiv preprint arXiv:2502.13923}, 2025.

\bibitem[Brier(1950)]{brier1950verification}
Glenn~W. Brier.
\newblock Verification of forecasts expressed in terms of probability.
\newblock \emph{Monthly Weather Review}, 78\penalty0 (1):\penalty0 1--3, 1950.

\bibitem[Chang et~al.(2022)Chang, Narang, Suzuki, Cao, Gao, and Bisk]{webqa2022}
Yingshan Chang, Mridu Narang, Hisami Suzuki, Guihong Cao, Jianfeng Gao, and Yonatan Bisk.
\newblock \href{https://openaccess.thecvf.com/content/CVPR2022/html/Chang_WebQA_Multihop_and_Multimodal_QA_CVPR_2022_paper.html}{{WebQA}: Multihop and Multimodal {QA}}.
\newblock In \emph{Proceedings of the IEEE/CVF Conference on Computer Vision and Pattern Recognition (CVPR)}, pages 16495--16504, 2022.

\bibitem[Chen et~al.(2022)Chen, Hu, Chen, Verga, and Cohen]{chen2022murag}
Wenhu Chen, Hexiang Hu, Xi Chen, Pat Verga, and William Cohen.
\newblock \href{https://aclanthology.org/2022.emnlp-main.375/}{{MuRAG}: Multimodal Retrieval-Augmented Generator for Open Question Answering over Images and Text}.
\newblock In \emph{Proceedings of the 2022 Conference on Empirical Methods in Natural Language Processing}, pages 5558--5570. Association for Computational Linguistics, 2022.

\bibitem[Chng et~al.(2026)Chng, Hu, Tong, Li, Chen, Yu, Lu, Guo, Deng, Xie, Huang, and Lu]{chng2026sensesearch}
Yong~Xien Chng, Tao Hu, Wenwen Tong, Xueheng Li, Jiandong Chen, Haojia Yu, Jiefan Lu, Hewei Guo, Hanming Deng, Chengjun Xie, Gao Huang, and Lewei Lu.
\newblock \href{https://openaccess.thecvf.com/content/CVPR2026/html/Chng_SenseSearch_Empowering_Vision-Language_Models_with_High-Resolution_Agentic_Search-Reasoning_via_Reinforcement_CVPR_2026_paper.html}{{SenseSearch}: Empowering Vision-Language Models with High-Resolution Agentic Search-Reasoning via Reinforcement Learning}.
\newblock In \emph{Proceedings of the IEEE/CVF Conference on Computer Vision and Pattern Recognition (CVPR)}, pages 26508--26517, 2026.

\bibitem[Duan et~al.(2025)Duan, Diffenderfer, Madireddy, Chen, Kailkhura, and Xu]{duan2025uprop}
Jinhao Duan, James Diffenderfer, Sandeep Madireddy, Tianlong Chen, Bhavya Kailkhura, and Kaidi Xu.
\newblock \href{https://arxiv.org/abs/2506.17419}{{UProp}: Investigating the Uncertainty Propagation of {LLMs} in Multi-Step Agentic Decision-Making}.
\newblock \emph{arXiv preprint arXiv:2506.17419}, 2025.

\bibitem[Faysse et~al.(2025)Faysse, Sibille, Wu, Omrani, Viaud, Hudelot, and Colombo]{faysse2025colpali}
Manuel Faysse, Hugues Sibille, Tony Wu, Bilel Omrani, Gautier Viaud, C{\'e}line Hudelot, and Pierre Colombo.
\newblock \href{https://proceedings.iclr.cc/paper_files/paper/2025/hash/99e9e141aafc314f76b0ca3dd66898b3-Abstract-Conference.html}{{ColPali}: Efficient Document Retrieval with Vision Language Models}.
\newblock In \emph{International Conference on Learning Representations}, 2025.

\bibitem[Gao et~al.(2026)Gao, Xu, Zhang, Mei, Yang, and Metaxas]{gao2026climb}
Hang Gao, Wujiang Xu, Zhixing Zhang, Kai Mei, Jingyi Yang, and Dimitris~N. Metaxas.
\newblock \href{https://aclanthology.org/2026.findings-emnlp.660/}{{CLIMB}: Confidence-Guided Complementary Evidence for Multimodal Retrieval-Augmented Generation}.
\newblock In \emph{Findings of the Association for Computational Linguistics: EMNLP 2026}, pages 13842--13857, Budapest, Hungary, 2026. Association for Computational Linguistics.

\bibitem[Geng et~al.(2026)Geng, Xia, Zhang, Wang, Wang, Ding, Wang, Wu, Li, Zhao, Yin, Jiang, Xie, Huang, Yao, Fung, and Zhou]{geng2026webwatcher}
Xinyu Geng, Peng Xia, Zhen Zhang, Xinyu Wang, Qiuchen Wang, Ruixue Ding, Chenxi Wang, Jialong Wu, Kuan Li, Yida Zhao, Huifeng Yin, Yong Jiang, Pengjun Xie, Fei Huang, Huaxiu Yao, Yi~R. Fung, and Jingren Zhou.
\newblock \href{https://openreview.net/forum?id=8jsaazdAb3}{{WebWatcher}: Breaking New Frontiers of Vision-Language Deep Research Agent}.
\newblock In \emph{International Conference on Learning Representations}, 2026.

\bibitem[Guo et~al.(2017)Guo, Pleiss, Sun, and Weinberger]{guo2017calibration}
Chuan Guo, Geoff Pleiss, Yu Sun, and Kilian~Q. Weinberger.
\newblock On calibration of modern neural networks.
\newblock In \emph{International Conference on Machine Learning (ICML)}, 2017.

\bibitem[Guu et~al.(2020)Guu, Lee, Tung, Pasupat, and Chang]{guu2020realm}
Kelvin Guu, Kenton Lee, Zora Tung, Panupong Pasupat, and Mingwei Chang.
\newblock \href{https://proceedings.mlr.press/v119/guu20a.html}{{REALM}: Retrieval Augmented Language Model Pre-Training}.
\newblock In \emph{Proceedings of the 37th International Conference on Machine Learning}, pages 3929--3938. PMLR, 2020.

\bibitem[Hu et~al.(2022)Hu, Shen, Wallis, Allen-Zhu, Li, Wang, Wang, and Chen]{hu2022lora}
Edward~J. Hu, Yelong Shen, Phillip Wallis, Zeyuan Allen-Zhu, Yuanzhi Li, Shean Wang, Lu Wang, and Weizhu Chen.
\newblock \href{https://arxiv.org/abs/2106.09685}{{LoRA}: Low-Rank Adaptation of Large Language Models}.
\newblock In \emph{International Conference on Learning Representations (ICLR)}, 2022.

\bibitem[Jin et~al.(2025)Jin, Zeng, Yue, Yoon, Arik, Wang, Zamani, and Han]{searchr1_2025}
Bowen Jin, Hansi Zeng, Zhenrui Yue, Jinsung Yoon, Sercan Arik, Dong Wang, Hamed Zamani, and Jiawei Han.
\newblock \href{https://proceedings.colm.cc/paper_files/paper/2025/hash/1377f76686d56439a2bd7a91859972f5-Abstract-Conference.html}{{Search-R1}: Training {LLMs} to Reason and Leverage Search Engines with Reinforcement Learning}.
\newblock In \emph{Conference on Language Modeling}, 2025.

\bibitem[Kadavath et~al.(2022)Kadavath, Conerly, Askell, Henighan, et~al.]{kadavath2022know}
Saurav Kadavath, Tom Conerly, Amanda Askell, Tom Henighan, et~al.
\newblock \href{https://arxiv.org/abs/2207.05221}{Language Models (Mostly) Know What They Know}.
\newblock \emph{arXiv preprint arXiv:2207.05221}, 2022.

\bibitem[Karpukhin et~al.(2020)Karpukhin, Oguz, Min, Lewis, Wu, Edunov, Chen, and Yih]{karpukhin2020dpr}
Vladimir Karpukhin, Barlas Oguz, Sewon Min, Patrick Lewis, Ledell Wu, Sergey Edunov, Danqi Chen, and Wen-tau Yih.
\newblock \href{https://aclanthology.org/2020.emnlp-main.550/}{Dense Passage Retrieval for Open-Domain Question Answering}.
\newblock In \emph{Proceedings of the 2020 Conference on Empirical Methods in Natural Language Processing (EMNLP)}, pages 6769--6781. Association for Computational Linguistics, 2020.

\bibitem[Kwon et~al.(2023)Kwon, Li, Zhuang, Sheng, Zheng, Yu, Gonzalez, Zhang, and Stoica]{kwon2023vllm}
Woosuk Kwon, Zhuohan Li, Siyuan Zhuang, Ying Sheng, Lianmin Zheng, Cody~Hao Yu, Joseph~E. Gonzalez, Hao Zhang, and Ion Stoica.
\newblock \href{https://arxiv.org/abs/2309.06180}{Efficient Memory Management for Large Language Model Serving with {PagedAttention}}.
\newblock In \emph{Proceedings of the 29th Symposium on Operating Systems Principles (SOSP)}, 2023.

\bibitem[Lewis et~al.(2020)Lewis, Perez, Piktus, Petroni, Karpukhin, Goyal, K{\"u}ttler, Lewis, Yih, Rockt{\"a}schel, Riedel, and Kiela]{lewis2020rag}
Patrick Lewis, Ethan Perez, Aleksandra Piktus, Fabio Petroni, Vladimir Karpukhin, Naman Goyal, Heinrich K{\"u}ttler, Mike Lewis, Wen-tau Yih, Tim Rockt{\"a}schel, Sebastian Riedel, and Douwe Kiela.
\newblock \href{https://papers.nips.cc/paper/2020/hash/6b493230205f780e1bc26945df7481e5-Abstract.html}{Retrieval-Augmented Generation for Knowledge-Intensive {NLP} Tasks}.
\newblock In \emph{Advances in Neural Information Processing Systems}, pages 9459--9474, 2020.

\bibitem[Liu et~al.(2024)Liu, Dou, Wang, Peng, and Yue]{liu2024uncertaintycalibration}
Hao Liu, Zi-Yi Dou, Yixin Wang, Nanyun Peng, and Yisong Yue.
\newblock \href{https://aclanthology.org/2024.findings-emnlp.978/}{Uncertainty Calibration for Tool-Using Language Agents}.
\newblock In \emph{Findings of the Association for Computational Linguistics: EMNLP 2024}, pages 16781--16805, Miami, Florida, USA, 2024. Association for Computational Linguistics.

\bibitem[Liu et~al.(2025)]{liu2025visualarft}
Ziyu Liu et~al.
\newblock \href{https://arxiv.org/abs/2505.14246}{Visual Agentic Reinforcement Fine-Tuning}.
\newblock \emph{arXiv preprint arXiv:2505.14246}, 2025.

\bibitem[Narayan et~al.(2025)Narayan, Xu, Cao, Nerella, Patel, Shiee, Grasch, Jia, Yang, and Gan]{narayan2025deepmmsearch}
Kartik Narayan, Yang Xu, Tian Cao, Kavya Nerella, Vishal~M. Patel, Navid Shiee, Peter Grasch, Chao Jia, Yinfei Yang, and Zhe Gan.
\newblock \href{https://arxiv.org/abs/2510.12801}{{DeepMMSearch-R1}: Empowering Multimodal {LLMs} in Multimodal Web Search}.
\newblock arXiv preprint arXiv:2510.12801, 2025.

\bibitem[Ning et~al.(2026)Ning, Fu, Wei, Ai, Zou, Li, Tong, Zhu, Hamann, and He]{ning2026mcsearch}
Xuying Ning, Dongqi Fu, Tianxin Wei, Mengting Ai, Jiaru Zou, Ting-Wei Li, Hanghang Tong, Yada Zhu, Hendrik Hamann, and Jingrui He.
\newblock \href{https://openreview.net/forum?id=JEGDp1E4OH}{{MC-Search}: Evaluating and Enhancing Multimodal Agentic Search with Structured Long Reasoning Chains}.
\newblock In \emph{International Conference on Learning Representations}, 2026.

\bibitem[Platt(1999)]{platt1999probabilistic}
John~C. Platt.
\newblock Probabilistic outputs for support vector machines and comparisons to regularized likelihood methods.
\newblock In \emph{Advances in Large Margin Classifiers}, pages 61--74. MIT Press, 1999.

\bibitem[{Qwen Team}(2025)]{qwen3vl2025}
{Qwen Team}.
\newblock \href{https://github.com/QwenLM/Qwen3-VL}{{Qwen3-VL}}.
\newblock \url{https://github.com/QwenLM/Qwen3-VL}, 2025.

\bibitem[Shao et~al.(2024)Shao, Wang, Zhu, Xu, Song, Bi, Zhang, Zhang, Li, Wu, and Guo]{grpo2024}
Zhihong Shao, Peiyi Wang, Qihao Zhu, Runxin Xu, Junxiao Song, Xiao Bi, Haowei Zhang, Mingchuan Zhang, Y.~K. Li, Y. Wu, and Daya Guo.
\newblock \href{https://arxiv.org/abs/2402.03300}{{DeepSeekMath}: Pushing the Limits of Mathematical Reasoning in Open Language Models}.
\newblock arXiv preprint arXiv:2402.03300, 2024.

\bibitem[{SST}(2026)]{opencode2026}
{SST}.
\newblock \href{https://github.com/sst/opencode}{{OpenCode}: An Open-Source {AI} Coding Agent for the Terminal}.
\newblock GitHub repository, 2026.
\newblock Version 1.19.5 used in ViBES.

\bibitem[Talmor et~al.(2021)Talmor, Yoran, Catav, Lahav, Wang, Asai, Ilharco, Hajishirzi, and Berant]{talmor2021mmqa}
Alon Talmor, Ori Yoran, Amnon Catav, Dan Lahav, Yizhong Wang, Akari Asai, Gabriel Ilharco, Hannaneh Hajishirzi, and Jonathan Berant.
\newblock \href{https://arxiv.org/abs/2104.06039}{{MultiModalQA}: Complex Question Answering over Text, Tables and Images}.
\newblock In \emph{International Conference on Learning Representations (ICLR)}, 2021.

\bibitem[Wang et~al.(2025)Wang, Ding, Chen, Wu, Wang, Xie, and Zhao]{wang2025vidorag}
Qiuchen Wang, Ruixue Ding, Zehui Chen, Weiqi Wu, Shihang Wang, Pengjun Xie, and Feng Zhao.
\newblock \href{https://doi.org/10.18653/v1/2025.emnlp-main.464}{{ViDoRAG}: Visual Document Retrieval-Augmented Generation via Dynamic Iterative Reasoning Agents}.
\newblock In \emph{Proceedings of the 2025 Conference on Empirical Methods in Natural Language Processing}, pages 9113--9134, 2025.

\bibitem[Wu et~al.(2026)Wu, Deng, Li, Liu, You, Li, Ma, and Liu]{wu2026mmsearch}
Jinming Wu, Zihao Deng, Wei Li, Yiding Liu, Bo You, Bo Li, Zejun Ma, and Ziwei Liu.
\newblock \href{https://aclanthology.org/2026.acl-long.114/}{{MMSearch-R1}: Incentivizing {LMMs} to Search}.
\newblock In \emph{Proceedings of the 64th Annual Meeting of the Association for Computational Linguistics (Volume 1: Long Papers)}, pages 2456--2487. Association for Computational Linguistics, 2026.

\bibitem[Yasunaga et~al.(2023)Yasunaga, Aghajanyan, Shi, James, Leskovec, Liang, Lewis, Zettlemoyer, and Yih]{racm3_2023}
Michihiro Yasunaga, Armen Aghajanyan, Weijia Shi, Richard James, Jure Leskovec, Percy Liang, Mike Lewis, Luke Zettlemoyer, and Wen-Tau Yih.
\newblock \href{https://proceedings.mlr.press/v202/yasunaga23a.html}{Retrieval-Augmented Multimodal Language Modeling}.
\newblock In \emph{Proceedings of the 40th International Conference on Machine Learning}, pages 39755--39769. PMLR, 2023.

\bibitem[Yu et~al.(2025)Yu, Tang, Xu, Cui, Ran, Yan, Liu, Wang, Han, Liu, and Sun]{yu2025visrag}
Shi Yu, Chaoyue Tang, Bokai Xu, Junbo Cui, Junhao Ran, Yukun Yan, Zhenghao Liu, Shuo Wang, Xu Han, Zhiyuan Liu, and Maosong Sun.
\newblock \href{https://proceedings.iclr.cc/paper_files/paper/2025/hash/3640a1997a4c9571cea9db2c82e1fc35-Abstract-Conference.html}{{VisRAG}: Vision-Based Retrieval-Augmented Generation on Multi-Modality Documents}.
\newblock In \emph{International Conference on Learning Representations}, 2025.

\bibitem[Zechner(2026)]{zechner2026pi}
Mario Zechner.
\newblock \href{https://github.com/badlogic/pi-mono}{pi: An Open-Source Terminal Coding Agent}.
\newblock GitHub repository, 2026.
\newblock Version 0.85.0 used in ViBES.

\bibitem[Zhang et~al.(2026{\natexlab{a}})Zhang, Xiong, and Wu]{zhang2026agenticconfidence}
Jiaxin Zhang, Caiming Xiong, and Chien-Sheng Wu.
\newblock \href{https://proceedings.mlr.press/v306/zhang26gy.html}{Agentic Confidence Calibration}.
\newblock In \emph{Proceedings of the 43rd International Conference on Machine Learning}, pages 159488--159525. PMLR, 2026{\natexlab{a}}.

\bibitem[Zhang et~al.(2026{\natexlab{b}})Zhang, Sun, Song, Qi, and Zheng]{zhang2026vsearcher}
Ruiyang Zhang, Qianguo Sun, Chao Song, Yiyan Qi, and Zhedong Zheng.
\newblock \href{https://arxiv.org/abs/2603.02795}{{VSearcher}: Long-Horizon Multimodal Search Agent via Reinforcement Learning}.
\newblock \emph{arXiv preprint arXiv:2603.02795}, 2026{\natexlab{b}}.

\bibitem[Zhao et~al.(2026)Zhao, Liu, Lei, Sun, Gao, and Li]{zhao2026mciterl}
Suifeng Zhao, Zida Liu, Xinyu Lei, Lei Sun, Jun Gao, and Sujian Li.
\newblock \href{https://arxiv.org/abs/2608.21808}{{MCite-RL}: Towards Reliable Multimodal {RAG} via Citation-enhanced Agentic Reinforcement Learning}.
\newblock \emph{arXiv preprint arXiv:2608.21808}, 2026.

\bibitem[Zheng et~al.(2026)Zheng, Yang, Hong, Zhao, Xu, Yang, Shen, and Yu]{deepeyes2025}
Ziwei Zheng, Michael Yang, Jack Hong, Chenxiao Zhao, Guohai Xu, Le Yang, Chao Shen, and Xing Yu.
\newblock \href{https://proceedings.iclr.cc/paper_files/paper/2026/hash/cdb347d7516d52a9280ea9d6708911f6-Abstract-Conference.html}{{DeepEyes}: Incentivizing ``Thinking with Images'' via Reinforcement Learning}.
\newblock In \emph{International Conference on Learning Representations}, 2026.

\end{thebibliography}
}

\clearpage
\twocolumn[{\centering\Large\bfseries Supplementary Material\par\vspace{1.5em}}]
\setcounter{figure}{0}\setcounter{table}{0}\setcounter{equation}{0}
\renewcommand{\thefigure}{S\arabic{figure}}\renewcommand{\thetable}{S\arabic{table}}\renewcommand{\theequation}{S\arabic{equation}}
\appendix
\makeatletter\def\pre@section{Appendix~}\def\@seccntformat#1{\csname pre@#1\endcsname\csname the#1\endcsname\quad}\makeatother

\section{Dataset Description}
\label{app:v2data}
\begin{figure*}[!t]
\centering
\dqcard{dqMCS}{MC-Search}{image-initiated chain, 3 hops}
{\dqtiles{\dqtile{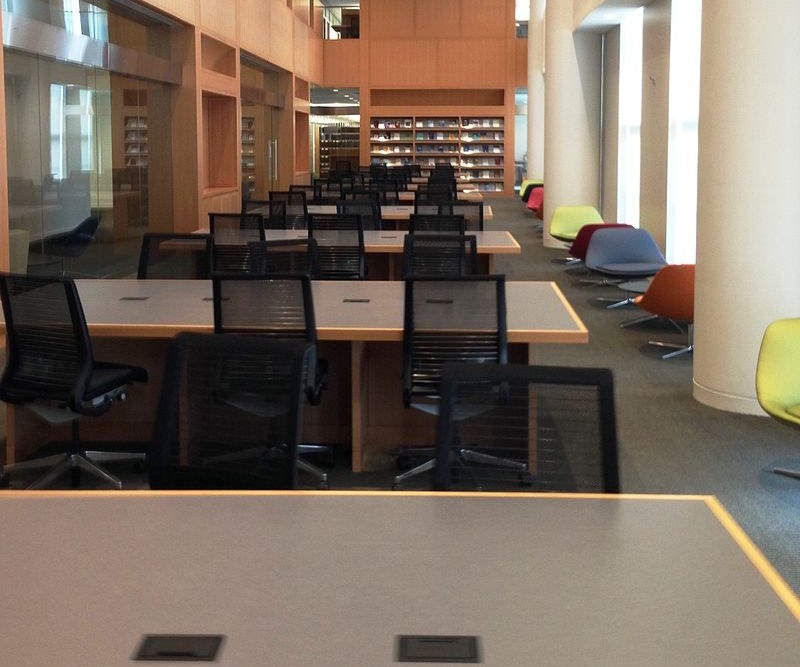}}{\dqtext{vsMCS}{``Originally located in New York's downtown Financial District, Fordham Law is currently [\ldots]''}}{input image}{gold passage}}
{\dqline{Question}{Where was the law school, whose library reading room is depicted in the image, originally located?}
\dqline{Answer}{It was originally located in New York's downtown Financial District.}
\dqline{Gold evidence}{The input image, whose caption names the reading room, and two passages on the law school \dqid{doc:d5c0aeb4}.}}\hfill
\dqcard{dqWQ}{WebQA}{image question, 1 gold image}
{\dqtiles{\dqtile{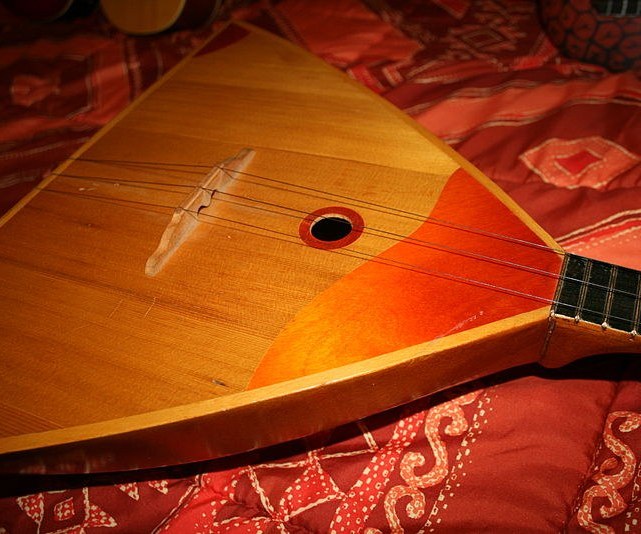}}{\dqtext{vsWQ}{``Balala\"{\i}ka [\ldots] Balalaika - 1981 - Model of 1980 made for the Olympic Games of Moscow''}}{gold image}{its caption}}
{\dqline{Question}{How many holes does a Balalaika have in the body on the front of the instrument?}
\dqline{Answer}{In the body on the front of the instrument, a Balalaika has one hole.}
\dqline{Gold evidence}{One image \dqid{img:30380590}. Its caption does not give the count, so the answer must be read from the image.}}\hfill
\dqcard{dqSAGE}{SAGE}{soybean, 27 classes}
{\dqtiles{\dqtile{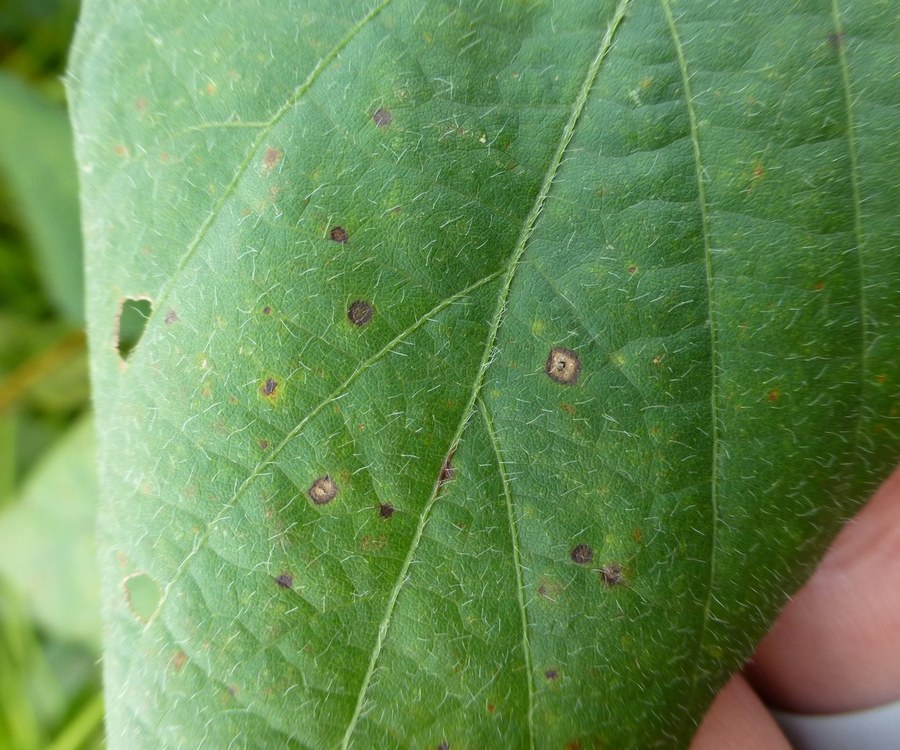}}{\dqtile{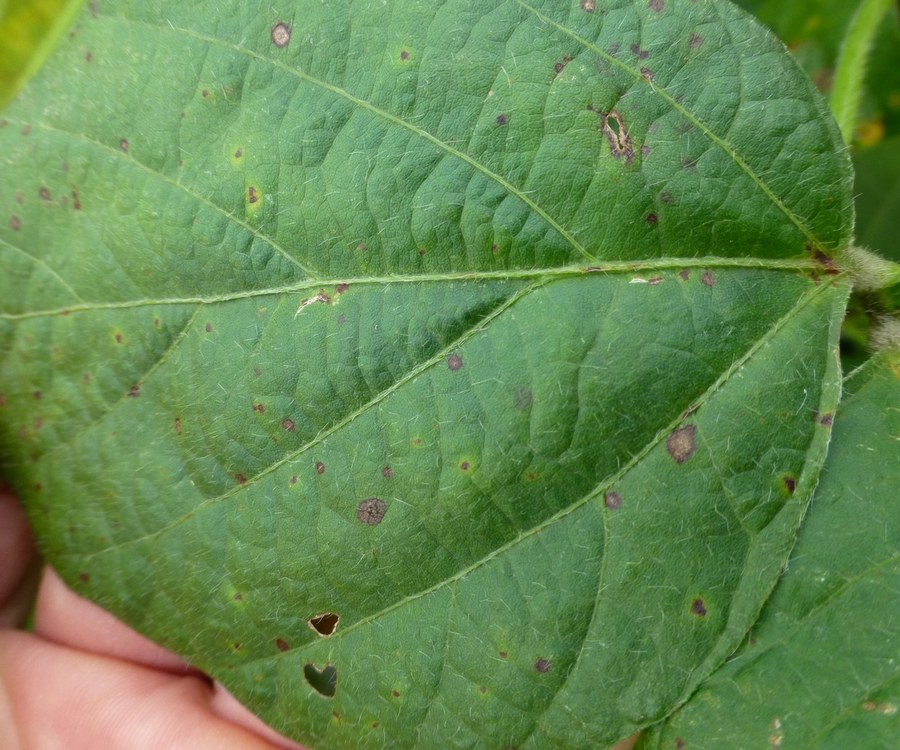}}{input photograph}{reference image, gold class}}
{\dqline{Question}{Which disease does the soybean plant in the input image have? Choose exactly one of these 27 soybean classes: [\ldots]}
\dqline{Answer}{Frogeye leaf spot}
\dqline{Gold evidence}{The class passage, ``The reddish-brown or purple ring surrounding round leaf spots [\ldots]'', and any reference image of the class.}}\hfill
\dqcard{dqMMQA}{MultiModalQA}{image + text, 2 gold}
{\dqtiles{\dqtext{vsMMQA}{``Angelique Kerber won her third Grand Slam singles title [\ldots]''}}{\dqtile{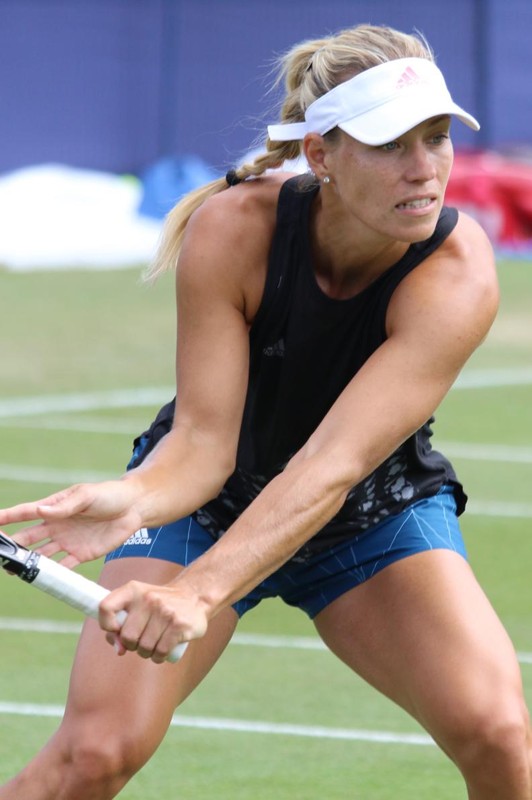}}{gold passage}{gold image}}
{\dqline{Question}{The woman who won in singles in the 2018 Wimbledon tennis tournament is wearing what type of hat?}
\dqline{Answer}{visor}
\dqline{Gold evidence}{The passage on the 2018 women's singles \dqid{doc:497bcd5b} and the winner's image \dqid{img:9dee9412}, captioned ``Angelique Kerber''.}}\\[5pt]
\dqcard{dqMCS}{MC-Search}{parallel multi-image fork, 3 hops}
{\dqtiles{\dqtile{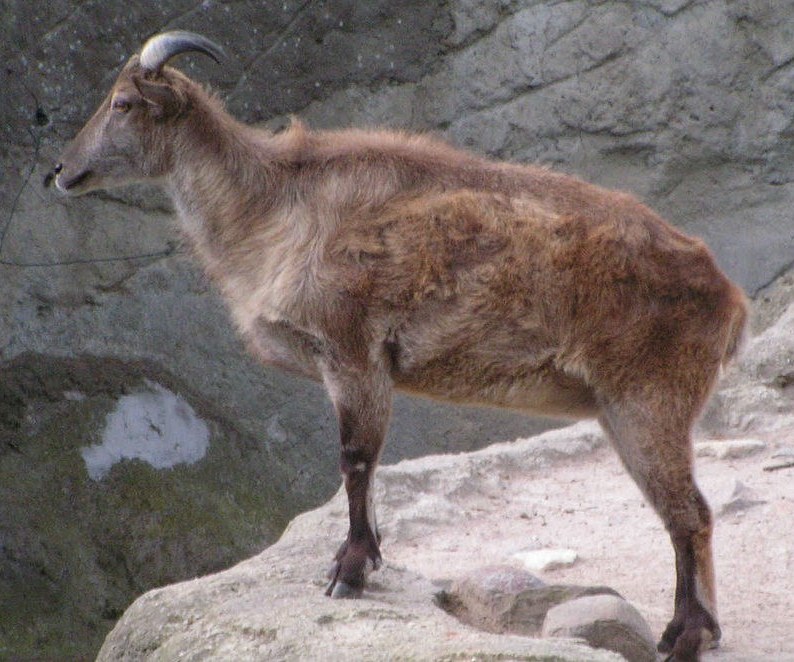}}{\dqtile{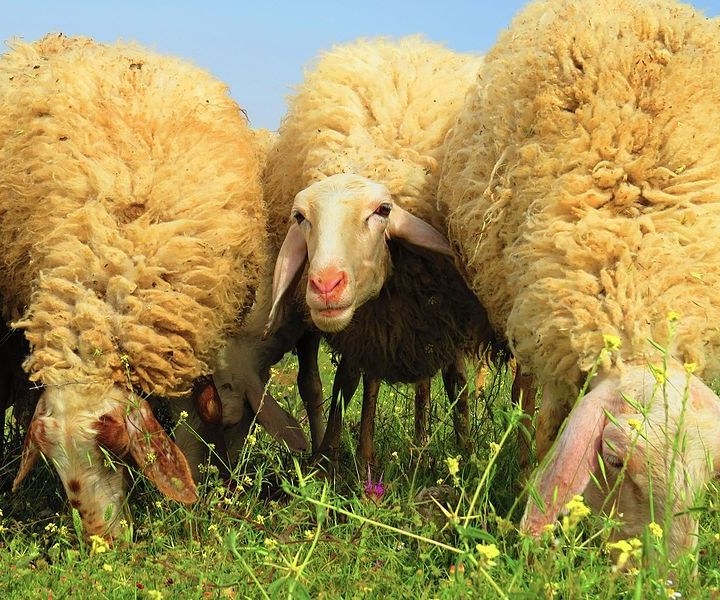}}{gold image}{gold image}}
{\dqline{Question}{[\ldots] what is a key physical distinction regarding horns between the Himalaja-Tahr and the Assaf sheep [\ldots]?}
\dqline{Answer}{The Himalaja-Tahr is visually depicted with horns [\ldots] while the Assaf sheep in the image do not have horns.}
\dqline{Gold evidence}{Two images \dqid{img:30216441, img:30097337} and a passage: ``Their horns reach a maximum length of 46 centimetres (18 in).''}}\hfill
\dqcard{dqWQ}{WebQA}{image question, 2 gold images}
{\dqtiles{\dqtile{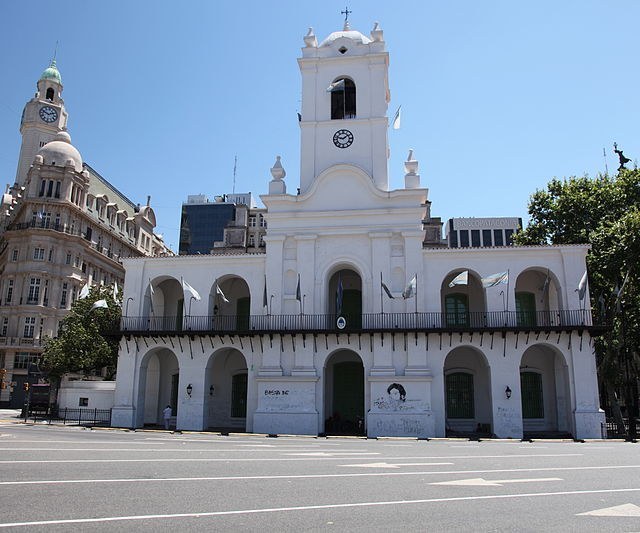}}{\dqtile{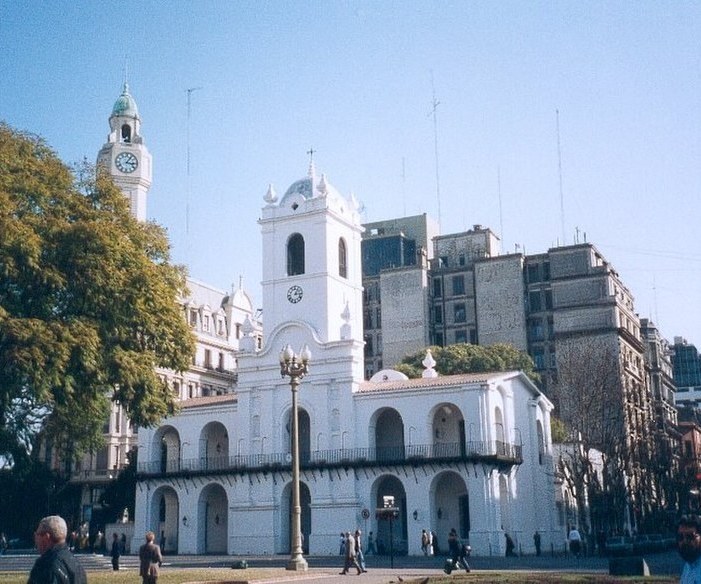}}{gold image}{gold image}}
{\dqline{Question}{How many arches are on the facade of the Cabildo in Buenos Aires, Argentina?}
\dqline{Answer}{Eleven arches are on the facade of the Cabildo in Buenos Aires, Argentina.}
\dqline{Gold evidence}{Two images \dqid{img:30175447, img:30206241}, captioned ``Buenos Aires Cabildo (5462698283)'' and ``Cabildo argentin''.}}\hfill
\dqcard{dqSAGE}{SAGE}{corn, 34 classes}
{\dqtiles{\dqtile{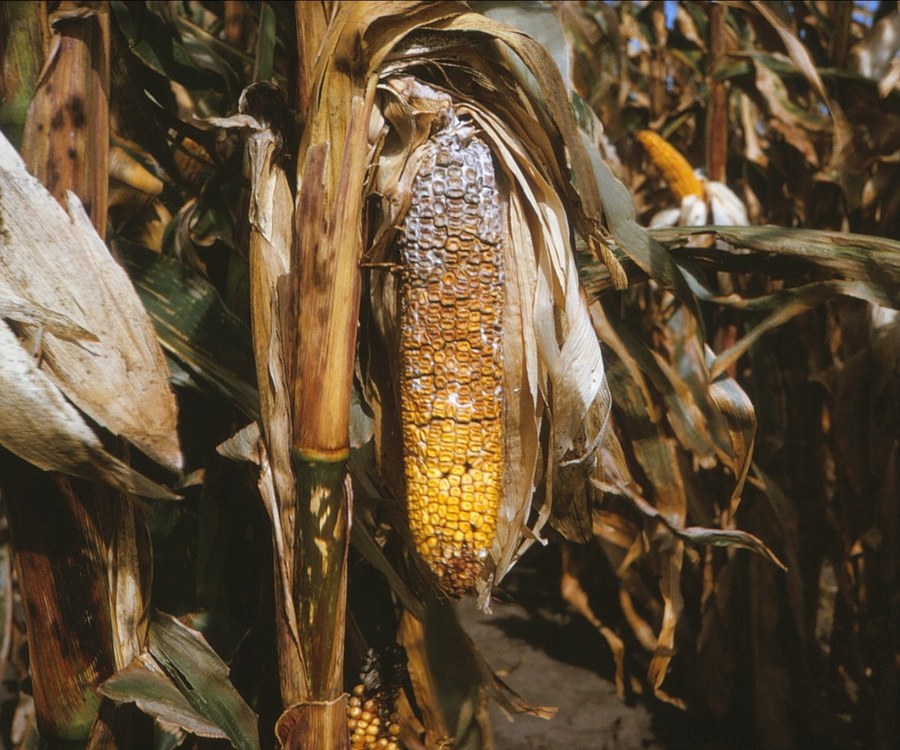}}{\dqtile{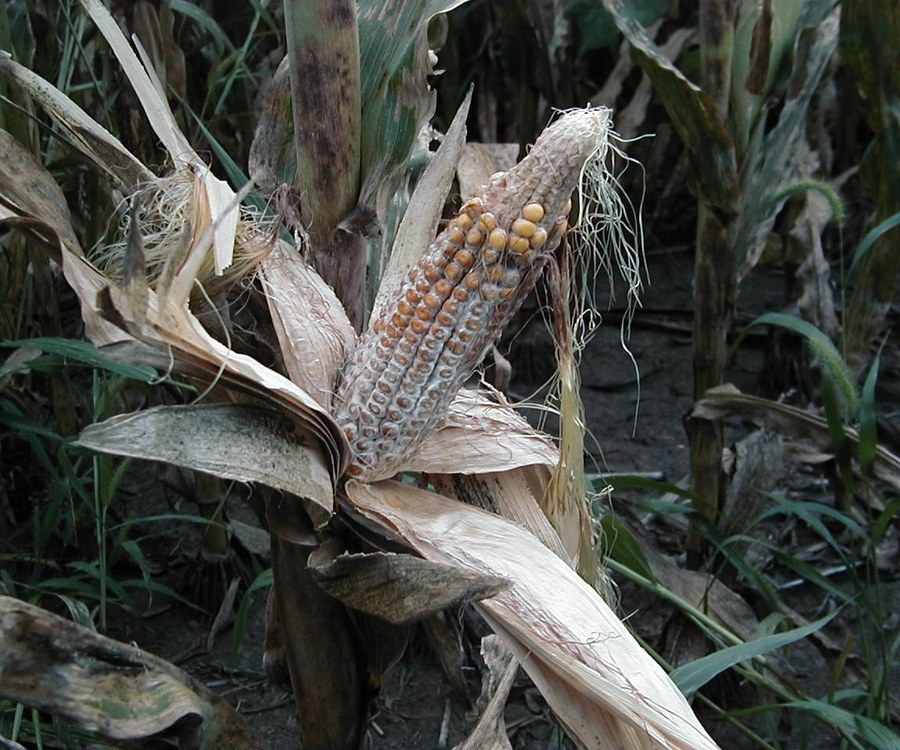}}{input photograph}{reference image, gold class}}
{\dqline{Question}{Which disease does the corn plant in the input image have? Choose exactly one of these 34 corn classes: [\ldots]}
\dqline{Answer}{Diplodia Ear Rot}
\dqline{Gold evidence}{The class passage, ``White mycelium grows over and between kernels, usually starting from the base of the ear.'', and any reference image of the class.}}\hfill
\dqcard{dqMMQA}{MultiModalQA}{image list, list answer}
{\dqtiles{\dqtile{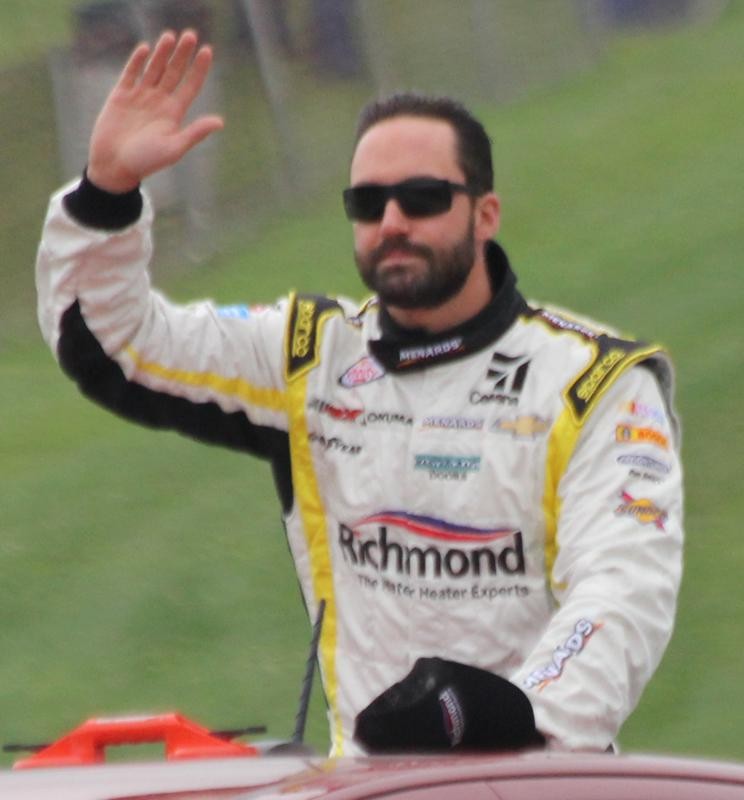}}{\dqtile{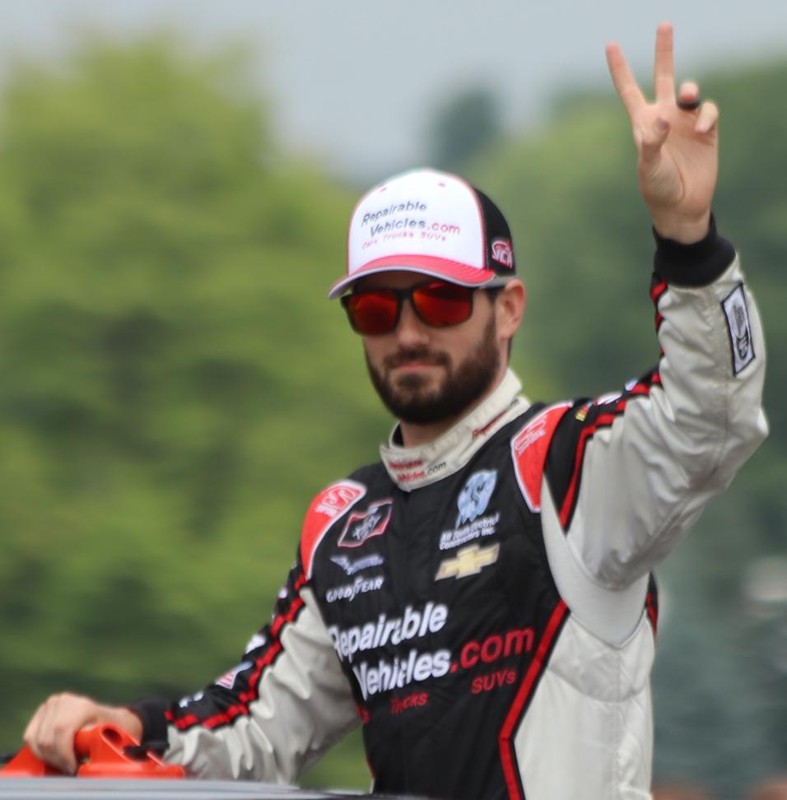}}{gold image}{gold image}}
{\dqline{Question}{Which Driver(s), in Past winners of Henry 180, has a mustache and beard?}
\dqline{Answer}{Paul Menard, Jeremy Clements}
\dqline{Gold evidence}{Two images \dqid{img:1d7549af, img:c082056b}, captioned ``Paul Menard'' and ``Jeremy Clements''.}}
\caption{\textbf{Representative test questions, two per dataset.} Each card shows the question, its reference answer
and its gold evidence as stored in the question manifests; tiles are image crops or quoted record text.}
\label{fig:dqexamples}
\end{figure*}

An example is a question with an optional input image, a reference answer and a list of gold evidence records, where a
record is one corpus entry, a text passage or a captioned image. Figure~\ref{fig:dqexamples} shows two test questions per dataset.

\paragraph{MC-Search~\cite{ning2026mcsearch}} asks long-form, multi-fact questions over reasoning graphs of mostly four
annotated hops. We report its \emph{primary} test questions: those that are not text-only chains, cite at least two
gold records with at least one image hop, and whose gold records all exist in the knowledge base. Training uses the
training questions whose gold records all exist, with the benchmark's answers and hop annotations unchanged.

\paragraph{WebQA~\cite{webqa2022}} asks short-answer questions whose gold evidence is images (image questions) or
passages (text questions). Test questions come from the official validation set, whose answers are public, and the
image questions form the subset of the main paper. Training and development questions come from the official
training data, half image and half text questions.

\paragraph{SAGE~\cite{sage2026}} asks for the disease label of a plant photograph over nine plants. Every question names
the plant and lists all its candidate classes, and the answer is the class name. The test photographs are fixed by the
benchmark; training and development questions use labeled reference photographs as queries, with the query and its near
duplicates hidden from retrieval in that episode.

\paragraph{MultiModalQA~\cite{talmor2021mmqa}} asks questions that combine image and text evidence. We keep the
questions whose type involves no table and whose supporting context cites at least one image and at least two distinct
documents. Questions come from the official training and development sets, whose answers are public, and are divided
into training, development and test questions by split group. Some questions have a list answer; answers are scored by
the official exact match.

\paragraph{Corpora.} MC-Search and WebQA search one shared corpus, the MC-Search knowledge base of text passages and
captioned images, which is built from the WebQA source pool. SAGE has its own corpus of reference photographs, each
captioned with plant, class and plant part, and one knowledge-base passage per class, and an episode searches only the
records of its own plant. MultiModalQA has its own corpus of text passages and images captioned with their titles,
from the MultiModalQA source pool; tables are never indexed. Test photographs are never indexed, and their exact copies and near duplicates are removed
from the corpus. An input image and its near duplicates are removed from the retrieval results of its own episode.

\paragraph{Splits.} Splits are assigned by \emph{split group}: a group joins the questions that share a source
question, a gold record or a gold image with the same perceptual hash (on SAGE, near-duplicate clusters of photographs),
and training, development and test never share a group. WebQA questions linked to MC-Search development or test
questions in this way are removed from every WebQA split. All three training stages use training questions only,
development questions serve for monitoring and selection, and test questions are never seen during training.

\paragraph{Evidence and recall.} The gold evidence of an MC-Search question is the set of records its hops cite, that
of a WebQA question its positive facts, that of a SAGE question the passage of the true class plus any one reference
photograph of that class, and that of a MultiModalQA question the passages and images of its supporting context. A record is acquired when a retrieval returns it or the agent inspects it, and gold recall is
the share of a question's retrievable gold records that are acquired.

\section{Training Details}
\label{app:v2train}
\begin{figure*}[!t]
\centering
\begin{tikzpicture}[
  stage/.style={draw=rqTeal, line width=0.7pt, rounded corners=3pt, fill=rqPale, text width=3.05cm, minimum height=2.15cm,
    align=center, font=\scriptsize, inner sep=3pt},
  model/.style={draw=black!55, line width=0.6pt, rounded corners=3pt, fill=black!4, text width=2.0cm, minimum height=2.15cm,
    align=center, font=\scriptsize, inner sep=3pt},
  data/.style={font=\tiny\itshape, text=black!60, align=center, text width=3.05cm},
  arr/.style={-{Stealth[length=2.2mm]}, line width=0.8pt, draw=rqTeal}]
\node[model] (base) {\textbf{Base model}\\[2pt]Qwen3-VL-8B or 4B-Instruct\\[2pt]vision tower frozen};
\node[stage, right=0.42cm of base] (s1) {\textbf{Stage 1: teacher}\\[2pt]base model runs the agent loop with a private hint (gold captions or hop sub-questions)\\[2pt]keep rule: all gold acquired, F1 $\ge$ 0.5};
\node[stage, right=0.42cm of s1] (s2) {\textbf{Stage 2: SFT}\\[2pt]LoRA (rank 16) on the kept episodes, hint removed\\[2pt]loss on assistant turns only, 2 epochs};
\node[stage, right=0.42cm of s2] (s3) {\textbf{Stage 3: GRPO}\\[2pt]32 questions $\times$ 8 episodes per step, $T{=}0.8$\\[2pt]$R=\mathrm{F1}+0.5\,\mathrm{Recall}$\\[1pt]evaluation budgets};
\node[model, right=0.42cm of s3, draw=vsWQ!80!black, fill=vsWQ!8] (agent) {\textbf{ViBES agent}\\[2pt]merged weights, greedy decoding};
\node[stage, below=0.55cm of s3, text width=6.9cm, minimum height=1.25cm, xshift=-1.95cm] (s4) {\textbf{Stage 4: ViBES-Lens selectors} (agent frozen)\\[2pt]attention network over the agent's state and image patches, trained on crop-request states of the
  training split: one box and mask from the teacher-forced gain, $N$ boxes from the frozen agent's answer correctness};
\node[data, above=0.08cm of s1] {training questions};
\node[data, above=0.08cm of s2] {kept teacher episodes};
\node[data, above=0.08cm of s3] {training questions (resampled)};
\draw[arr] (base) -- (s1);
\draw[arr] (s1) -- (s2);
\draw[arr] (s2) -- (s3);
\draw[arr] (s3) -- (agent);
\draw[arr] (agent.south) |- (s4.east);
\end{tikzpicture}
\caption{\textbf{Training stages of ViBES and ViBES-Lens.} Stages 1 to 3 train the agent from training questions only;
each stage merges its LoRA adapter into the weights before the next starts a new one. Stage 4 trains the region
selectors of ViBES-Lens against the frozen agent.}
\label{fig:v2stages}
\end{figure*}
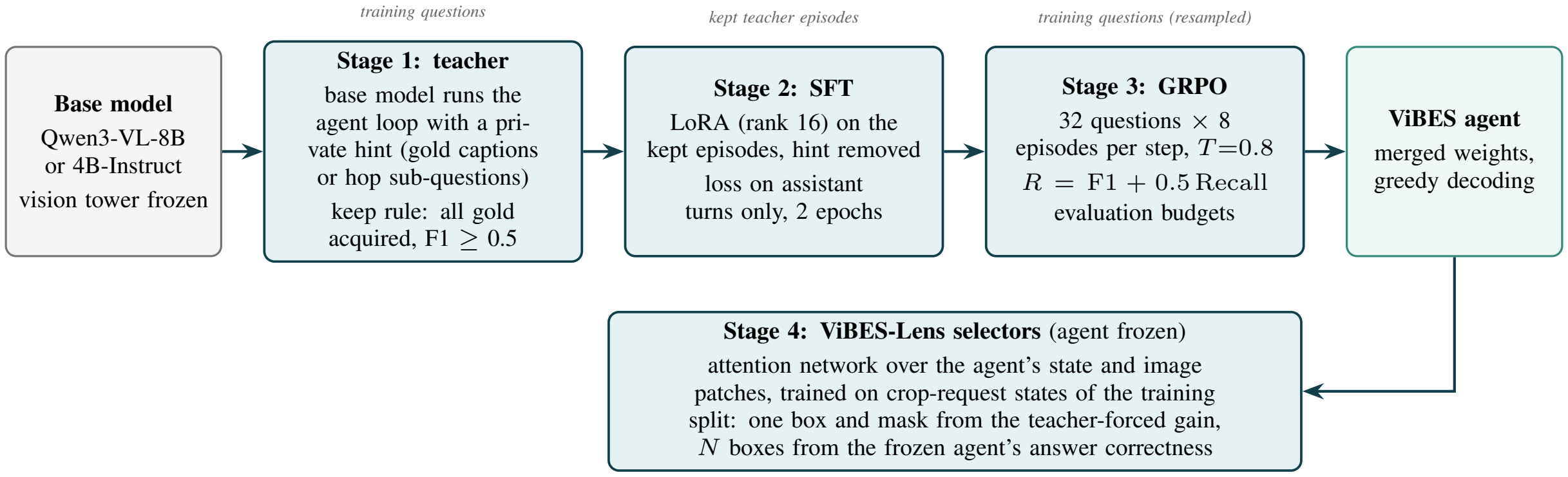

Figure~\ref{fig:v2stages} summarizes how the agent of ViBES and the selectors of ViBES-Lens are trained
(Section~\ref{sec:policy} and Section~\ref{sec:module}). One agent is trained per dataset and model size.

\paragraph{Model.} The policy is Qwen3-VL-8B-Instruct or Qwen3-VL-4B-Instruct~\cite{qwen3vl2025}, served with
vLLM~\cite{kwon2023vllm}. Training updates LoRA adapters~\cite{hu2022lora} (rank 16, $\alpha{=}32$) on the attention
and feed-forward projections of the language model; the vision tower stays frozen. Retrieval uses
Qwen3-VL-Embedding-2B with reciprocal-rank fusion of the text, image and joint views.

\paragraph{Stage 1: teacher.} The base model runs the agent loop on training questions with a private hint
(Figure~\ref{fig:p-teacher}): the captions or titles of the gold records on WebQA and SAGE, and the hop sub-questions
on MC-Search. A trajectory is kept if it answers correctly under the dataset-specific evaluation protocol and cites only visible records.

\paragraph{Stage 2: SFT.} Token cross-entropy on the assistant turns only (the action JSON and its end token), with
AdamW at a peak learning rate of $10^{-4}$, 5\% warm-up and cosine decay, gradient accumulation over 16 conversations,
clipping at 1.0, LoRA dropout 0.05 and two epochs.

\paragraph{Stage 3: GRPO.} Each step samples 8 episodes for each of 32 training questions at temperature 0.8 under the
evaluation prompts and budgets. The reward is $R=\mathrm{F1}+0.5\,\mathrm{Recall}$ (answer token F1 and gold-evidence
recall), with no per-step term or penalty. Advantages are $(R-\bar R)/\max(\sigma_R,0.1)$ within a question's episodes,
and the loss is the advantage-weighted negative log-probability of the sampled assistant tokens (AdamW, learning rate
$2{\times}10^{-5}$, 5 warm-up steps, clipping at 1.0, no KL term). Training runs 100 steps on WebQA, 53 on SAGE and 60
on MC-Search, and the last step is evaluated. The main
paper uses GRPO seed 17; Appendix~\ref{app:v2seedvibes} adds seeds 18 and 19 at 8B from the same SFT checkpoint.

\paragraph{Stage 4: ViBES-Lens selectors.} At a crop request, one forward pass of the frozen agent gives the
final-layer state of the last token and the vision tokens of the image, both reduced to 256 dimensions by principal
components, and a small attention network maps them to a distribution over patches. The \emph{one-box} selector adds
size and skip heads and is trained by GRPO on stored crop-request states with the teacher-forced gain in gold-answer
log-probability over no crop as reward, minus 0.1 times the box area. The \emph{mask} selector keeps the top $k\%$ of
the smoothed attention map, trained with the same gain or, where that gain does not track the decoded answer score on
development data, with the decoded answer score of the masked view (at 4B on MC-Search, SAGE and MultiModalQA). The \emph{$N$-box} selector places $N$ boxes at greedy
attention peaks and is rewarded by the correctness of the frozen agent's answer after seeing them. All use 16 states per
step and Adam (learning rate $3{\times}10^{-4}$); the checkpoint, $k$ and $N$ are selected on development data.

\section{Evidence Acquisition and Cost}
\label{app:v2evidence}
This section supports the evidence and cost statements of Section~\ref{sec:mainres} and Section~\ref{sec:portability}
with the 8B test runs of Table~\ref{tab:indomain}.

\paragraph{ViBES retrieves more of the gold records.} ViBES raises gold-evidence recall by 20.2 points over one-shot RAG
on matched questions. Table~\ref{tab:v2recall} gives the recall of every system, in the layout of
Table~\ref{tab:indomain}.

\begin{figure}[!t]
\centering
\includegraphics[width=0.92\columnwidth]{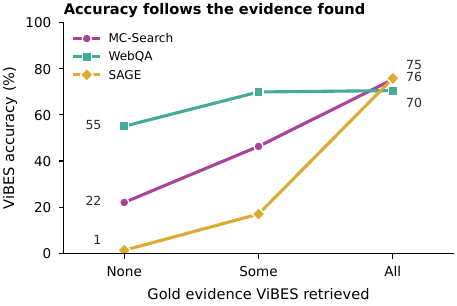}
\caption{\textbf{ViBES accuracy by the share of gold records it retrieves} (none, some or all; 8B, test).}
\label{fig:v2evidence}
\end{figure}

\paragraph{Accuracy follows the retrieved evidence.} Figure~\ref{fig:v2evidence} groups the questions by the share of
their gold records that ViBES retrieves. On MC-Search and SAGE, its accuracy rises from at most one in five questions
without any gold record (22\% and 1\%) to about three in four with all of them (75\% and 76\%); on WebQA, from 55\% to
70\%.

\paragraph{The collected observations matter more than the answerer.} On SAGE, we freeze the observations that the 8B SFT
and GRPO agents collected on all test questions (question, input image, result cards and inspected records, in logged
order, without the agent's notes or crops) and let each checkpoint answer from each set with the tools disabled
(exact-label accuracy). From the SFT agent's observations, the SFT and GRPO checkpoints answer 36.9\% and 37.6\% of the
questions correctly; from the GRPO agent's observations, 44.4\% and 46.7\%. With the answerer fixed, the GRPO agent's
observations raise accuracy by 7.5 and 9.1 points; with the observations fixed, swapping the answerer changes it by 0.7
and 2.2 points. 

\paragraph{ViBES reaches the evidence earlier, at a higher token cost.} After three executed tool calls (the seeded
retrieval is call 1), ViBES has
retrieved 64.4\% of the WebQA gold records and 54.3\% on SAGE, more than the zero-shot agent after ten (61.7\% and
48.6\%). ViBES processes 10 to 19 times as many prompt tokens per question as one-shot RAG, counting every re-read of the
context.

\begin{table*}[!t]
\centering\scriptsize\setlength{\tabcolsep}{5pt}\setlength{\aboverulesep}{0pt}\setlength{\belowrulesep}{0pt}\setlength{\extrarowheight}{1.2pt}
\caption{\textbf{Gold-evidence recall (\%) of every system} (test, one run per cell): the share of the retrievable gold
records that a system retrieves, in the layout of Table~\ref{tab:indomain}. 8B and 4B: our loop; Pi and OC: 8B models
in the pi and OpenCode harnesses. $^{\ast}$32B model in the 8B column. --: no model of that size exists for the setting. Bold: best per column; underline: second best.}
\label{tab:v2recall}
\begin{tabular}{@{}cl|rrrr|rrrr|rrrr@{}}
\toprule
& & \multicolumn{4}{>{\columncolor{tMCS}}c|}{\textbf{MC-Search}} & \multicolumn{4}{>{\columncolor{tWQ}}c|}{\textbf{WebQA}} & \multicolumn{4}{>{\columncolor{tSAGE}}c}{\textbf{SAGE}} \\
\cmidrule(lr){3-6}\cmidrule(lr){7-10}\cmidrule(l){11-14}
\textbf{Group} & \textbf{System} & \multicolumn{1}{>{\columncolor{tMCS}}r}{8B} & \multicolumn{1}{>{\columncolor{tMCS}}r}{4B} & \multicolumn{1}{>{\columncolor{tMCS}}r}{Pi} & \multicolumn{1}{>{\columncolor{tMCS}}r|}{OC} & \multicolumn{1}{>{\columncolor{tWQ}}r}{8B} & \multicolumn{1}{>{\columncolor{tWQ}}r}{4B} & \multicolumn{1}{>{\columncolor{tWQ}}r}{Pi} & \multicolumn{1}{>{\columncolor{tWQ}}r|}{OC} & \multicolumn{1}{>{\columncolor{tSAGE}}r}{8B} & \multicolumn{1}{>{\columncolor{tSAGE}}r}{4B} & \multicolumn{1}{>{\columncolor{tSAGE}}r}{Pi} & \multicolumn{1}{>{\columncolor{tSAGE}}r}{OC} \\
\midrule
\multirow{4}{*}{\rotatebox[origin=c]{90}{\textbf{In-house}}} & One-shot RAG & 27.5 & 27.5 & 27.3 & 27.3 & 49.0 & 49.0 & 51.1 & 51.1 & 28.1 & 28.4 & 28.3 & 28.3 \\
 & RAG-SFT & 27.5 & 27.5 & 27.3 & 27.3 & 49.0 & 49.0 & 51.1 & 51.1 & 28.1 & 28.1 & 28.3 & 28.3 \\
 & Agent, zero-shot & 39.3 & 38.3 & 38.8 & 37.1 & 63.6 & 61.7 & 60.5 & 56.3 & 48.6 & 48.3 & 47.1 & 52.0 \\
 & Agent + SFT & 39.6 & \underline{39.1} & \underline{39.1} & \underline{37.4} & 62.7 & 61.1 & 59.6 & 55.4 & 46.4 & 45.3 & 44.9 & 49.8 \\
\midrule
\multirow{6}{*}{\rotatebox[origin=c]{90}{\textbf{External}}} & DeepEyes + cards~\cite{deepeyes2025} & 27.5 & -- & 27.3 & 27.3 & 49.0 & -- & 51.3 & 51.3 & 15.7 & -- & 28.3 & 28.3 \\
 & Qwen2.5-VL~\cite{qwen25vl} & 33.4 & -- & 27.3 & 27.3 & 55.2 & -- & 51.3 & 51.3 & 30.9 & -- & 28.3 & 28.3 \\
 & Qwen2.5-VL-32B-AWQ~\cite{qwen25vl} & 27.3$^{\ast}$ & -- & 27.3 & 27.3 & 51.3$^{\ast}$ & -- & 51.3 & 51.3 & 28.3$^{\ast}$ & -- & 28.3 & 28.3 \\
 & DeepEyes weights~\cite{deepeyes2025} & 35.8 & -- & 30.2 & 27.9 & 58.9 & -- & 57.1 & 56.6 & 35.4 & -- & 39.4 & 37.5 \\
 & MMSearch-R1 weights~\cite{wu2026mmsearch} & 35.3 & -- & 28.1 & 27.6 & 57.8 & -- & 52.4 & 51.6 & 31.9 & -- & 29.7 & 28.4 \\
 & SenseNova-MARS weights~\cite{chng2026sensesearch} & \textbf{41.0} & -- & 38.1 & 36.7 & \underline{64.2} & -- & 62.5 & 60.3 & 52.1 & -- & 54.5 & 51.8 \\
\midrule
 & \textbf{ViBES} (ours) & \underline{40.4} & \textbf{39.9} & \textbf{39.5} & \textbf{38.7} & \textbf{65.9} & \textbf{66.3} & \underline{65.5} & \underline{61.5} & \textbf{62.3} & \textbf{60.0} & \underline{61.7} & \underline{57.8} \\
 & \textbf{ViBES-Lens} (ours) & \underline{40.4} & \textbf{39.9} & \textbf{39.5} & \textbf{38.7} & \textbf{65.9} & \underline{66.2} & \textbf{66.1} & \textbf{61.9} & \underline{61.5} & \underline{59.9} & \textbf{63.0} & \textbf{58.1} \\
\bottomrule
\end{tabular}
\par\smallskip\footnotesize The ViBES and ViBES-Lens rows use the WebQA image questions.
\end{table*}

\section{Multi-Seed ViBES Accuracy}
\label{app:v2seedvibes}
At 8B, we retrain GRPO with seeds 18 and 19 from the SFT checkpoint of seed 17 (the main paper), with the same data,
settings and steps, and evaluate each checkpoint once in our loop and in the pi and OpenCode harnesses with greedy decoding
(Table~\ref{tab:v2seedvibes}). The 4B agents have seed 17 only (Table~\ref{tab:indomain}). Across seeds, the standard
deviation is 0.2--1.6 points in every dataset and harness, and every seed stays above one-shot RAG in the loop by a wide
margin (mean gain $+12.1$ on MC-Search, $+18.7$ on WebQA and $+17.4$ on SAGE). Seed 17 is the best of the three seeds on
MC-Search in the loop and in pi, and the lowest on WebQA and SAGE in the loop; it lies within 1.7 points of the
three-seed mean in every cell.

\begin{table*}[!t]
\centering\scriptsize\setlength{\tabcolsep}{3.4pt}\setlength{\aboverulesep}{0pt}\setlength{\belowrulesep}{0pt}\setlength{\extrarowheight}{1.2pt}
\caption{\textbf{ViBES-8B accuracy (\%) over three GRPO seeds.} Seed 17: the checkpoint of the main paper; seeds 18 and
19: GRPO retrained from the same SFT checkpoint. Loop: our execution loop; Pi and OC: the pi and OpenCode harnesses.
Mean $\pm$ sample standard deviation over seeds. $\Delta$ RAG: mean over seeds minus the mean over three runs of one-shot RAG in the loop.
Question sets as in Table~\ref{tab:indomain}.}
\label{tab:v2seedvibes}
\begin{tabular*}{\textwidth}{@{\extracolsep{\fill}}l|rrrc|rrrc|rrrc|c@{}}
\toprule
& \multicolumn{4}{c|}{\textbf{Loop}} & \multicolumn{4}{c|}{\textbf{Pi}} & \multicolumn{4}{c|}{\textbf{OC}} & \\
\cmidrule(lr){2-5}\cmidrule(lr){6-9}\cmidrule(lr){10-13}
\textbf{Dataset} & s17 & s18 & s19 & Mean $\pm$ sd & s17 & s18 & s19 & Mean $\pm$ sd & s17 & s18 & s19 & Mean $\pm$ sd & $\Delta$ RAG (Loop) \\
\midrule
\cellcolor{tMCS}MC-Search & 42.7 & 39.8 & 40.5 & 41.0\,$\pm$\,1.5 & 43.1 & 40.4 & 40.9 & 41.5\,$\pm$\,1.4 & 40.1 & 40.5 & 40.5 & 40.4\,$\pm$\,0.2 & \cellcolor{dGood}$+$12.1 \\
\midrule
\cellcolor{tWQ}WebQA & 66.4 & 68.4 & 67.8 & 67.5\,$\pm$\,1.0 & 67.2 & 67.2 & 66.0 & 66.8\,$\pm$\,0.7 & 67.2 & 67.4 & 64.6 & 66.4\,$\pm$\,1.6 & \cellcolor{dGood}$+$18.7 \\
\midrule
\cellcolor{tSAGE}SAGE & 42.5 & 44.1 & 44.3 & 43.6\,$\pm$\,1.0 & 42.7 & 42.5 & 43.4 & 42.9\,$\pm$\,0.5 & 42.4 & 42.7 & 43.6 & 42.9\,$\pm$\,0.6 & \cellcolor{dGood}$+$17.4 \\
\bottomrule
\end{tabular*}
\end{table*}

\begin{table*}[!t]
\centering\scriptsize\setlength{\tabcolsep}{3.1pt}\setlength{\aboverulesep}{0pt}\setlength{\belowrulesep}{0pt}\setlength{\extrarowheight}{1.2pt}
\caption{\textbf{In-house 8B systems and ViBES over three runs (accuracy, \%).} Same checkpoints in every run; runs 2 and 3 repeat
inference with the test questions in a different random order. Mean $\pm$ sample
standard deviation over runs. ViBES: the seed-17 checkpoint. Bold: best per column within a dataset; underline: second best.}
\label{tab:v2reruns}
\begin{tabular}{@{}ll|rrrc|rrrc|rrrc@{}}
\toprule
& & \multicolumn{4}{c|}{\textbf{Loop}} & \multicolumn{4}{c|}{\textbf{Pi}} & \multicolumn{4}{c}{\textbf{OC}} \\
\cmidrule(lr){3-6}\cmidrule(lr){7-10}\cmidrule(l){11-14}
\textbf{Dataset} & \textbf{System} & r1 & r2 & r3 & Mean $\pm$ sd & r1 & r2 & r3 & Mean $\pm$ sd & r1 & r2 & r3 & Mean $\pm$ sd \\
\midrule
\cellcolor{tMCS}MC-Search & Closed-book & 6.6 & 7.5 & 6.9 & 7.0\,$\pm$\,0.4 & 9.4 & 9.1 & 9.2 & 9.2\,$\pm$\,0.1 & 10.7 & 10.1 & 10.2 & 10.3\,$\pm$\,0.3 \\
\cellcolor{tMCS} & One-shot RAG & 28.9 & 29.4 & 28.4 & 28.9\,$\pm$\,0.5 & 28.9 & 29.1 & 28.1 & 28.7\,$\pm$\,0.5 & 27.8 & 26.6 & 27.2 & 27.2\,$\pm$\,0.6 \\
\cellcolor{tMCS} & RAG-SFT & 27.9 & 28.2 & 28.8 & 28.3\,$\pm$\,0.5 & 28.9 & 29.2 & 28.9 & 29.0\,$\pm$\,0.2 & 28.9 & 27.0 & 26.5 & 27.5\,$\pm$\,1.3 \\
\cellcolor{tMCS} & Agent, zero-shot & 37.1 & \underline{38.7} & 37.0 & 37.6\,$\pm$\,1.0 & \underline{41.5} & \textbf{42.3} & \underline{40.9} & \underline{41.6\,$\pm$\,0.7} & \underline{39.9} & \textbf{40.9} & \underline{39.5} & \textbf{40.1\,$\pm$\,0.8} \\
\cellcolor{tMCS} & Agent + SFT & \underline{37.9} & 37.7 & \underline{38.5} & \underline{38.0\,$\pm$\,0.4} & 40.2 & 40.4 & 40.1 & 40.2\,$\pm$\,0.1 & 39.3 & 37.9 & 38.6 & 38.6\,$\pm$\,0.7 \\
\cellcolor{tMCS} & ViBES & \textbf{42.7} & \textbf{42.5} & \textbf{40.9} & \textbf{42.1\,$\pm$\,1.0} & \textbf{43.1} & \underline{42.1} & \textbf{41.2} & \textbf{42.2\,$\pm$\,1.0} & \textbf{40.1} & \underline{38.6} & \textbf{39.6} & \underline{39.4\,$\pm$\,0.8} \\
\midrule
\cellcolor{tWQ}WebQA & Closed-book & 42.9 & 42.9 & 43.0 & 42.9\,$\pm$\,0.1 & 42.0 & 41.4 & 41.6 & 41.7\,$\pm$\,0.3 & 37.2 & 37.4 & 37.0 & 37.2\,$\pm$\,0.2 \\
\cellcolor{tWQ} & One-shot RAG & 55.7 & 55.4 & 56.1 & 55.7\,$\pm$\,0.4 & 55.8 & 55.6 & 54.2 & 55.2\,$\pm$\,0.9 & 55.8 & 57.2 & 56.2 & 56.4\,$\pm$\,0.7 \\
\cellcolor{tWQ} & RAG-SFT & 63.5 & 63.3 & 63.3 & 63.4\,$\pm$\,0.1 & 62.8 & \underline{63.4} & \underline{63.6} & \underline{63.3\,$\pm$\,0.4} & \underline{63.2} & \underline{62.2} & \underline{62.8} & \underline{62.7\,$\pm$\,0.5} \\
\cellcolor{tWQ} & Agent, zero-shot & 64.3 & 64.6 & \underline{66.0} & 65.0\,$\pm$\,0.9 & \underline{63.0} & 63.2 & 63.0 & 63.1\,$\pm$\,0.1 & 60.8 & 61.6 & 61.8 & 61.4\,$\pm$\,0.5 \\
\cellcolor{tWQ} & Agent + SFT & \underline{65.9} & \underline{65.9} & 65.8 & \underline{65.9\,$\pm$\,0.1} & 61.2 & 62.0 & 62.8 & 62.0\,$\pm$\,0.8 & 61.8 & 60.6 & 60.8 & 61.1\,$\pm$\,0.6 \\
\cellcolor{tWQ} & ViBES & \textbf{66.4} & \textbf{66.8} & \textbf{66.6} & \textbf{66.6\,$\pm$\,0.2} & \textbf{67.2} & \textbf{66.8} & \textbf{67.0} & \textbf{67.0\,$\pm$\,0.2} & \textbf{67.2} & \textbf{66.6} & \textbf{68.2} & \textbf{67.3\,$\pm$\,0.8} \\
\midrule
\cellcolor{tSAGE}SAGE & Closed-book & 24.9 & 24.5 & 24.5 & 24.7\,$\pm$\,0.2 & 24.5 & 24.2 & 23.5 & 24.1\,$\pm$\,0.5 & 21.8 & 22.0 & 21.8 & 21.8\,$\pm$\,0.1 \\
\cellcolor{tSAGE} & One-shot RAG & 26.2 & 26.2 & 26.2 & 26.2\,$\pm$\,0.0 & 23.7 & 23.7 & 23.3 & 23.6\,$\pm$\,0.2 & 20.2 & 20.6 & 21.1 & 20.6\,$\pm$\,0.5 \\
\cellcolor{tSAGE} & RAG-SFT & \underline{41.3} & \underline{41.5} & \underline{40.5} & \underline{41.1\,$\pm$\,0.5} & 24.2 & 24.0 & 22.3 & 23.5\,$\pm$\,1.0 & 20.6 & 20.6 & 20.8 & 20.6\,$\pm$\,0.1 \\
\cellcolor{tSAGE} & Agent, zero-shot & 36.7 & 34.0 & 34.3 & 35.0\,$\pm$\,1.5 & 31.7 & 33.6 & 31.9 & 32.4\,$\pm$\,1.1 & 33.6 & \underline{34.8} & 36.0 & 34.8\,$\pm$\,1.2 \\
\cellcolor{tSAGE} & Agent + SFT & 33.1 & 33.1 & 34.1 & 33.4\,$\pm$\,0.6 & \underline{39.1} & \underline{38.8} & \underline{39.3} & \underline{39.1\,$\pm$\,0.3} & \underline{40.8} & \textbf{40.7} & \textbf{41.5} & \underline{41.0\,$\pm$\,0.5} \\
\cellcolor{tSAGE} & ViBES & \textbf{42.5} & \textbf{43.2} & \textbf{42.7} & \textbf{42.8\,$\pm$\,0.4} & \textbf{42.7} & \textbf{42.0} & \textbf{41.9} & \textbf{42.2\,$\pm$\,0.5} & \textbf{42.4} & \textbf{40.7} & \underline{41.3} & \textbf{41.5\,$\pm$\,0.9} \\
\bottomrule
\end{tabular}
\end{table*}

\section{Repeated Runs of the In-House Systems and ViBES}
\label{app:v2reruns}
The in-house systems are not retrained across seeds. To measure how much their accuracy varies between runs of the same
checkpoint, we repeat inference twice more, with the test questions in a different random order, in our loop and in both
harnesses, at 8B (Table~\ref{tab:v2reruns}). Rerunning the seed-17 ViBES checkpoint in the same way changes its accuracy by a standard deviation of
0.2--1.0 points (largest: MC-Search in the loop).

\section{Multi-Seed ViBES-Lens Accuracy}
\label{app:v2seedlens}
For each seed of Appendix~\ref{app:v2seedvibes}, the ViBES-Lens selectors are retrained against that seed's agent with
the recipe and selection rule of seed 17 (Table~\ref{tab:v2seedlens}). With $N$ boxes, ViBES-Lens shows more of the image at a
crop request: in the seed-17 test runs, the union of its boxes covers a median of 64\% of the image on WebQA and 65\% on
SAGE, against 17\% and 44\% for the agent's own box. Averaged over the three seeds, ViBES-Lens is within 1.4 points of the
own box in the loop ($\Delta$ own from $-1.4$ to $+0.6$), and the $N$ boxes are within 0.6 points on every dataset.

\begin{table*}[!t]
\centering\scriptsize\setlength{\tabcolsep}{3.1pt}\setlength{\aboverulesep}{0pt}\setlength{\belowrulesep}{0pt}\setlength{\extrarowheight}{1.2pt}
\caption{\textbf{ViBES-Lens accuracy (\%) at 8B over three seeds.} For each seed, the agent (ViBES, own box) and the
ViBES-Lens selectors trained against it; seed 17: Table~\ref{tab:crop_ablation}. Mean $\pm$ sample standard deviation
over seeds. $\Delta$ own: mean over seeds of ViBES-Lens minus the own box of the same seed in the loop. Bold: best per column within a dataset; underline: second best.}
\label{tab:v2seedlens}
\begin{tabular*}{\textwidth}{@{\extracolsep{\fill}}ll|rrrc|rrrc|rrrc|c@{}}
\toprule
& & \multicolumn{4}{c|}{\textbf{Loop}} & \multicolumn{4}{c|}{\textbf{Pi}} & \multicolumn{4}{c|}{\textbf{OC}} & \\
\cmidrule(lr){3-6}\cmidrule(lr){7-10}\cmidrule(lr){11-14}
\textbf{Dataset} & \textbf{Crop regions} & s17 & s18 & s19 & Mean $\pm$ sd & s17 & s18 & s19 & Mean $\pm$ sd & s17 & s18 & s19 & Mean $\pm$ sd & $\Delta$ own (Loop) \\
\midrule
\cellcolor{tMCS}MC-Search & own box (ViBES) & \textbf{38.9} & 38.5 & \underline{38.5} & 38.6\,$\pm$\,0.3 & 40.2 & \textbf{41.7} & \textbf{42.5} & \underline{41.5\,$\pm$\,1.2} & \underline{38.5} & \underline{38.0} & \underline{37.4} & \underline{38.0\,$\pm$\,0.5} & -- \\
\cellcolor{tMCS} & one box & \underline{38.6} & \underline{39.0} & \underline{38.5} & \underline{38.7\,$\pm$\,0.3} & \underline{42.1} & \underline{40.2} & 41.5 & 41.3\,$\pm$\,1.0 & 38.2 & \textbf{38.7} & \textbf{38.5} & \textbf{38.5\,$\pm$\,0.3} & \cellcolor{dGood}$+$0.1 \\
\cellcolor{tMCS} & $N$ boxes & 38.5 & \textbf{40.1} & \textbf{39.2} & \textbf{39.2\,$\pm$\,0.8} & \textbf{42.7} & \textbf{41.7} & \underline{42.0} & \textbf{42.1\,$\pm$\,0.5} & \textbf{39.5} & 37.4 & \textbf{38.5} & \textbf{38.5\,$\pm$\,1.0} & \cellcolor{dGood}$+$0.6 \\
\midrule
\cellcolor{tWQ}WebQA & own box (ViBES) & \underline{66.4} & \textbf{68.4} & \textbf{67.8} & \textbf{67.5\,$\pm$\,1.0} & \underline{67.2} & 67.2 & \underline{66.0} & \underline{66.8\,$\pm$\,0.7} & \underline{67.2} & \textbf{67.4} & \underline{64.6} & \textbf{66.4\,$\pm$\,1.6} & -- \\
\cellcolor{tWQ} & one box & 66.0 & 66.8 & 65.6 & 66.1\,$\pm$\,0.6 & 66.4 & \underline{67.6} & 65.0 & 66.3\,$\pm$\,1.3 & 65.8 & 64.2 & \textbf{65.2} & 65.1\,$\pm$\,0.8 & \cellcolor{dBad}$-$1.4 \\
\cellcolor{tWQ} & $N$ boxes & \textbf{66.8} & \underline{67.2} & \underline{67.6} & \underline{67.2\,$\pm$\,0.4} & \textbf{68.4} & \textbf{68.0} & \textbf{66.6} & \textbf{67.7\,$\pm$\,0.9} & \textbf{67.4} & \underline{65.0} & \textbf{65.2} & \underline{65.9\,$\pm$\,1.3} & \cellcolor{dBad}$-$0.3 \\
\midrule
\cellcolor{tSAGE}SAGE & own box (ViBES) & \underline{42.5} & 44.1 & \textbf{44.3} & \underline{43.6\,$\pm$\,1.0} & 42.7 & \underline{42.5} & \textbf{43.4} & \underline{42.9\,$\pm$\,0.5} & 42.4 & \textbf{42.7} & \textbf{43.6} & \textbf{42.9\,$\pm$\,0.6} & -- \\
\cellcolor{tSAGE} & one box & \textbf{43.1} & \underline{44.4} & \underline{43.6} & \textbf{43.7\,$\pm$\,0.7} & \underline{43.4} & 40.8 & \underline{41.9} & 42.0\,$\pm$\,1.3 & \underline{42.5} & \underline{40.7} & \underline{42.5} & 41.9\,$\pm$\,1.1 & \cellcolor{dGood}$+$0.1 \\
\cellcolor{tSAGE} & $N$ boxes & 42.2 & \textbf{45.1} & 42.7 & 43.3\,$\pm$\,1.6 & \textbf{43.6} & \textbf{43.4} & \textbf{43.4} & \textbf{43.5\,$\pm$\,0.1} & \textbf{42.9} & \textbf{42.7} & 42.2 & \underline{42.6\,$\pm$\,0.4} & \cellcolor{dBad}$-$0.3 \\
\bottomrule
\end{tabular*}
\end{table*}

\section{ViBES-Lens Correctness Prediction per Configuration and Seed}
\label{app:v2seedauroc}
Table~\ref{tab:v2seedauroc} reports the absolute correctness-prediction metrics underlying Table~\ref{tab:confidence}
for seed 17 at 8B and 4B. All metrics use the probe score $c(\tau)$
(Eq.~\ref{eq:probe}) on the questions where both ViBES and ViBES-Lens crop, each system scored against its own answer
correctness; $\Delta$AUROC (Eq.~\ref{eq:dauroc}) is ViBES-Lens minus ViBES (own box). We repeat the one-box comparison
at 8B for GRPO seeds 18 and 19, with the selectors of Appendix~\ref{app:v2seedlens}. Averaged over the three datasets and
then over seeds 17, 18 and 19, the one-box selector improves AUROC by $+0.037\pm0.012$ (mean $\pm$ sample standard
deviation over seeds), and it improves AUROC in 8 of the 9 dataset--seed evaluations, indicating that the discrimination
gains extend beyond the primary seed.

\begin{table*}[!t]
\centering\scriptsize\setlength{\tabcolsep}{4pt}\setlength{\aboverulesep}{0pt}\setlength{\belowrulesep}{0pt}\setlength{\extrarowheight}{0.4pt}
\caption{\textbf{Correctness prediction of ViBES and ViBES-Lens.} $\Delta$ = ViBES-Lens $-$ ViBES (own box); a positive
$\Delta$AUROC and a negative $\Delta$ECE or $\Delta$Brier favour ViBES-Lens. \swatch{dGood}{Teal}: ViBES-Lens better; \swatch{dBad}{orange}: worse. Seed 17, the values behind
Table~\ref{tab:confidence}: AUROC, ECE and Brier score (0--1 scale) of the probe score of each system against its own
answer correctness, on the questions where both systems crop, as ViBES / ViBES-Lens.}
\label{tab:v2seedauroc}
\begin{tabular}{@{}lll|ccc|ccc@{}}
\toprule
& & & \textbf{AUROC}$\uparrow$ & \textbf{ECE}$\downarrow$ & \textbf{Brier}$\downarrow$ & & & \\
\textbf{Dataset} & \textbf{Size} & \textbf{Crop regions} & ViBES / Lens & ViBES / Lens & ViBES / Lens & $\Delta$AUROC$\uparrow$ & $\Delta$ECE$\downarrow$ & $\Delta$Brier$\downarrow$ \\
\midrule
\cellcolor{tMCS}MC-Search & 8B & one box & 0.599 / 0.629 & 0.032 / 0.015 & 0.227 / 0.222 & \cellcolor{dGood}$+$0.029 & \cellcolor{dGood}$-$0.018 & \cellcolor{dGood}$-$0.005 \\
\cellcolor{tMCS} &  & mask & 0.566 / 0.674 & 0.055 / 0.051 & 0.235 / 0.220 & \cellcolor{dGood}$+$0.108 & \cellcolor{dGood}$-$0.003 & \cellcolor{dGood}$-$0.015 \\
\cellcolor{tMCS} &  & $N$ boxes & 0.557 / 0.560 & 0.072 / 0.035 & 0.231 / 0.224 & \cellcolor{dGood}$+$0.003 & \cellcolor{dGood}$-$0.037 & \cellcolor{dGood}$-$0.007 \\
\addlinespace[2pt]
\cellcolor{tMCS} & 4B & one box & 0.482 / 0.618 & 0.054 / 0.073 & 0.204 / 0.209 & \cellcolor{dGood}$+$0.136 & \cellcolor{dBad}$+$0.020 & \cellcolor{dBad}$+$0.004 \\
\cellcolor{tMCS} &  & mask & 0.576 / 0.594 & 0.088 / 0.047 & 0.223 / 0.225 & \cellcolor{dGood}$+$0.018 & \cellcolor{dGood}$-$0.041 & \cellcolor{dBad}$+$0.002 \\
\cellcolor{tMCS} &  & $N$ boxes & 0.573 / 0.563 & 0.041 / 0.039 & 0.234 / 0.229 & \cellcolor{dBad}$-$0.010 & \cellcolor{dGood}$-$0.002 & \cellcolor{dGood}$-$0.005 \\
\midrule
\cellcolor{tWQ}WebQA & 8B & one box & 0.564 / 0.601 & 0.006 / 0.030 & 0.216 / 0.220 & \cellcolor{dGood}$+$0.036 & \cellcolor{dBad}$+$0.024 & \cellcolor{dBad}$+$0.004 \\
\cellcolor{tWQ} &  & mask & 0.558 / 0.611 & 0.014 / 0.025 & 0.216 / 0.215 & \cellcolor{dGood}$+$0.052 & \cellcolor{dBad}$+$0.011 & \cellcolor{dGood}$-$0.001 \\
\cellcolor{tWQ} &  & $N$ boxes & 0.568 / 0.580 & 0.018 / 0.025 & 0.217 / 0.217 & \cellcolor{dGood}$+$0.011 & \cellcolor{dBad}$+$0.008 & 0.000 \\
\addlinespace[2pt]
\cellcolor{tWQ} & 4B & one box & 0.576 / 0.598 & 0.005 / 0.026 & 0.223 / 0.224 & \cellcolor{dGood}$+$0.022 & \cellcolor{dBad}$+$0.022 & \cellcolor{dBad}$+$0.002 \\
\cellcolor{tWQ} &  & mask & 0.503 / 0.565 & 0.005 / 0.019 & 0.225 / 0.225 & \cellcolor{dGood}$+$0.062 & \cellcolor{dBad}$+$0.014 & 0.000 \\
\cellcolor{tWQ} &  & $N$ boxes & 0.497 / 0.511 & 0.000 / 0.011 & 0.227 / 0.224 & \cellcolor{dGood}$+$0.014 & \cellcolor{dBad}$+$0.011 & \cellcolor{dGood}$-$0.003 \\
\midrule
\cellcolor{tSAGE}SAGE & 8B & one box & 0.635 / 0.704 & 0.098 / 0.049 & 0.238 / 0.218 & \cellcolor{dGood}$+$0.069 & \cellcolor{dGood}$-$0.049 & \cellcolor{dGood}$-$0.020 \\
\cellcolor{tSAGE} &  & mask & 0.567 / 0.632 & 0.025 / 0.084 & 0.238 / 0.235 & \cellcolor{dGood}$+$0.065 & \cellcolor{dBad}$+$0.058 & \cellcolor{dGood}$-$0.003 \\
\cellcolor{tSAGE} &  & $N$ boxes & 0.586 / 0.578 & 0.037 / 0.021 & 0.233 / 0.234 & \cellcolor{dBad}$-$0.008 & \cellcolor{dGood}$-$0.016 & \cellcolor{dBad}$+$0.001 \\
\addlinespace[2pt]
\cellcolor{tSAGE} & 4B & one box & 0.643 / 0.716 & 0.069 / 0.059 & 0.238 / 0.209 & \cellcolor{dGood}$+$0.073 & \cellcolor{dGood}$-$0.010 & \cellcolor{dGood}$-$0.029 \\
\cellcolor{tSAGE} &  & mask & 0.645 / 0.661 & 0.154 / 0.139 & 0.235 / 0.222 & \cellcolor{dGood}$+$0.016 & \cellcolor{dGood}$-$0.015 & \cellcolor{dGood}$-$0.013 \\
\cellcolor{tSAGE} &  & $N$ boxes & 0.706 / 0.747 & 0.176 / 0.134 & 0.224 / 0.201 & \cellcolor{dGood}$+$0.041 & \cellcolor{dGood}$-$0.042 & \cellcolor{dGood}$-$0.023 \\
\bottomrule
\end{tabular}
\par\smallskip\footnotesize ECE and Brier score use probabilities from Platt scaling of the probe score, cross-fitted
within each cell; AUROC uses the score itself. $\Delta$: from unrounded values.

\end{table*}

\begin{table*}[!t]
\centering\scriptsize\setlength{\tabcolsep}{3.2pt}\setlength{\aboverulesep}{0pt}\setlength{\belowrulesep}{0pt}\setlength{\extrarowheight}{0.4pt}
\caption{\textbf{MultiModalQA accuracy (\%)} (official exact match, one run per cell). Bold: best per column; underline: second best.}
\label{tab:v2mmqa}
\begin{tabular}{@{}l|rrr|rrr@{}}
\toprule
& \multicolumn{3}{c|}{\textbf{8B}} & \multicolumn{3}{c}{\textbf{4B}} \\
\cmidrule(lr){2-4}\cmidrule(l){5-7}
\textbf{System} & Loop & Pi & OC & Loop & Pi & OC \\
\midrule
Closed-book & 11.1 & 12.7 & 11.8 & 9.8 & 12.4 & 12.7 \\
One-shot RAG & 16.0 & 14.7 & 13.7 & 9.5 & 14.4 & 13.1 \\
RAG-SFT & 15.0 & 14.1 & 13.7 & 16.3 & 12.7 & 14.4 \\
Agent, zero-shot & 14.7 & 15.4 & \textbf{16.7} & 14.1 & 16.3 & 13.7 \\
Agent + SFT & 14.4 & 18.0 & 14.4 & 17.6 & \textbf{18.6} & 15.4 \\
\midrule
ViBES (own box) & \underline{19.9} & \underline{19.6} & 14.7 & 17.6 & \underline{17.3} & \textbf{16.7} \\
ViBES-Lens, one box & \textbf{20.9} & \textbf{20.3} & 15.0 & \textbf{19.6} & 16.3 & \underline{16.3} \\
ViBES-Lens, mask & 19.0 & 19.3 & 14.7 & \underline{19.3} & 17.0 & \textbf{16.7} \\
ViBES-Lens, $N$ boxes & \textbf{20.9} & 19.0 & \underline{16.0} & \underline{19.3} & 17.0 & \textbf{16.7} \\
\bottomrule
\end{tabular}
\par\smallskip\footnotesize The 8B loop cells of the mask and $N$ boxes come from a job that also reran the own box
(19.6).
\end{table*}

\begin{table*}[!t]
\centering\scriptsize\setlength{\tabcolsep}{4pt}\setlength{\aboverulesep}{0pt}\setlength{\belowrulesep}{0pt}\setlength{\extrarowheight}{0.4pt}\definecolor{diag}{gray}{0.9}
\caption{\textbf{Cross-dataset transfer accuracy (\%), own box / $N$ boxes.} Rows: dataset the agent and the
selector were trained on; columns: test dataset, scored under its own protocol. Grey: in-domain. One run per cell.}
\label{tab:v2ood}
\begin{tabular}{@{}l|rrrr|rrrr@{}}
\toprule
& \multicolumn{4}{c|}{\textbf{8B}} & \multicolumn{4}{c}{\textbf{4B}} \\
\cmidrule(lr){2-5}\cmidrule(l){6-9}
\textbf{Trained on} & \multicolumn{1}{>{\columncolor{tWQ}}r}{WebQA} & MMQA & \multicolumn{1}{>{\columncolor{tSAGE}}r}{SAGE} & \multicolumn{1}{>{\columncolor{tMCS}}r|}{MC-Search} & \multicolumn{1}{>{\columncolor{tWQ}}r}{WebQA} & MMQA & \multicolumn{1}{>{\columncolor{tSAGE}}r}{SAGE} & \multicolumn{1}{>{\columncolor{tMCS}}r}{MC-Search} \\
\midrule
WebQA & \cellcolor{diag}66.4/66.8 & 19.0/18.3 & 32.4/31.7 & 41.8/40.8 & \cellcolor{diag}65.0/65.4 & 18.3/18.3 & 30.2/30.9 & 37.6/35.5 \\
MultiModalQA & 64.4/63.4 & \cellcolor{diag}19.6/20.9 & 32.6/31.4 & 42.0/40.9 & 60.0/60.6 & \cellcolor{diag}17.6/19.3 & 29.2/28.8 & 37.6/37.4 \\
SAGE & 64.2/61.6 & 14.4/14.4 & \cellcolor{diag}42.5/42.2 & 36.8/38.9 & 56.4/54.8 & 14.7/15.0 & \cellcolor{diag}36.9/39.1 & 31.9/32.5 \\
MC-Search & 59.3/56.7 & 14.1/13.4 & 35.5/35.0 & \cellcolor{diag}38.9/38.5 & 59.5/61.1 & 18.6/18.0 & 28.1/28.6 & \cellcolor{diag}41.8/41.8 \\
\bottomrule
\end{tabular}
\end{table*}

\section{Additional Results on MultiModalQA}
\label{app:v2mmqa}
MultiModalQA~\cite{talmor2021mmqa} asks image--text questions; we use its questions without tables, scored by official
exact match. It has no external baselines in our setting and is not part of the averages of the main paper.
Table~\ref{tab:v2mmqa} gives the in-house systems, ViBES and ViBES-Lens.
$\Delta$AUROC of ViBES-Lens over ViBES (8B, 4B) is $+0.034$ and $+0.023$ for the one box, $+0.018$ and $-0.053$ for
the mask, and $+0.120$ and $-0.002$ for the $N$ boxes.

\newtcolorbox{promptbox}[1]{enhanced, colback=black!6, colframe=black!70, colbacktitle=black!70,
  coltitle=white, fonttitle=\bfseries\footnotesize, title={#1}, boxrule=0.5pt, arc=2pt, left=5pt, right=5pt, top=3pt,
  bottom=3pt, before skip=4pt, after skip=2pt, width=\textwidth}
\newcommand{\promptfile}[2]{\begin{promptbox}{#1}\VerbatimInput[fontsize=\scriptsize,baselinestretch=0.95]{supp/prompts/#2.txt}\end{promptbox}}

\begin{figure*}[!tp]
\promptfile{System Message (template)}{p01_system_template}
\promptfile{Notes appended to the system message}{p01b_notes}
\promptfile{Answer instruction per dataset}{p02b_answer_formats}
\caption{\textbf{System message of the agent loop.} The template, the notes appended to it and the answer instruction that ends it on each dataset. The action list and crop clauses depend on whether the crop tool is on; the budget is 10 tool calls, 3 retrievals, 3 inspections and 4 crops in every reported run.}
\label{fig:p-template}
\end{figure*}

\begin{figure*}[!tp]
\promptfile{Prompt (user turns)}{p06_observations}
\promptfile{Output schema}{p05_action_schema}
\caption{\textbf{User-turn templates and output constraint.} The first message, the observation after each tool with the remaining-budget line, the error messages, and the JSON schema that constrains every action. Images are passed as image inputs at the marked positions.}
\label{fig:p-obs}
\end{figure*}

\begin{figure*}[!tp]
\promptfile{System Message (closed-book)}{p03_single_none}
\promptfile{System Message (one-shot RAG, RAG-SFT)}{p03_single_retrieved}
\caption{\textbf{Single-call baselines.} For one-shot RAG and RAG-SFT, the five retrieved records follow the question as numbered blocks (WebQA answer format shown).}
\label{fig:p-single}
\promptfile{Hint appended to the system message of the teacher run}{p07_teacher_hints}
\caption{\textbf{Teacher hints} (Stage 1 of Figure~\ref{fig:v2stages}), removed from the stored training conversations.}
\label{fig:p-teacher}
\promptfile{System Message and Prompt}{p08_judge}
\caption{\textbf{Judge prompt} (Qwen2.5-14B-Instruct) for WebQA and MC-Search accuracy.}
\label{fig:p-judge}
\end{figure*}

\section{Cross-Dataset Transfer}
\label{app:v2ood}
We run each ViBES agent on the test questions of the other three datasets, under the target's prompt, corpus and
scoring, once with its own box and once with the $N$-box ViBES-Lens selector trained on the same source dataset
(Table~\ref{tab:v2ood}). For MC-Search $\to$ WebQA we drop the WebQA test questions linked to MC-Search training
questions (388 of 500 remain). Transfer between the general benchmarks costs little: on WebQA, the 8B agents trained on
MultiModalQA and SAGE (64.4, 64.2) match the zero-shot agent (64.3), and on MC-Search the 8B agents trained on WebQA and
MultiModalQA (41.8, 42.0) exceed the in-domain cell (38.9). SAGE needs in-domain training:
agents trained elsewhere reach 32.4--35.5 at 8B, 7.0--10.1 points below ViBES and below the zero-shot agent (36.7), and
28.1--30.2 at 4B, 6.7--8.7 points below ViBES. ViBES-Lens with $N$ boxes stays within $-2.6$ to $+2.0$ points of the own
box in every transferred cell (mean $-0.8$ at 8B, $0.0$ at 4B). The same runs in pi and OpenCode give mean differences
over the 12 transfer pairs of $-0.3$ and $-0.2$ (8B) and $+0.3$ and $+0.1$ (4B).
\section{Prompts}
\label{app:prompts}
All agents in our loop (zero-shot agent, Agent + SFT, ViBES, ViBES-Lens and the external weights) share the system
message of Figure~\ref{fig:p-template} and the observation templates of Figure~\ref{fig:p-obs}, and training uses the
same text; only the teacher run adds the hint of Figure~\ref{fig:p-teacher}. The single-call baselines use
Figure~\ref{fig:p-single}, and the judge Figure~\ref{fig:p-judge}. No prompt contains a few-shot example. Long lines
are wrapped with a two-space indent, and names in braces are filled at run time.

\FloatBarrier
\section{Agent Trajectories}\label{app:traces}
\definecolor{trRet}{HTML}{2E9E80}\definecolor{trIns}{HTML}{8F7A16}\definecolor{trCrop}{HTML}{AA4499}
\definecolor{trAns}{HTML}{3F78B4}\definecolor{trRef}{HTML}{7A7A7A}\definecolor{trInsP}{HTML}{C9B44A}
\newlength{\trth}\setlength{\trth}{1.05cm}%
\newlength{\tric}\setlength{\tric}{0.55cm}%
\newsavebox{\trsave}
\newcommand{\trcapfig}[1]{\captionof{figure}{#1}}%
\newlength{\trmaxh}
\newenvironment{trcasecol}[3][1]{\begin{minipage}[t]{0.493\textwidth}\vspace{0pt}\centering
  \gdef\trcap{#2}\gdef\trlabel{#3}\setlength{\trmaxh}{0.94\textheight}%
  \begin{lrbox}{\trsave}\begin{minipage}[b]{#1\linewidth}\setlength{\parindent}{0pt}}
  {\end{minipage}\end{lrbox}%
  \ifdim\dimexpr(\ht\trsave+\dp\trsave)*\numexpr\linewidth/1000\relax/\numexpr\trmaxh/1000\relax\relax>\wd\trsave
    \resizebox{!}{\trmaxh}{\usebox{\trsave}}%
  \else\ifdim\wd\trsave>\linewidth\resizebox{\linewidth}{!}{\usebox{\trsave}}\else\usebox{\trsave}\fi\fi
  \par\vspace{3pt}\expandafter\trcapfig\expandafter{\trcap}\label{\trlabel}\end{minipage}}
\newlength{\trpad}\setlength{\trpad}{1.4pt}\newlength{\trgap}\setlength{\trgap}{1.8pt}
\newcommand{\trtight}{\setlength{\trth}{0.95cm}\setlength{\trpad}{0.9pt}\setlength{\trgap}{1.2pt}}
\providecommand{\trdot}{$\cdot$}
\providecommand{\trgold}{$\bigstar$\,}
\providecommand{\trtimes}[1]{$\times$#1}
\providecommand{\trthumb}[1]{\hspace{2pt}\raisebox{-0.5\height}{{\setlength{\fboxsep}{0pt}\setlength{\fboxrule}{0.3pt}\fcolorbox{black!30}{white}{\includegraphics[width=\trth,height=\trth]{#1}}}}}
\providecommand{\trcut}{\hspace{1.5pt}\raisebox{-0.3\height}{\color{trCrop}\footnotesize\faCut}\hspace{-0.5pt}}
\providecommand{\trbudget}[1]{\par{\fontsize{6.4}{7.2}\selectfont\color{black!50}left: #1}}
\providecommand{\trfooter}[1]{\par\vspace{1pt}\noindent\hspace{\tric}\parbox[t]{\dimexpr\linewidth-\tric\relax}{\raggedright\sffamily\fontsize{6.4}{7.4}\selectfont\color{black!60}#1}\par}
\newtcolorbox{trbub}{enhanced, colback=white, colframe=black!80, boxrule=0.5pt, arc=3pt, left=3pt, right=3pt, top=\trpad,
  bottom=\trpad, fontupper=\sffamily\fontsize{7.4}{8.6}\selectfont, before skip=0pt, after skip=0pt, width=\linewidth}
\newtcolorbox{trtool}{enhanced, colback=white, frame hidden, borderline={0.5pt}{0pt}{black!45,dashed}, arc=3pt, left=3pt,
  right=2pt, top=\trpad, bottom=\trpad, fontupper=\sffamily\fontsize{7.4}{8.6}\selectfont, before skip=0pt, after skip=0pt, width=\linewidth}
\newtcolorbox{trpanel}[1]{enhanced, colback=#1!9, frame hidden, arc=2pt, left=2pt, right=2pt, top=\trpad, bottom=\trpad,
  fontupper=\sffamily\fontsize{7}{8.1}\selectfont, before skip=0pt, after skip=0pt, width=\linewidth}
\newcommand{\trrow}[2]{\noindent\parbox[c]{\tric}{\centering #1}\parbox[c]{\dimexpr\linewidth-\tric\relax}{#2}\par\vspace{\trgap}}
\newcommand{\truser}[1]{\trrow{\normalsize\faUser}{\begin{trbub}#1\end{trbub}}}
\newcommand{\tragent}[1]{\trrow{\normalsize\faRobot}{\begin{trbub}\raggedright #1\end{trbub}}}
\newcommand{\trcall}[4]{\trrow{\normalsize\color{black!55}\faCog}{\begin{trtool}%
  \begin{minipage}[c]{0.27\linewidth}\raggedright #2\trbudget{#4}\end{minipage}\hfill
  \begin{minipage}[c]{0.715\linewidth}\begin{trpanel}{#1}#3\end{trpanel}\end{minipage}\end{trtool}}}

In this section, we show eight logged test episodes of the 8B ViBES agent, two per dataset (SAGE, WebQA, MC-Search
and MultiModalQA), from the question to the answer. For each dataset, we take the episodes with a correct answer (on
MultiModalQA, by official exact match), every gold record retrieved and no refused call, and a seeded shuffle picks the
first episode of two groups defined by the logged counters: little or no gold in the seeded retrieval, and a crop or a
partly complete seeded retrieval. The agent bubbles quote the note that the policy
writes with its next action, shortened with [...]. Each tool box shows a call, the records it returns ($\bigstar$:
gold; the other records are counted) and the budget left ($K$ tool calls, R retrievals, I inspections, C crops). Crops
are re-cut from the logged boxes, drawn in magenta on their source image.

\begin{figure*}[!p]
\begin{trcasecol}[1.45]{Case study 1 of ViBES (SAGE): an image search on the input returns four gold-class images after a seeded retrieval without gold.}{fig:v2case1}\trtight
\truser{\begin{minipage}[c]{\dimexpr\linewidth-2.37cm\relax}\raggedright Which disease does the soybean plant in the input image have? Choose exactly one of these 27 soybean classes: Alfalfa Mosaic Virus; Anthracnose; Bacterial Blight; Bacterial Pustule; [...].\end{minipage}\hfill\raisebox{-0.5\height}{\includegraphics[height=1.55cm]{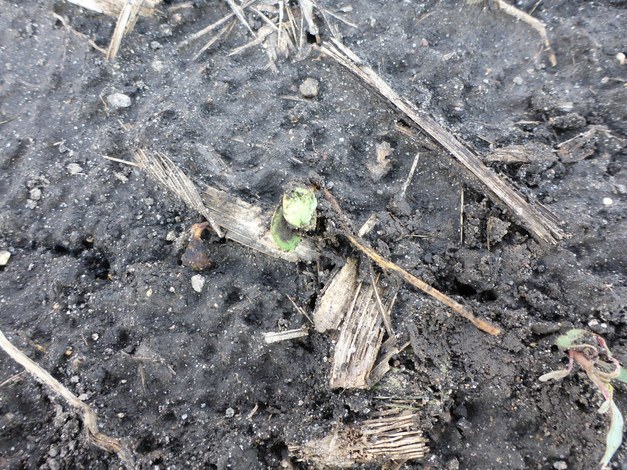}}}
\trcall{trRet}{Tool: \textbf{\color{trRet}Retrieve} (seeded)\\ multimodal: the question and the input image}{\raggedright \#1 \texttt{doc:\allowbreak{}kb\_\allowbreak{}soybean\_\allowbreak{}sudden\_\allowbreak{}death\_\allowbreak{}syndrome} \underline{Soybean - Sudden death syndrome}: symptom knowledge base entry. [...]\par + 4 other records}{$K$\,9 \trdot{} R\,2 \trdot{} I\,3 \trdot{} C\,4}
\tragent{``[...] {\color{trRet}I need to search for images of 'Soybean - Pythium damping off' to compare with the input image.}''}
\trcall{trRet}{Tool: \textbf{\color{trRet}Retrieve} (image)\\ image: \texttt{input}}{\begin{minipage}[c]{\dimexpr\linewidth-2\trth-6pt\relax}\raggedright \trgold{}\#2 \texttt{img:\allowbreak{}0963ba07} Soybean - Pythium damping off (merged, stem)\par \trgold{}\#3 \texttt{img:\allowbreak{}362fee74} Soybean - Pythium damping off (merged, stem)\par \trgold{}+ 2 more gold records (\#4, \#5)\par + 1 other record\end{minipage}\hfill\trthumb{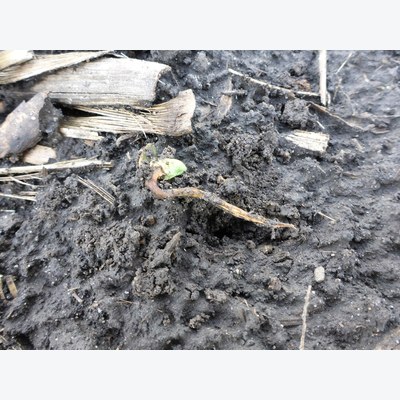}\trthumb{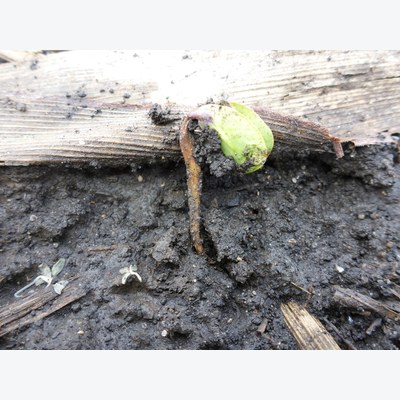}}{$K$\,8 \trdot{} R\,1 \trdot{} I\,3 \trdot{} C\,4}
\tragent{``[...] {\color{trIns}I will inspect one of the images to confirm the visual match.}''}
\trcall{trInsP}{Tool: \textbf{\color{trIns}Inspect}\\ record: \texttt{img:\allowbreak{}0963ba07}}{\begin{minipage}[c]{\dimexpr\linewidth-1\trth-3pt\relax}\raggedright \trgold{}\texttt{img:\allowbreak{}0963ba07} Soybean - Pythium damping off (merged, stem)\end{minipage}\hfill\trthumb{figs/traces/t01_0963ba07.jpg}}{$K$\,7 \trdot{} R\,1 \trdot{} I\,2 \trdot{} C\,4}
\tragent{``[...] {\color{trCrop}I will now crop the input image to focus on the affected stem area for a closer comparison.}''}
\trcall{trCrop}{Tool: \textbf{\color{trCrop}Crop}\\ \texttt{input}, box [350, 380, 750, 600]}{\raggedright\setlength{\baselineskip}{\dimexpr\trth+5pt\relax}\trthumb{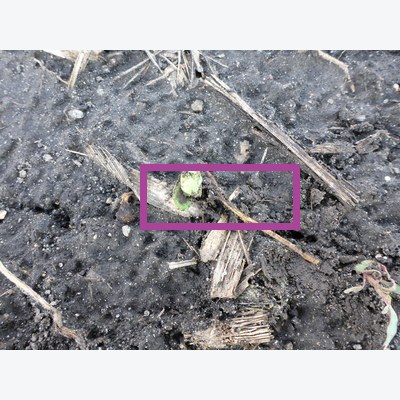}\trcut\trthumb{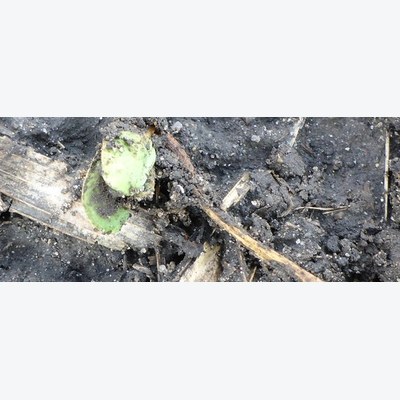}}{$K$\,6 \trdot{} R\,1 \trdot{} I\,2 \trdot{} C\,3}
\tragent{``[...] {\color{trIns}I will now inspect another image of Pythium damping off to further confirm the diagnosis.}''}
\trcall{trInsP}{Tool: \textbf{\color{trIns}Inspect}\\ record: \texttt{img:\allowbreak{}362fee74}}{\begin{minipage}[c]{\dimexpr\linewidth-1\trth-3pt\relax}\raggedright \trgold{}\texttt{img:\allowbreak{}362fee74} Soybean - Pythium damping off (merged, stem)\end{minipage}\hfill\trthumb{figs/traces/t01_362fee74.jpg}}{$K$\,5 \trdot{} R\,1 \trdot{} I\,1 \trdot{} C\,3}
\tragent{``[...] {\color{trCrop}I will now crop the input image to focus on the affected stem area for a closer comparison.}''}
\trcall{trCrop}{Tool: \textbf{\color{trCrop}Crop}\\ \texttt{input}, box [350, 380, 750, 600]}{\raggedright\setlength{\baselineskip}{\dimexpr\trth+5pt\relax}\trthumb{figs/traces/t01_input_b350-380-750-600.jpg}\trcut\trthumb{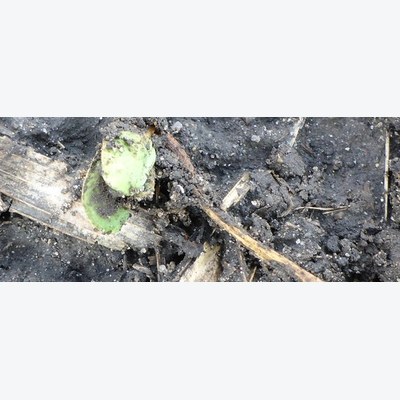}}{$K$\,4 \trdot{} R\,1 \trdot{} I\,1 \trdot{} C\,2}
\tragent{``[...] {\color{trIns}I will now inspect another image of Pythium damping off to further confirm the diagnosis.}''}
\trcall{trInsP}{Tool: \textbf{\color{trIns}Inspect}\\ record: \texttt{img:\allowbreak{}14fcc085}}{\begin{minipage}[c]{\dimexpr\linewidth-1\trth-3pt\relax}\raggedright \trgold{}\texttt{img:\allowbreak{}14fcc085} Soybean - Pythium damping off (merged, seed)\end{minipage}\hfill\trthumb{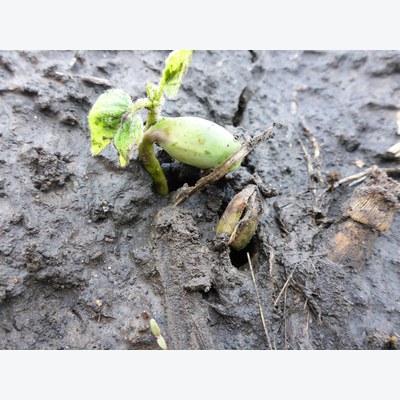}}{$K$\,3 \trdot{} R\,1 \trdot{} I\,0 \trdot{} C\,2}
\tragent{``[...] {\color{trCrop}I will now crop the input image to focus on the affected stem area for a closer comparison.}''}
\trcall{trCrop}{Tool: \textbf{\color{trCrop}Crop}\\ \texttt{input}, box [350, 380, 750, 600]}{\raggedright\setlength{\baselineskip}{\dimexpr\trth+5pt\relax}\trthumb{figs/traces/t01_input_b350-380-750-600.jpg}\trcut\trthumb{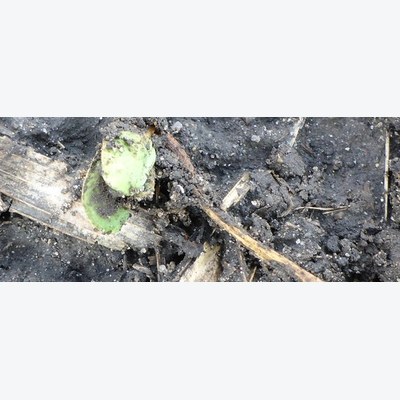}}{$K$\,2 \trdot{} R\,1 \trdot{} I\,0 \trdot{} C\,1}
\tragent{``[...] {\color{trRet}I will now perform a final search to confirm this diagnosis with a text passage.}''}
\trcall{trRet}{Tool: \textbf{\color{trRet}Retrieve} (text)\\ query: ``Soybean - Pythium damping off''}{\begin{minipage}[c]{\dimexpr\linewidth-1\trth-3pt\relax}\raggedright \trgold{}\#1 \texttt{doc:\allowbreak{}kb\_\allowbreak{}soybean\_\allowbreak{}pythium\_\allowbreak{}damping\_\allowbreak{}off} \underline{Soybean - Pythium damping off}: symptom knowledge base entry. [...]\par \trgold{}\#2 \texttt{img:\allowbreak{}cbcf2857} Soybean - Pythium damping off (merged, seed)\par \trgold{}+ 3 more gold records (\#3, \#4, \#5)\end{minipage}\hfill\trthumb{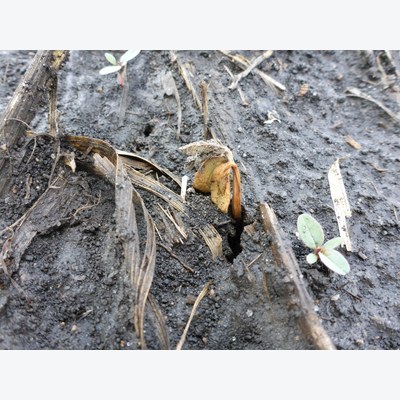}}{$K$\,1 \trdot{} R\,0 \trdot{} I\,0 \trdot{} C\,1}
\tragent{``[...] {\color{trCrop}I will now perform the final crop to focus on the affected stem area for a closer comparison.}''}
\trcall{trCrop}{Tool: \textbf{\color{trCrop}Crop}\\ \texttt{input}, box [350, 380, 750, 600]}{\raggedright\setlength{\baselineskip}{\dimexpr\trth+5pt\relax}\trthumb{figs/traces/t01_input_b350-380-750-600.jpg}\trcut\trthumb{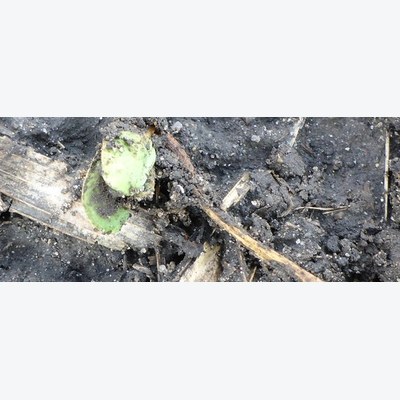}}{$K$\,0 \trdot{} R\,0 \trdot{} I\,0 \trdot{} C\,0}
\tragent{``The visual evidence from the input image and the retrieved images of Pythium damping off is consistent. The affected stem and the small green shoot are characteristic of Pythium damping off. The text passage also confirms the [...]''\par\vspace{1.5pt}\textbf{Answer:} \textbf{\color{trAns}Pythium damping off}\par\vspace{0.5pt}{\color{black!60}Cited: \texttt{img:\allowbreak{}0963ba07}, \texttt{img:\allowbreak{}362fee74}, \texttt{img:\allowbreak{}14fcc085}, \texttt{doc:\allowbreak{}kb\_\allowbreak{}soybean\_\allowbreak{}pythium\_\allowbreak{}damping\_\allowbreak{}off}}}
\trfooter{Reference: Pythium damping off \trdot{} Evaluator: correct (exact label) \trdot{} gold recall 1.0 \trdot{} 10 of 10 tool calls \trdot{} example \texttt{sage:\allowbreak{}Soybean:\allowbreak{}d0e6d484}}
\end{trcasecol}
\hfill
\begin{trcasecol}[1.3]{Case study 2 of ViBES (SAGE): the gold class is in the seeded retrieval, and the agent checks it with three inspections and two crops of the input.}{fig:v2case2}
\truser{\begin{minipage}[c]{\dimexpr\linewidth-2.63cm\relax}\raggedright Which disease does the cauliflower plant in the input image have? Choose exactly one of these 4 cauliflower classes: Alternaria Leaf Spot; Bacterial Soft Rot; Black Rot; Downy Mildew.\end{minipage}\hfill\raisebox{-0.5\height}{\includegraphics[height=1.55cm]{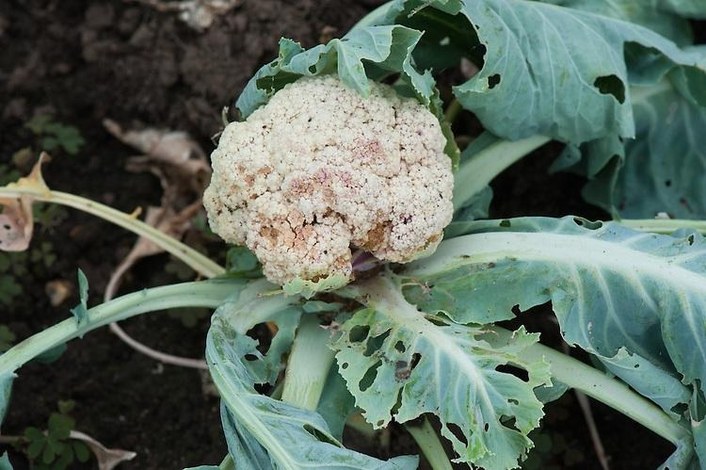}}}
\trcall{trRet}{Tool: \textbf{\color{trRet}Retrieve} (seeded)\\ multimodal: the question and the input image}{\raggedright \trgold{}\#3 \texttt{doc:\allowbreak{}kb\_\allowbreak{}cauliflower\_\allowbreak{}bacterial\_\allowbreak{}soft\_\allowbreak{}rot} \underline{Cauliflower - Bacterial Soft Rot}: symptom knowledge base entry. Pathogen: Pectobacterium [...]\par + 4 other records}{$K$\,9 \trdot{} R\,2 \trdot{} I\,3 \trdot{} C\,4}
\tragent{``The input image shows a cauliflower head with visible browning and decay, which matches symptoms of Bacterial Soft Rot. {\color{trRet}I will search for a text passage titled 'Cauliflower - Bacterial Soft Rot' to confirm.}''}
\trcall{trRet}{Tool: \textbf{\color{trRet}Retrieve} (text)\\ query: ``Cauliflower - Bacterial Soft Rot''}{\begin{minipage}[c]{\dimexpr\linewidth-2\trth-6pt\relax}\raggedright \trgold{}\#1 \texttt{doc:\allowbreak{}kb\_\allowbreak{}cauliflower\_\allowbreak{}bacterial\_\allowbreak{}soft\_\allowbreak{}rot} \underline{Cauliflower - Bacterial Soft Rot}: symptom knowledge base entry. Pathogen: Pectobacterium [...]\par \trgold{}\#3 \texttt{img:\allowbreak{}6e92c5a0} Cauliflower - Bacterial Soft Rot (whole plant)\par \trgold{}\#4 \texttt{img:\allowbreak{}0669b532} Cauliflower - Bacterial Soft Rot (whole plant)\par \trgold{}+ 1 more gold record (\#5)\par + 1 other record\end{minipage}\hfill\trthumb{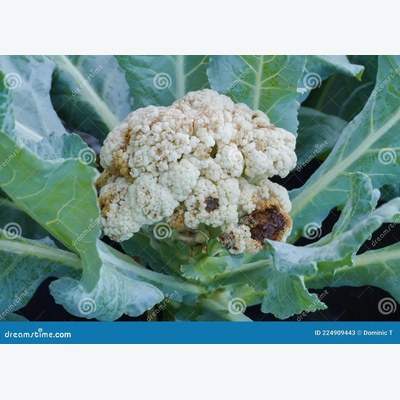}\trthumb{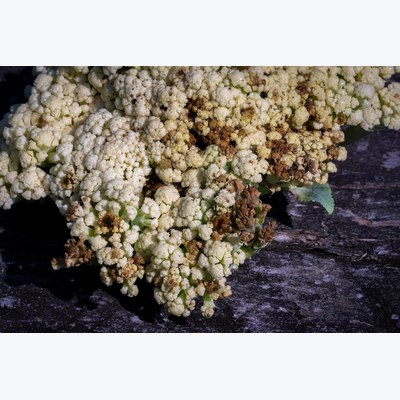}}{$K$\,8 \trdot{} R\,1 \trdot{} I\,3 \trdot{} C\,4}
\tragent{``The text passage and images for 'Cauliflower - Bacterial Soft Rot' describe symptoms matching the input image [...] {\color{trIns}I will now inspect the reference image for 'Cauliflower - Bacterial Soft Rot' to confirm visual similarity.}''}
\trcall{trInsP}{Tool: \textbf{\color{trIns}Inspect}\\ record: \texttt{img:\allowbreak{}6e92c5a0}}{\begin{minipage}[c]{\dimexpr\linewidth-1\trth-3pt\relax}\raggedright \trgold{}\texttt{img:\allowbreak{}6e92c5a0} Cauliflower - Bacterial Soft Rot (whole plant)\end{minipage}\hfill\trthumb{figs/traces/t03_6e92c5a0.jpg}}{$K$\,7 \trdot{} R\,1 \trdot{} I\,2 \trdot{} C\,4}
\tragent{``The inspected image of 'Cauliflower - Bacterial Soft Rot' shows similar browning and decay on the curd as the input [...] {\color{trCrop}I will now crop the input image to focus on the affected area of the cauliflower head for closer comparison.}''}
\trcall{trCrop}{Tool: \textbf{\color{trCrop}Crop}\\ \texttt{input}, box [280, 140, 660, 580]}{\raggedright\setlength{\baselineskip}{\dimexpr\trth+5pt\relax}\trthumb{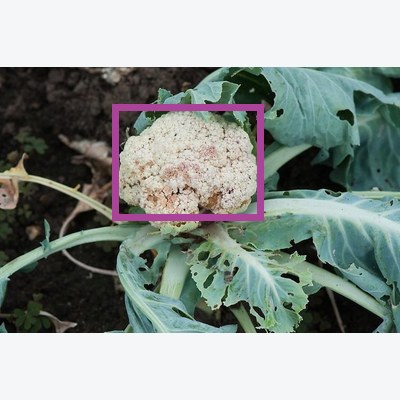}\trcut\trthumb{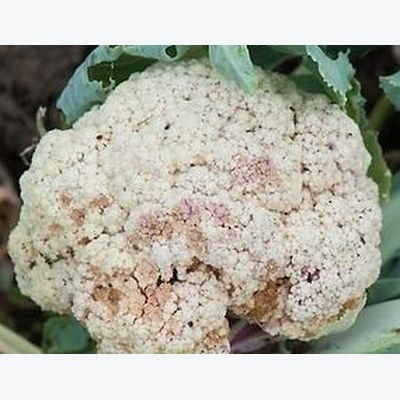}}{$K$\,6 \trdot{} R\,1 \trdot{} I\,2 \trdot{} C\,3}
\tragent{``The cropped image shows browning and decay on the cauliflower head, consistent with Bacterial Soft Rot. {\color{trIns}I will now inspect another reference image for 'Cauliflower - Bacterial Soft Rot' to further confirm.}''}
\trcall{trInsP}{Tool: \textbf{\color{trIns}Inspect}\\ record: \texttt{img:\allowbreak{}0669b532}}{\begin{minipage}[c]{\dimexpr\linewidth-1\trth-3pt\relax}\raggedright \trgold{}\texttt{img:\allowbreak{}0669b532} Cauliflower - Bacterial Soft Rot (whole plant)\end{minipage}\hfill\trthumb{figs/traces/t03_0669b532.jpg}}{$K$\,5 \trdot{} R\,1 \trdot{} I\,1 \trdot{} C\,3}
\tragent{``The inspected image shows advanced decay and browning on the cauliflower curd, matching the symptoms of Bacterial Soft [...] {\color{trCrop}I will now crop the input image to focus on the base of the cauliflower head where decay is most apparent.}''}
\trcall{trCrop}{Tool: \textbf{\color{trCrop}Crop}\\ \texttt{input}, box [350, 480, 600, 700]}{\raggedright\setlength{\baselineskip}{\dimexpr\trth+5pt\relax}\trthumb{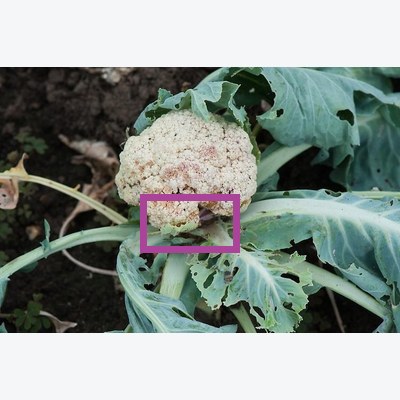}\trcut\trthumb{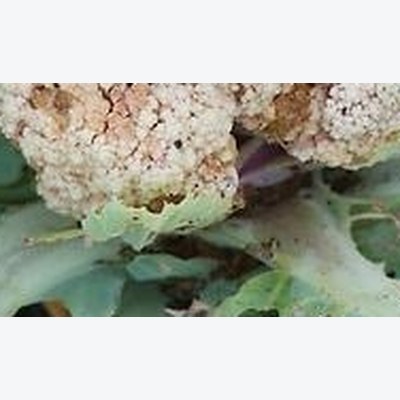}}{$K$\,4 \trdot{} R\,1 \trdot{} I\,1 \trdot{} C\,2}
\tragent{``The cropped image shows the base of the cauliflower head with visible decay and browning, consistent with Bacterial [...] {\color{trIns}I will now inspect the third reference image for 'Cauliflower - Bacterial Soft Rot' to finalize confirmation.}''}
\trcall{trInsP}{Tool: \textbf{\color{trIns}Inspect}\\ record: \texttt{img:\allowbreak{}79c354d5}}{\begin{minipage}[c]{\dimexpr\linewidth-1\trth-3pt\relax}\raggedright \trgold{}\texttt{img:\allowbreak{}79c354d5} Cauliflower - Bacterial Soft Rot (whole plant)\end{minipage}\hfill\trthumb{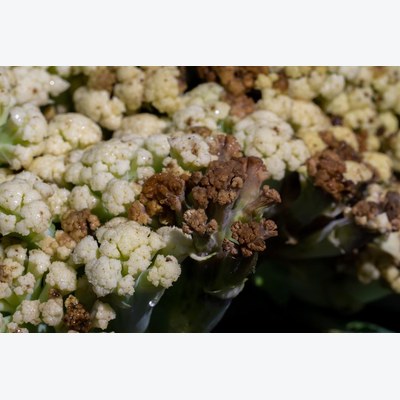}}{$K$\,3 \trdot{} R\,1 \trdot{} I\,0 \trdot{} C\,2}
\tragent{``All visual evidence and textual descriptions confirm the disease is Bacterial Soft Rot. I will now finish with the answer.''\par\vspace{1.5pt}\textbf{Answer:} \textbf{\color{trAns}Bacterial Soft Rot}\par\vspace{0.5pt}{\color{black!60}Cited: \texttt{doc:\allowbreak{}kb\_\allowbreak{}cauliflower\_\allowbreak{}bacterial\_\allowbreak{}soft\_\allowbreak{}rot}, \texttt{img:\allowbreak{}6e92c5a0}, \texttt{img:\allowbreak{}0669b532}, \texttt{img:\allowbreak{}79c354d5}}}
\trfooter{Reference: Bacterial Soft Rot \trdot{} Evaluator: correct (exact label) \trdot{} gold recall 1.0 \trdot{} 7 of 10 tool calls \trdot{} example \texttt{sage:\allowbreak{}Cauliflower:\allowbreak{}f3fefc94}}
\end{trcasecol}
\end{figure*}

\begin{figure*}[!p]
\begin{trcasecol}[1]{Case study 3 of ViBES (WebQA): the third retrieval returns the gold image that the first two miss.}{fig:v2case3}
\truser{\raggedright What four letter word is inscribed at the bottom of the grave of Babe Ruth and his wife?}
\trcall{trRet}{Tool: \textbf{\color{trRet}Retrieve} (seeded)\\ text: the question}{\raggedright \#1 \texttt{doc:\allowbreak{}d5cb5800\_\allowbreak{}14} \underline{The Babe}: The Babe is still his hero and he gives back the signed ball [...]\par + 4 other records}{$K$\,9 \trdot{} R\,2 \trdot{} I\,3 \trdot{} C\,4}
\tragent{``{\color{trRet}Search for the four-letter word inscribed at the bottom of Babe Ruth's grave.}''}
\trcall{trRet}{Tool: \textbf{\color{trRet}Retrieve} (text)\\ query: ``four letter word inscribed at the bottom of Babe Ruth's grave''}{\raggedright \#1 \texttt{doc:\allowbreak{}d5cb5800\_\allowbreak{}14} \underline{The Babe}: The Babe is still his hero and he gives back the signed ball [...]\par + 4 other records}{$K$\,8 \trdot{} R\,1 \trdot{} I\,3 \trdot{} C\,4}
\tragent{``{\color{trIns}Inspect the image of the Babe Ruth plaque to find the four-letter word at the bottom.}''}
\trcall{trInsP}{Tool: \textbf{\color{trIns}Inspect}\\ record: \texttt{img:\allowbreak{}30298623}}{\begin{minipage}[c]{\dimexpr\linewidth-1\trth-3pt\relax}\raggedright \texttt{img:\allowbreak{}30298623} Babe Ruth Plaque commons\end{minipage}\hfill\trthumb{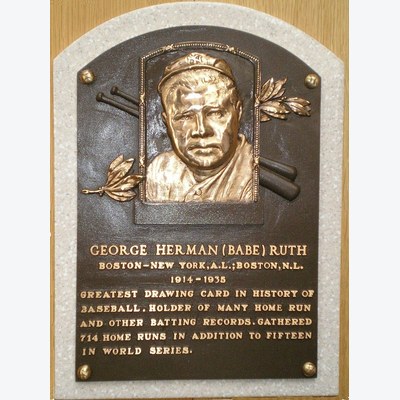}}{$K$\,7 \trdot{} R\,1 \trdot{} I\,2 \trdot{} C\,4}
\tragent{``{\color{trRet}Search for information about Babe Ruth's wife's grave inscription.}''}
\trcall{trRet}{Tool: \textbf{\color{trRet}Retrieve} (text)\\ query: ``inscription on Babe Ruth's wife's grave''}{\begin{minipage}[c]{\dimexpr\linewidth-1\trth-3pt\relax}\raggedright \trgold{}\#5 \texttt{img:\allowbreak{}30035295} Babe ruth grave Grave site of Babe Ruth and [...]\par + 4 other records\end{minipage}\hfill\trthumb{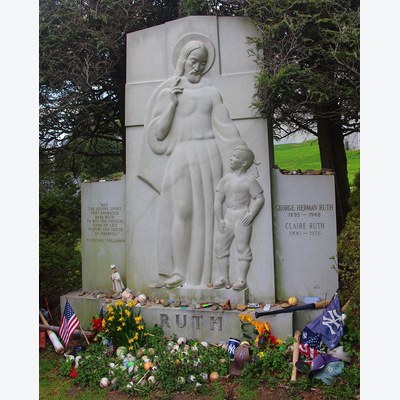}}{$K$\,6 \trdot{} R\,0 \trdot{} I\,2 \trdot{} C\,4}
\tragent{``{\color{trIns}Inspect the image of Babe Ruth's grave to find the inscription.}''}
\trcall{trInsP}{Tool: \textbf{\color{trIns}Inspect}\\ record: \texttt{img:\allowbreak{}30035295}}{\begin{minipage}[c]{\dimexpr\linewidth-1\trth-3pt\relax}\raggedright \trgold{}\texttt{img:\allowbreak{}30035295} Babe ruth grave Grave site of Babe Ruth and his [...]\end{minipage}\hfill\trthumb{figs/traces/t05_30035295.jpg}}{$K$\,5 \trdot{} R\,0 \trdot{} I\,1 \trdot{} C\,4}
\tragent{``The four-letter word inscribed at the bottom of the grave is 'RUTH'.''\par\vspace{1.5pt}\textbf{Answer:} \textbf{\color{trAns}The four-letter word inscribed at the bottom of the grave of Babe Ruth and his wife is RUTH.}\par\vspace{0.5pt}{\color{black!60}Cited: \texttt{img:\allowbreak{}30035295}}}
\trfooter{Reference: Ruth is the four-letter word inscribed at the bottom of the grave of Babe Ruth and his wife. \trdot{} Evaluator: correct (judge) \trdot{} gold recall 1.0 \trdot{} 5 of 10 tool calls \trdot{} example \texttt{webqa:\allowbreak{}d5cc688a}}
\end{trcasecol}
\hfill
\begin{trcasecol}[1.15]{Case study 4 of ViBES (WebQA): both gold images are in the seeded retrieval, and the agent inspects and crops one image of each painting.}{fig:v2case4}
\truser{\raggedright Is the headcovering worn by the central figures of the painting Mademoiselle Guimard as Terpsichore by Jacques-Louis David the same color as the one worn by the central figure of Camille Pissarro's painting Washerwoman, Study?}
\trcall{trRet}{Tool: \textbf{\color{trRet}Retrieve} (seeded)\\ text: the question}{\begin{minipage}[c]{\dimexpr\linewidth-2\trth-6pt\relax}\raggedright \trgold{}\#4 \texttt{img:\allowbreak{}30381956} Camille Pissarro Washerwoman, Study The [...]\par \trgold{}\#5 \texttt{img:\allowbreak{}30340481} Mademoiselle Guimard as Terpsichore, by [...]\par + 3 other records\end{minipage}\hfill\trthumb{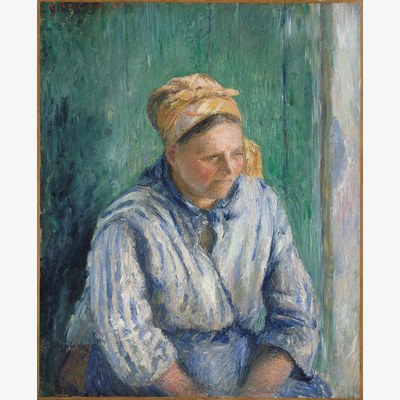}\trthumb{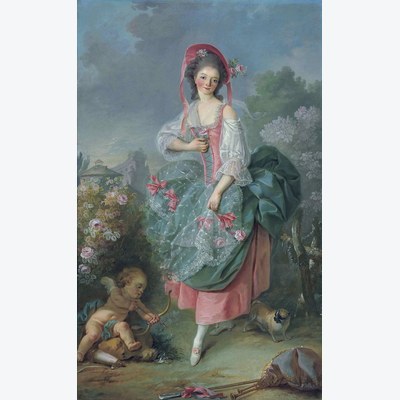}}{$K$\,9 \trdot{} R\,2 \trdot{} I\,3 \trdot{} C\,4}
\tragent{``{\color{trRet}Search for the color of the headcovering in Mademoiselle Guimard as Terpsichore.}''}
\trcall{trRet}{Tool: \textbf{\color{trRet}Retrieve} (text)\\ query: ``color of headcovering in Mademoiselle Guimard as Terpsichore by Jacques-Louis David''}{\raggedright \#1 \texttt{doc:\allowbreak{}d5c07c1e\_\allowbreak{}6} \underline{Portrait of Madame Récamier}: Portrait of Madame Récamier is an 1800 portrait of the [...]\par + 4 other records}{$K$\,8 \trdot{} R\,1 \trdot{} I\,3 \trdot{} C\,4}
\tragent{``{\color{trRet}Search for the color of the headcovering in Camille Pissarro's Washerwoman, Study.}''}
\trcall{trRet}{Tool: \textbf{\color{trRet}Retrieve} (text)\\ query: ``color of headcovering in Camille Pissarro's Washerwoman, Study''}{\begin{minipage}[c]{\dimexpr\linewidth-1\trth-3pt\relax}\raggedright \trgold{}\#4 \texttt{img:\allowbreak{}30381956} Camille Pissarro Washerwoman, Study The [...]\par + 4 other records\end{minipage}\hfill\trthumb{figs/traces/t07_30381956.jpg}}{$K$\,7 \trdot{} R\,0 \trdot{} I\,3 \trdot{} C\,4}
\tragent{``{\color{trIns}Inspect the image of Mademoiselle Guimard as Terpsichore to determine the color of the headcovering.}''}
\trcall{trInsP}{Tool: \textbf{\color{trIns}Inspect} \trtimes{2}\\ record: \texttt{img:\allowbreak{}30147060}, \texttt{img:\allowbreak{}30381956}}{\begin{minipage}[c]{\dimexpr\linewidth-2\trth-6pt\relax}\raggedright \texttt{img:\allowbreak{}30147060} Mademoiselle Guimard as Terpsichore, by [...]\par \trgold{}\texttt{img:\allowbreak{}30381956} Camille Pissarro Washerwoman, Study The [...]\end{minipage}\hfill\trthumb{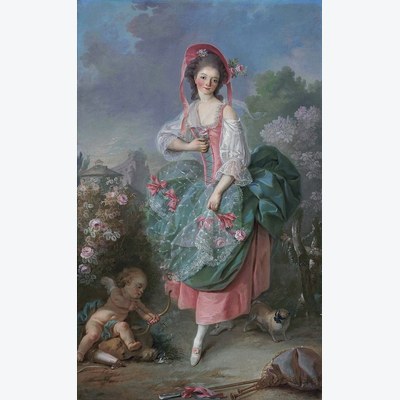}\trthumb{figs/traces/t07_30381956.jpg}}{$K$\,5 \trdot{} R\,0 \trdot{} I\,1 \trdot{} C\,4}
\tragent{``{\color{trCrop}Crop the image of Mademoiselle Guimard as Terpsichore to focus on the headcovering.}''}
\trcall{trCrop}{Tool: \textbf{\color{trCrop}Crop} \trtimes{2}\\ \texttt{img:\allowbreak{}30147060}, box [400, 110, 650, 250]\\ \texttt{img:\allowbreak{}30381956}, box [320, 175, 520, 350]}{\raggedright\setlength{\baselineskip}{\dimexpr\trth+5pt\relax}\trthumb{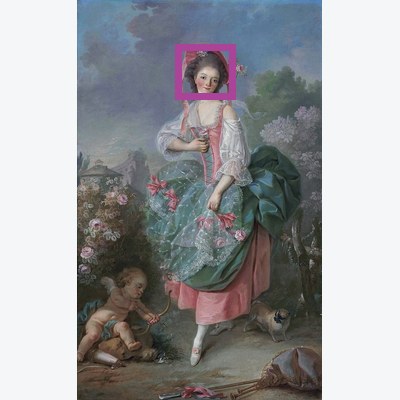}\trcut\trthumb{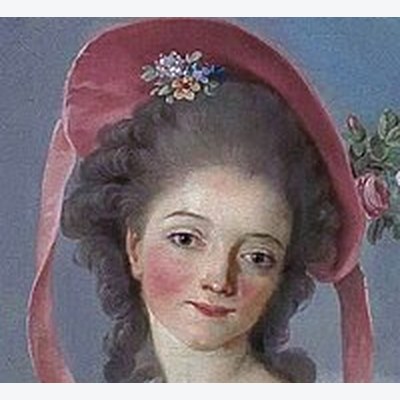}\hspace{6pt}\trthumb{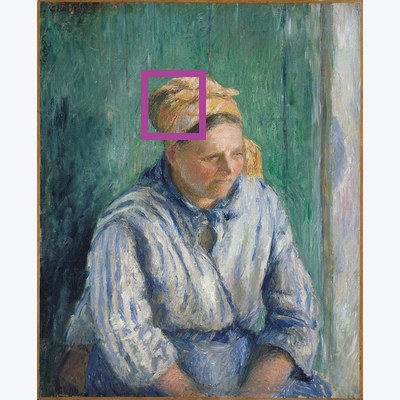}\trcut\trthumb{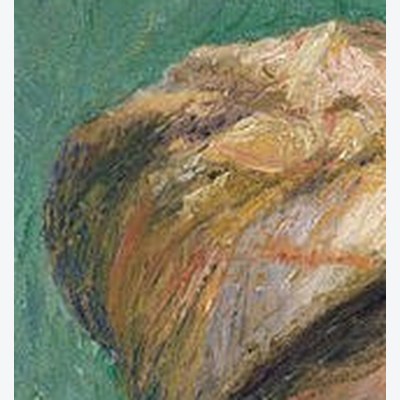}}{$K$\,3 \trdot{} R\,0 \trdot{} I\,1 \trdot{} C\,2}
\tragent{``The headcovering in Mademoiselle Guimard as Terpsichore is red, while the headcovering in Camille Pissarro's Washerwoman, Study is yellowish-brown. Therefore, they are not the same color.''\par\vspace{1.5pt}\textbf{Answer:} \textbf{\color{trAns}No, the headcoverings are not the same color.}\par\vspace{0.5pt}{\color{black!60}Cited: \texttt{img:\allowbreak{}30147060}, \texttt{img:\allowbreak{}30381956}}}
\trfooter{Reference: The headcovering worn by the central figures of the painting Mademoiselle Guimard as Terpsichore by Jacques-Louis David is in a different color. \trdot{} Evaluator: correct (judge) \trdot{} gold recall 1.0 \trdot{} 7 of 10 tool calls \trdot{} example \texttt{webqa:\allowbreak{}d5cb78bc}}
\end{trcasecol}
\end{figure*}

\begin{figure*}[!p]
\begin{trcasecol}[1.15]{Case study 5 of ViBES (MC-Search): one of four gold records is in the seeded retrieval, and two queries written by the agent return the other three.}{fig:v2case5}
\truser{\raggedright Who is the player in jersey number 55 for the Miami Heat, what nickname was he given due to his playing style, what position did he play during his 2006 NBA championship, and which former teammate facilitated his move to Miami?}
\trcall{trRet}{Tool: \textbf{\color{trRet}Retrieve} (seeded)\\ text: the question}{\begin{minipage}[c]{\dimexpr\linewidth-1\trth-3pt\relax}\raggedright \trgold{}\#4 \texttt{img:\allowbreak{}30324469} Gary Payton and Jason Williams Gary Payton [...]\par + 4 other records\end{minipage}\hfill\trthumb{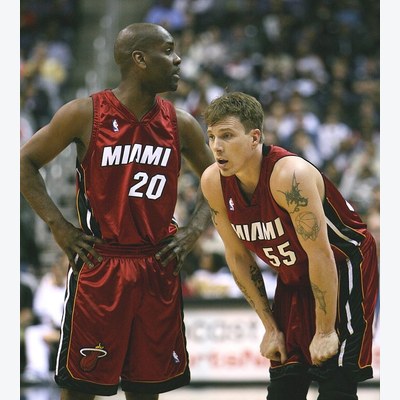}}{$K$\,9 \trdot{} R\,2 \trdot{} I\,3}
\tragent{``{\color{trIns}Identify the player in jersey number 55 from the image.}''}
\trcall{trInsP}{Tool: \textbf{\color{trIns}Inspect}\\ record: \texttt{img:\allowbreak{}30324469}}{\begin{minipage}[c]{\dimexpr\linewidth-1\trth-3pt\relax}\raggedright \trgold{}\texttt{img:\allowbreak{}30324469} Gary Payton and Jason Williams Gary Payton and [...]\end{minipage}\hfill\trthumb{figs/traces/t09_30324469.jpg}}{$K$\,8 \trdot{} R\,2 \trdot{} I\,2}
\tragent{``{\color{trRet}Confirm the player in jersey number 55 is Jason Williams.}''}
\trcall{trRet}{Tool: \textbf{\color{trRet}Retrieve} (text)\\ query: ``Jason Williams Miami Heat jersey number''}{\begin{minipage}[c]{\dimexpr\linewidth-1\trth-3pt\relax}\raggedright \trgold{}\#2 \texttt{doc:\allowbreak{}d5cbae54\_\allowbreak{}6} \underline{Jason Williams (basketball, born 1975)}: Jason Chandler Williams (born November 18, 1975) is an [...]\par \trgold{}\#3 \texttt{doc:\allowbreak{}d5cbae54\_\allowbreak{}3} \underline{Jason Williams (basketball, born 1975)}: Williams is a close friend of former Miami Heat teammate and [...]\par \trgold{}\#5 \texttt{img:\allowbreak{}30324469} Gary Payton and Jason Williams Gary Payton [...]\par + 2 other records\end{minipage}\hfill\trthumb{figs/traces/t09_30324469.jpg}}{$K$\,7 \trdot{} R\,1 \trdot{} I\,2}
\tragent{``{\color{trRet}Find the nickname Jason Williams was given due to his playing style.}''}
\trcall{trRet}{Tool: \textbf{\color{trRet}Retrieve} (text)\\ query: ``nickname Jason Williams playing style''}{\raggedright \trgold{}\#1 \texttt{doc:\allowbreak{}d5cbae54\_\allowbreak{}6} \underline{Jason Williams (basketball, born 1975)}: Jason Chandler Williams (born November 18, 1975) is an [...]\par \trgold{}\#2 \texttt{doc:\allowbreak{}d5cbae54\_\allowbreak{}15} \underline{Jason Williams (basketball, born 1975)}: In 2006, Williams won an NBA championship as the starting [...]\par \trgold{}\#3 \texttt{doc:\allowbreak{}d5cbae54\_\allowbreak{}3} \underline{Jason Williams (basketball, born 1975)}: Williams is a close friend of former Miami Heat teammate and [...]\par + 2 other records}{$K$\,6 \trdot{} R\,0 \trdot{} I\,2}
\tragent{``{\color{trIns}Confirm Jason Williams played as a point guard during the 2006 NBA championship.}''}
\trcall{trInsP}{Tool: \textbf{\color{trIns}Inspect} \trtimes{2}\\ record: \texttt{doc:\allowbreak{}d5cbae54\_\allowbreak{}15}, \texttt{doc:\allowbreak{}d5cbae54\_\allowbreak{}3}}{\raggedright \trgold{}\texttt{doc:\allowbreak{}d5cbae54\_\allowbreak{}15} \underline{Jason Williams (basketball, born 1975)}: In 2006, Williams won an NBA championship as the starting point guard for the Miami Heat. Due to his flashy style of play, and [...]\par \trgold{}\texttt{doc:\allowbreak{}d5cbae54\_\allowbreak{}3} \underline{Jason Williams (basketball, born 1975)}: Williams is a close friend of former Miami Heat teammate and center Shaquille O'Neal, having been neighbors in Orlando for three [...]}{$K$\,4 \trdot{} R\,0 \trdot{} I\,0}
\tragent{``Compile the final answer with all identified facts.''\par\vspace{1.5pt}\textbf{Answer:} \textbf{\color{trAns}The player in jersey number 55 for the Miami Heat is Jason Williams, who was nicknamed 'White Chocolate' due to his flashy playing style. He played as a point guard during the 2006 NBA championship, and his move to Miami was facilitated by former teammate Shaquille O'Neal.}\par\vspace{0.5pt}{\color{black!60}Cited: \texttt{img:\allowbreak{}30324469}, \texttt{doc:\allowbreak{}d5cbae54\_\allowbreak{}15}, \texttt{doc:\allowbreak{}d5cbae54\_\allowbreak{}3}}}
\trfooter{Reference: The player in jersey number 55 is Jason Williams, who was nicknamed 'White Chocolate', played point guard during the 2006 NBA championship, and whose move to Miami was facilitated by Shaquille O'Neal. \trdot{} Evaluator: correct (judge) \trdot{} gold recall 1.0 \trdot{} 6 of 10 tool calls \trdot{} example \texttt{mcsearch:\allowbreak{}2592}}
\end{trcasecol}
\hfill
\begin{trcasecol}[1]{Case study 6 of ViBES (MC-Search): a follow-up query returns the one gold passage that the seeded retrieval misses.}{fig:v2case6}
\truser{\begin{minipage}[c]{\dimexpr\linewidth-2.37cm\relax}\raggedright What architectural landmark, known as the ''Oculus,'' is shown in the image, what station is situated near its head house, how is this station ADA-accessible, and what further elevator connections does it offer?\end{minipage}\hfill\raisebox{-0.5\height}{\includegraphics[height=1.55cm]{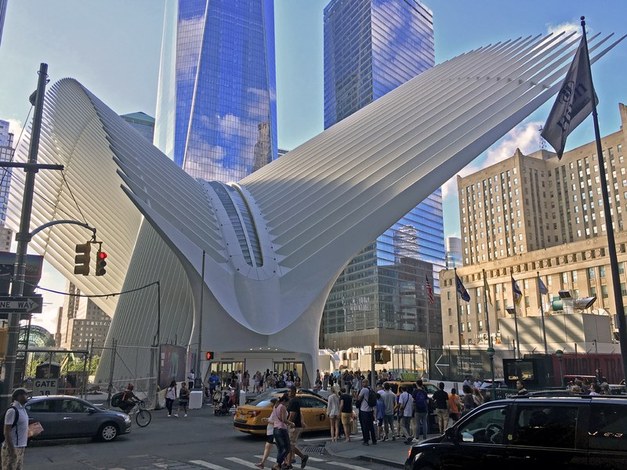}}}
\trcall{trRet}{Tool: \textbf{\color{trRet}Retrieve} (seeded)\\ multimodal: the question and the input image}{\raggedright \trgold{}\#1 \texttt{doc:\allowbreak{}d5cd705e\_\allowbreak{}4} \underline{WTC Cortlandt station}: The station is ADA-accessible via the elevator at Vesey [...]\par \trgold{}\#3 \texttt{doc:\allowbreak{}d5cd705e\_\allowbreak{}3} \underline{WTC Cortlandt station}: The station is ADA-accessible via the elevator at Vesey [...]\par + 3 other records}{$K$\,9 \trdot{} R\,2 \trdot{} I\,3}
\tragent{``{\color{trIns}Confirm the architectural landmark in the image is the Oculus.}''}
\trcall{trInsP}{Tool: \textbf{\color{trIns}Inspect}\\ record: \texttt{img:\allowbreak{}30028827}}{\begin{minipage}[c]{\dimexpr\linewidth-1\trth-3pt\relax}\raggedright \texttt{img:\allowbreak{}30028827} The WTC Transportation Hub The WTC Transportation [...]\end{minipage}\hfill\trthumb{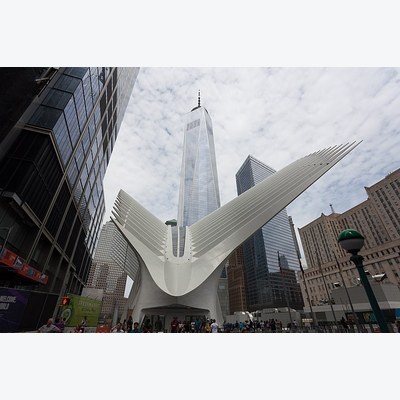}}{$K$\,8 \trdot{} R\,2 \trdot{} I\,2}
\tragent{``{\color{trRet}Identify the station situated near the Oculus's head house.}''}
\trcall{trRet}{Tool: \textbf{\color{trRet}Retrieve} (text)\\ query: ``station near WTC Transportation Hub head house''}{\raggedright \trgold{}\#2 \texttt{doc:\allowbreak{}d5cd705e\_\allowbreak{}5} \underline{WTC Cortlandt station}: The WTC Cortlandt station is located just west to the World [...]\par + 4 other records}{$K$\,7 \trdot{} R\,1 \trdot{} I\,2}
\tragent{``{\color{trIns}Confirm ADA accessibility of WTC Cortlandt station via elevator at Vesey Street.}''}
\trcall{trInsP}{Tool: \textbf{\color{trIns}Inspect} \trtimes{2}\\ record: \texttt{doc:\allowbreak{}d5cd705e\_\allowbreak{}4}, \texttt{doc:\allowbreak{}d5cd705e\_\allowbreak{}3}}{\raggedright \trgold{}\texttt{doc:\allowbreak{}d5cd705e\_\allowbreak{}4} \underline{WTC Cortlandt station}: The station is ADA-accessible via the elevator at Vesey Street, as well as existing elevators to the World Trade Center [...]\par \trgold{}\texttt{doc:\allowbreak{}d5cd705e\_\allowbreak{}3} \underline{WTC Cortlandt station}: The station is ADA-accessible via the elevator at Vesey Street, as well as existing elevators to the World Trade Center [...]}{$K$\,5 \trdot{} R\,1 \trdot{} I\,0}
\tragent{``Confirm the architectural landmark is the Oculus and the station is WTC Cortlandt.''\par\vspace{1.5pt}\textbf{Answer:} \textbf{\color{trAns}The architectural landmark shown is the Oculus, the head house of the World Trade Center Transportation Hub. The WTC Cortlandt station is situated near its head house. It is ADA-accessible via the elevator at Vesey Street and has additional elevators from each platform to the crossunder beneath the station.}\par\vspace{0.5pt}{\color{black!60}Cited: \texttt{img:\allowbreak{}30028827}, \texttt{doc:\allowbreak{}d5cb39e2\_\allowbreak{}10}, \texttt{doc:\allowbreak{}d5cd705e\_\allowbreak{}4}, \texttt{doc:\allowbreak{}d5cd705e\_\allowbreak{}3}}}
\trfooter{Reference: The image displays the World Trade Center Transportation Hub, also known as the ''Oculus.'' The WTC Cortlandt station is located west of its head house. This station is ADA-accessible via an elevator at Vesey [...] \trdot{} Evaluator: correct (judge) \trdot{} gold recall 1.0 \trdot{} 5 of 10 tool calls \trdot{} example \texttt{mcsearch:\allowbreak{}2967}}
\end{trcasecol}
\end{figure*}

\begin{figure*}[!p]
\begin{trcasecol}[1]{Case study 7 of ViBES (MultiModalQA): the seeded retrieval returns no gold record, and the agent's query ``Madrid'' returns the gold passage and image.}{fig:v2case7}
\truser{\raggedright The city from the 1949-50 La Liga with a poster featuring a tall missile shaped tower with a round cap is the capital of what country?}
\trcall{trRet}{Tool: \textbf{\color{trRet}Retrieve} (seeded)\\ text: the question}{\raggedright \#1 \texttt{doc:\allowbreak{}1b5ffff5} \underline{Torrespaña}: Torrespaña (literally ''Spain Tower'') is a steel-and-concrete [...]\par + 4 other records}{$K$\,9 \trdot{} R\,2 \trdot{} I\,3 \trdot{} C\,4}
\tragent{``{\color{trRet}Identify the city from the 1949-50 La Liga with a poster featuring a tall missile-shaped tower with a round cap.}''}
\trcall{trRet}{Tool: \textbf{\color{trRet}Retrieve} (text)\\ query: ``Madrid''}{\begin{minipage}[c]{\dimexpr\linewidth-1\trth-3pt\relax}\raggedright \trgold{}\#1 \texttt{doc:\allowbreak{}583bdf87} \underline{Madrid}: Madrid is the capital of Spain and the largest municipality [...]\par \trgold{}\#5 \texttt{img:\allowbreak{}5eb01238} Madrid\par + 3 other records\end{minipage}\hfill\trthumb{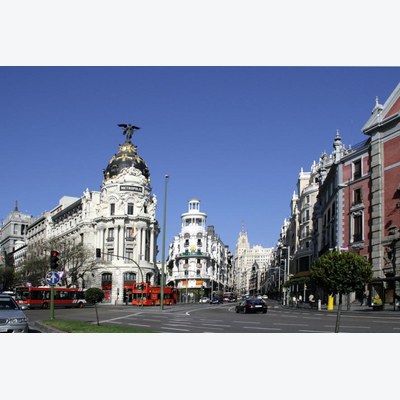}}{$K$\,8 \trdot{} R\,1 \trdot{} I\,3 \trdot{} C\,4}
\tragent{``{\color{trIns}Inspect the image of Madrid to confirm the presence of a tall missile-shaped tower with a round cap.}''}
\trcall{trInsP}{Tool: \textbf{\color{trIns}Inspect}\\ record: \texttt{img:\allowbreak{}5eb01238}}{\begin{minipage}[c]{\dimexpr\linewidth-1\trth-3pt\relax}\raggedright \trgold{}\texttt{img:\allowbreak{}5eb01238} Madrid\end{minipage}\hfill\trthumb{figs/traces/t13_mmqa_5eb012385a10b00b17dd46029050aea3.jpg}}{$K$\,7 \trdot{} R\,1 \trdot{} I\,2 \trdot{} C\,4}
\tragent{``{\color{trRet}Confirm that Madrid is the capital of Spain.}''}
\trcall{trRet}{Tool: \textbf{\color{trRet}Retrieve} (text)\\ query: ``capital of Spain''}{\begin{minipage}[c]{\dimexpr\linewidth-1\trth-3pt\relax}\raggedright \trgold{}\#1 \texttt{doc:\allowbreak{}583bdf87} \underline{Madrid}: Madrid is the capital of Spain and the largest municipality [...]\par \trgold{}\#5 \texttt{img:\allowbreak{}5eb01238} Madrid\par + 3 other records\end{minipage}\hfill\trthumb{figs/traces/t13_mmqa_5eb012385a10b00b17dd46029050aea3.jpg}}{$K$\,6 \trdot{} R\,0 \trdot{} I\,2 \trdot{} C\,4}
\tragent{``The city is Madrid, and it is the capital of Spain. Final answer.''\par\vspace{1.5pt}\textbf{Answer:} \textbf{\color{trAns}Spain}\par\vspace{0.5pt}{\color{black!60}Cited: \texttt{doc:\allowbreak{}583bdf87}, \texttt{img:\allowbreak{}5eb01238}, \texttt{img:\allowbreak{}6ed5317c}}}
\trfooter{Reference: Spain \trdot{} Evaluator: correct (official exact match) \trdot{} gold recall 1.0 \trdot{} 4 of 10 tool calls \trdot{} example \texttt{mmqa:\allowbreak{}95d0024e}}
\end{trcasecol}
\hfill
\begin{trcasecol}[1.12]{Case study 8 of ViBES (MultiModalQA): the seeded retrieval returns the gold passage naming the winner, a query on her name returns the gold image, and the agent crops the region of the hat.}{fig:v2case8}
\truser{\raggedright The winner of women's singles at wimbledon 2018 is wearing what type of hat?}
\trcall{trRet}{Tool: \textbf{\color{trRet}Retrieve} (seeded)\\ text: the question}{\raggedright \trgold{}\#3 \texttt{doc:\allowbreak{}497bcd5b} \underline{2018 Wimbledon Championships -- Women's Singles}: Angelique Kerber won her third Grand Slam singles title [...]\par + 4 other records}{$K$\,9 \trdot{} R\,2 \trdot{} I\,3 \trdot{} C\,4}
\tragent{``{\color{trRet}Identify the winner of women's singles at Wimbledon 2018 from the text results.}''}
\trcall{trRet}{Tool: \textbf{\color{trRet}Retrieve} (text)\\ query: ``Angelique Kerber''}{\begin{minipage}[c]{\dimexpr\linewidth-1\trth-3pt\relax}\raggedright \trgold{}\#4 \texttt{img:\allowbreak{}9dee9412} Angelique Kerber\par + 4 other records\end{minipage}\hfill\trthumb{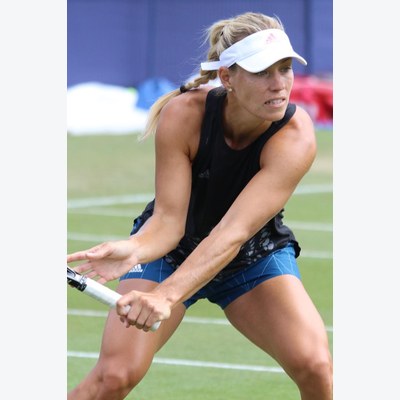}}{$K$\,8 \trdot{} R\,1 \trdot{} I\,3 \trdot{} C\,4}
\tragent{``{\color{trIns}Inspect the image of Angelique Kerber to identify the type of hat she is wearing.}''}
\trcall{trInsP}{Tool: \textbf{\color{trIns}Inspect}\\ record: \texttt{img:\allowbreak{}9dee9412}}{\begin{minipage}[c]{\dimexpr\linewidth-1\trth-3pt\relax}\raggedright \trgold{}\texttt{img:\allowbreak{}9dee9412} Angelique Kerber\end{minipage}\hfill\trthumb{figs/traces/t14_mmqa_9dee941260cd8ab74d030837e8ae9dc6.jpg}}{$K$\,7 \trdot{} R\,1 \trdot{} I\,2 \trdot{} C\,4}
\tragent{``{\color{trCrop}Crop the image to focus on the hat worn by Angelique Kerber.}''}
\trcall{trCrop}{Tool: \textbf{\color{trCrop}Crop}\\ \texttt{img:\allowbreak{}9dee9412}, box [499, 71, 906, 185]}{\raggedright\setlength{\baselineskip}{\dimexpr\trth+5pt\relax}\trthumb{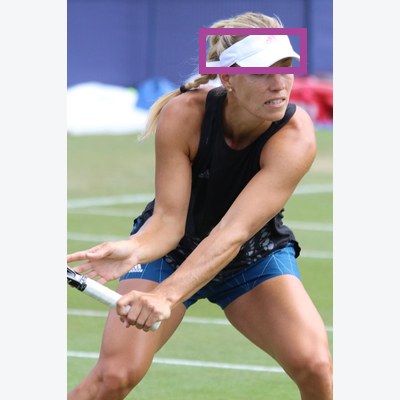}\trcut\trthumb{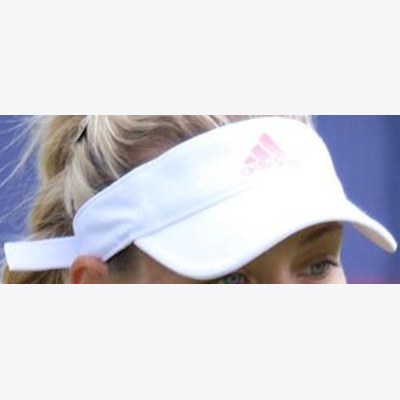}}{$K$\,6 \trdot{} R\,1 \trdot{} I\,2 \trdot{} C\,3}
\tragent{``The hat is a white visor with an Adidas logo. Confirm the type of hat.''\par\vspace{1.5pt}\textbf{Answer:} \textbf{\color{trAns}visor}\par\vspace{0.5pt}{\color{black!60}Cited: \texttt{img:\allowbreak{}9dee9412}, \texttt{crop:\allowbreak{}1}}}
\trfooter{Reference: visor \trdot{} Evaluator: correct (official exact match) \trdot{} gold recall 1.0 \trdot{} 4 of 10 tool calls \trdot{} example \texttt{mmqa:\allowbreak{}18010110}}
\end{trcasecol}
\end{figure*}

\FloatBarrier %

\end{document}